\documentclass[11pt,a4paper]{article}
\pdfoutput=1
\usepackage{amsfonts}
\usepackage{amssymb}
\usepackage[centertags]{amsmath}
\usepackage{graphicx}
\usepackage{hyperref}
\usepackage{booktabs}
\usepackage{theorem}
\usepackage{array}
\usepackage{epsfig}
\usepackage{colordvi}
\usepackage{color}
\usepackage{ulem}
\usepackage[small,bf]{caption}
\usepackage{cite}
\usepackage{color}
\usepackage{subfigure}
\usepackage{extpfeil}

\def\1{\mathbf{1}}
\def\3{\mathbf{3}}
\def\2{\mathbf{2}}

\def\pr{\prime}
\def\sec{\prime \prime}

\newcommand{\meff}{\mbox{$\left|  \langle \!\,  m  \,\!  \rangle \right| $}}
\newcommand{\betabeta}{\mbox{$(\beta \beta)_{0 \nu}  $}}
\def\ltap{\ \raisebox{-.4ex}{\rlap{$\sim$}} \raisebox{.4ex}{$<$}\ }

\allowdisplaybreaks[1]

\usepackage{a4wide}
\numberwithin{equation}{section}

\newcounter{mysubequation}[equation]

\newcommand{\bec}{\begin{cases}}
\newcommand{\eec}{\end{cases}}
\newcommand{\beq}{\begin{equation*}}
\newcommand{\eeq}{\end{equation*}}
\newcommand{\be}{\begin{equation}}
\newcommand{\ee}{\end{equation}}
\newcommand{\ba}{\begin{eqnarray}}
\newcommand{\ea}{\end{eqnarray}}

\newcommand{\si}{\textbf{1}}
\newcommand{\db}{\textbf{2}}
\newcommand{\tp}{\textbf{3}}
\newcommand{\TBM}{U_{\text{TBM}}}

\DeclareMathOperator{\diag}{Diag}
\DeclareMathOperator{\ci}{\text{i}}

\makeatletter

\newcommand{\Rmnum}[1]{\expandafter\@slowromancap\romannumeral #1@}
\makeatother

\begin{document}

\begin{titlepage}

\vspace*{-15mm}
\begin{flushright}
SISSA 56/2013/FISI\\
RM3-TH/13-10\\
TTP13-04
\end{flushright}
\vspace*{0.7cm}

\begin{center}
{ \bf\LARGE Generalised Geometrical CP Violation\\[1ex]
in a $\boldsymbol{T^{\prime}}$ Lepton Flavour Model}
\\[8mm]
Ivan Girardi$^{\, a,}$ \footnote{E-mail:
\texttt{igirardi@sissa.it}},
Aurora Meroni$^{\,a,b,}$
\footnote{E-mail: \texttt{ameroni@fis.uniroma3.it}},
S. T. Petcov$^{\, a,c,}$ \footnote{Also at:
 Institute of Nuclear Research and Nuclear Energy,
  Bulgarian Academy of Sciences, 1784 Sofia, Bulgaria.},
Martin~Spinrath$^{\,a,d,}$
\footnote{E-mail:
\texttt{martin.spinrath@kit.edu}}
\\[1mm]
\end{center}
\vspace*{0.50cm}
\centerline{$^{a}$ \it SISSA/INFN, Via Bonomea 265, I-34136 Trieste, Italy }
\vspace*{0.2cm}
\centerline{$^{b}$ \it  Dipartimento di Matematica e Fisica, 
Universit\`{a} di Roma Tre,}
\centerline{\it  Via della Vasca Navale 84, I-00146 Rome, and} 
\centerline{\it INFN, Laboratori Nazionali di Frascati, 
Via E. Fermi 40, I-00044 Frascati, Italy}
\vspace*{0.2cm}
\centerline{$^{c}$ \it Kavli IPMU (WPI), 
University of Tokyo, Tokyo, Japan}
\vspace*{0.2cm}
\centerline{$^{d}$ \it Institut f\"ur Theoretische Teilchenphysik, Karlsruhe Institute of Technology,}
\centerline{\it Engesserstra\ss{}e 7, D-76131 Karlsruhe, Germany}
\vspace*{1.20cm}

\begin{abstract}
\noindent
We analyse the interplay of generalised CP transformations
and the non-Abelian discrete  group $T^{\prime}$ and
use the semi-direct product $G_f= T^{\prime}\rtimes H_{\text{CP}}$,
as family symmetry acting in the lepton sector. The family symmetry
is shown to be spontaneously broken in a geometrical manner. In
the resulting flavour model, naturally small Majorana neutrino
masses for the light active neutrinos are
obtained through the type I see-saw mechanism.
The known masses of the charged leptons, lepton mixing angles
and the two neutrino mass squared differences
are reproduced by the model with a good accuracy.
The model allows for two neutrino mass spectra
with normal ordering (NO) and one with inverted ordering (IO).
For each of the three spectra the
absolute scale of neutrino masses is predicted
with relatively small uncertainty.
The value of the Dirac CP violation (CPV) phase
$\delta$ in the lepton mixing
matrix is predicted to be $\delta \cong \pi/2~{\rm or}~ 3\pi/2$.
Thus, the CP violating effects in neutrino oscillations
are predicted to be maximal (given the values of the
neutrino mixing angles) and experimentally observable.
We present also predictions
for the sum of the neutrino masses, 
for the Majorana CPV phases and
for the effective Majorana mass
in neutrinoless double beta decay.
The predictions of the model can be tested
in a variety of ongoing and future planned
neutrino experiments.
\end{abstract}

\end{titlepage}
\setcounter{footnote}{0}

\newpage

\section{Introduction}

  Understanding the origin of the patterns of neutrino
masses and mixing, emerging from the neutrino oscillation,
$^3H$ $\beta-$decay, cosmological, etc.\ data is one of the most
challenging problems in neutrino physics.
It is part of the more general fundamental problem
in particle physics of understanding the origins of
flavour, i.e., of the patterns
of the quark, charged lepton and neutrino masses
and of the quark and lepton mixing.

   At present we have compelling evidence for the
existence of mixing of three light massive neutrinos
$\nu_i$, $i=1,2,3$, in the weak charged lepton current
(see, e.g., \cite{PDG2012}). The masses $m_i$ of the three
light neutrinos $\nu_i$ do not exceed approximately
1 eV, $m_i \ltap 1$ eV, i.e., they are much smaller than the
masses of the charged leptons and quarks.
The three light neutrino mixing is described
(to a good approximation)
by the Pontecorvo, Maki, Nakagawa, Sakata (PMNS) $3\times 3$
unitary mixing matrix, $U_{\rm PMNS}$. In the widely used
standard parametrisation \cite{PDG2012},  $U_{\rm PMNS}$ is
expressed in terms of the solar, atmospheric and reactor
neutrino mixing angles $\theta_{12}$,  $\theta_{23}$ and
$\theta_{13}$, respectively, and one Dirac - $\delta$, and
two Majorana \cite{BHP80} - $\beta_{1}$ and $\beta_{2}$,
CP violation phases:
\begin{equation}
 U_{\text{PMNS}} \equiv U = V(\theta_{12},\theta_{23},\theta_{13},\delta)\,
Q(\beta_1,\beta_2)\,,
\label{eq:UPMNS}
\end{equation}
where
\begin{equation}
V = \begin{pmatrix}
       1 & 0 & 0 \\
       0 & c_{23} & s_{23} \\
       0 & -s_{23} & c_{23} \\
     \end{pmatrix}
     \begin{pmatrix}
       c_{13} & 0 & s_{13}e^{-\ci \delta} \\
       0 & 1 & 0 \\
       -s_{13}e^{\ci \delta} & 0 & c_{13} \\
     \end{pmatrix}
     \begin{pmatrix}
       c_{12} & s_{12} & 0 \\
       -s_{12} & c_{12} & 0 \\
       0 & 0 & 1 \\
     \end{pmatrix}\,,
\label{eq:V}
\end{equation}
%
and we have used the standard notation $c_{ij} \equiv
\cos\theta_{ij}$, $s_{ij} \equiv \sin\theta_{ij}$,
{\bf $0 \leq  \theta_{ij} \leq \pi/2$, $0 \leq \delta \leq 2\pi$.} 
The matrix $Q$ contains the two physical Majorana CP violation 
(CPV) phases:
\begin{equation}
Q = \diag(\text{e}^{-\ci \beta_1/2}, \text{e}^{-\ci \beta_2/2},1) \,.
\label{Q}
\end{equation}
%
The parametrization of the phase matrix $Q$
in Eq.~\eqref{Q} differs from the standard one \cite{PDG2012}
$Q = \diag(1, \text{e}^{\ci \alpha_{21}/2}, \text{e}^{\ci \alpha_{31}/2})$.
Obviously, one has $\alpha_{21} = (\beta_1 - \beta_2)$ and
$\alpha_{31} = \beta_1$. In the case of the seesaw mechanism of neutrino 
mass generation, which we are going to employ, the Majorana phases 
$\beta_1$ and $\beta_2$ (or $\alpha_{21}$ and $\alpha_{31}$) 
vary in the interval \cite{EMSPEJP09}
$0 \leq \beta_{1,2} \leq 4\pi$~ 
\footnote{The interval beyond $2\pi$, 
$2\pi \leq \beta_{1,2} \leq 4\pi$, is relevant, 
e.g., in the calculations of the baryon asymmetry within 
the leptogenesis scenario \cite{EMSPEJP09},  
in the calculation of the neutrinoless double beta decay 
effective Majorana mass in the TeV scale version of the 
type I seesaw model of neutrino mass generation 
\cite{Ibarra:2011xn}, etc.
}.
If CP invariance holds, we have
$\delta =0,\pi,2\pi$, and
\cite{LW81} $\beta_{1(2)} = k^{(')}\,\pi$, $k^{(')}=0,1,2,3,4$.

 All compelling neutrino oscillation data can be described
within the indicated 3-flavour neutrino mixing scheme.
These data allowed to determine
the angles $\theta_{12}$,  $\theta_{23}$ and
$\theta_{13}$ and the two neutrino mass squared
differences $\Delta m^2_{21}$ and
$\Delta m^2_{31}$ (or $\Delta m^2_{32}$),
which drive the observed oscillations involving
the three active flavour neutrinos and antineutrinos,
$\nu_l$ and $\bar{\nu}_l$, $l=e,\mu,\tau$,
with a relatively high precision
\cite{Fogli:2012ua,Gonzales-Garcia:2012}.
In Table 1 we give the values of the 3-flavour neutrino
oscillation parameters as determined in the global
analysis performed in \cite{Fogli:2012ua}.
\begin{table}
\centering
\renewcommand{\arraystretch}{1.1}
\begin{tabular}{lcc}
\toprule
 Parameter  &  best-fit ($\pm 1\sigma$) & 3$\sigma$ \\ \midrule
 $\Delta m^{2}_{21} \; [10^{-5}\text{ eV}^2]$  & 7.54$^{+0.26}_{-0.22}$ &
               6.99 - 8.18 \\
$ |\Delta m^{2}_{31}|~\text{ (NO)}  \; [10^{-3}\text{ eV}^2]$ &
   2.47$^{+0.06}_{-0.10}$  &
         2.19 - 2.62\\
$ |\Delta m^{2}_{32}|~\text{ (IO)}  \; [10^{-3}\text{ eV}^2]$   &
2.46$^{+0.07}_{-0.11}$
&  2.17 - 2.61\\
 $\sin^2\theta_{12}$ (NO or IO)  & 0.307$^{+0.018}_{-0.016}$
            & 0.259 - 0.359\\
$\sin^2\theta_{23}$  (NO) & 0.386$^{+0.024}_{-0.021}$ &  0.331 - 0.637 \\
\phantom{$\sin^2\theta_{23}$}  (IO)  &  0.392$^{+0.039}_{-0.022}$  & 0.335 - 0.663\\
$\sin^2\theta_{13}$ (NO)  &  0.0241$^{+0.0025}_{-0.0025}$  & 0.0169 - 0.0313\\
\phantom{$\sin^2\theta_{13}$}  (IO)  & 0.0244$^{+0.0023}_{-0.0025}$ & 0.0171 - 0.0315 \\
\bottomrule
\end{tabular}
\caption{The best-fit values and $3\sigma$
allowed ranges of the 3-flavour neutrino oscillation
parameters derived from a global fit of the current
neutrino oscillation data
(from \cite{Fogli:2012ua}).
If two values are given, the upper one corresponds to neutrino
mass spectrum with normal hierarchy (NO)
and the lower one - to spectrum with inverted hierarchy (IO)
(see text for further details).
}
 \label{tabNudata}
\end{table}
%

 An inspection of Table \ref{tabNudata}
shows that although $\theta_{13}\neq 0$, $\theta_{23} \neq \pi/4$
and $\theta_{12} \neq \pi/4$,
the deviations from these values are small, in fact
we have $\sin\theta_{13}\cong 0.16 \ll 1$,
$\pi/4 - \theta_{23} \cong 0.11$ and
\mbox{$\pi/4 - \theta_{12} \cong 0.20$},
where we have used the relevant best fit values in Table
\ref{tabNudata}.
The value of $\theta_{13}$ and the
magnitude of deviations of $\theta_{23}$
and $\theta_{12}$ from $\pi/4$
suggest that the observed values of
$\theta_{13}$, $\theta_{23}$ and
$\theta_{12}$ might originate from
certain ``symmetry'' values which
undergo relatively small (perturbative)
corrections as a result of
the corresponding symmetry breaking.
This idea was and continues to be
widely explored in attempts to
understand the pattern of
mixing in the lepton sector (see, e.g.,
\cite{GTani02,FPR04,SPWR04,Romanino:2004ww,HPR07,
Marzocca:2011dh,Marzocca:2013cr,Alta,Chao:2011sp}).
Given the fact that the PMNS matrix is a product
of two unitary matrices,
\begin{equation}
U = U_e^\dagger\, U_\nu\,,
\label{UeUnu1}
\end{equation}
%
where $U_e$ and $U_\nu$ result respectively from the diagonalisation
of the charged lepton and neutrino mass matrices, it is usually assumed that
$U_\nu$ has a specific form dictated by a symmetry which
fixes the values of the three mixing angles in
$U_\nu$ that would differ, in general, by perturbative
corrections from those measured in the
PMNS matrix, while $U_e$ (and symmetry breaking effects
that we assume to be subleading)
provide the requisite corrections.
A variety of potential ``symmetry'' forms of $U_\nu$,
have been explored in the literature
on the subject (see, e.g., \cite{AlbRode2010}).
Many of the phenomenologically acceptable
``symmetry'' forms of $U_\nu$,
as the tribimaximal (TBM) \cite{TBM} and
bimaximal (BM) \cite{SPPD82,BM} mixing,
can be obtained using discrete
flavour symmetries (see, e.g., the reviews
\cite{King:2013eh,Alta:2010ab,Tani:2010cd}
and the references quoted there in).
Discrete symmetries combined with GUT symmetries
have been used also in attempts to construct
realistic unified models of flavour
(see, e.g., \cite{King:2013eh}).

 In the present article we will exploit the
approximate flavour symmetry based on the group $T^{\pr}$,
which is the double covering of the better known
group $A_4$ (see, e.g., \cite{Tani:2010cd}),
with the aim to explain the observed pattern of
lepton (neutrino) mixing
and to obtain predictions for the CP violating phases
in the PMNS matrix and possibly
for the absolute neutrino mass scale and the type of
the neutrino mass spectrum.
Flavour models based on the discrete symmetry
$T^{\pr}$ have been proposed by a number of authors
\cite{Feruglio:2007uu,Chen:2007afa,Chen:2009gf}
before the angle $\theta_{13}$ was determined with a
high precision in the Day Bay \cite{DayaBay}
and RENO \cite{RENO} experiments (see also
\cite{Abe:2011sj,Abe:2011fz,Adamson:2011qu}).
All these models predicted values of $\theta_{13}$
which turned out to be much smaller
than the experimentally determined value.

In \cite{Chen:2007afa,Chen:2009gf}, in particular,
an attempt was made to construct a realistic
unified supersymmetric model of flavour,
based on the group $SU(5)\times T^{\pr}$,
which describes the quark masses, the quark mixing and
CP violation in the quark sector, the charged lepton
masses and the known mixing angles in the lepton sector,
and predicts the angle $\theta_{13}$ and possibly
the neutrino masses and the type of the neutrino
mass spectrum as well as the values of the
CPV phases in the PMNS matrix.
The light neutrino masses
are generated in the model  by the type I seesaw
mechanism \cite{seesaw} and are naturally small.
It was suggested in \cite{Chen:2007afa,Chen:2009gf}
that the complex Clebsch-Gordan (CG)
coefficients of $T^{\pr}$ \cite{cg} might be a
source of CP violation and hence that the CP symmetry
might be broken geometrically \cite{Branco:1983tn}
in models with approximate $T^{\pr}$ symmetry.
Since the phases of the CG coefficients of  $T^{\pr}$
are fixed, this leads to specific predictions for the
CPV phases in the quark and lepton mixing matrices.
Apart from the incorrect prediction for $\theta_{13}$,
the authors of \cite{Chen:2007afa, Chen:2009gf} did not
address the problem of vacuum alignment of the flavon vevs,
i.e., of demonstrating that the
flavon vevs, needed for the correct
description of the quark and lepton masses and
of the the mixing in both the quark and
lepton sectors, can be derived from a
flavon potential and that the latter
does not lead to additional
arbitrary flavon vev phases which
would destroy the predictivity, e.g., of
the leptonic CP violation of the model.

 A SUSY $SU(5)\times T^{\pr}$ model of flavour, which
reproduces the correct value of the
lepton mixing angle $\theta_{13}$
was proposed in \cite{Meroni:2012ty},
where the problem of  vacuum alignment
of the flavon vevs was also successfully
addressed
\footnote{A  modified version of
the model published in \cite{Chen:2007afa,Chen:2009gf},
which predicts a correct value of the angle
$\theta_{13}$, was constructed in \cite{Chen:2013wba},
but the authors of \cite{Chen:2013wba}
left open the issue of the vacuum alignment
of the flavon vevs.
}.
In \cite{Meroni:2012ty}
it was assumed that the CP violation in the quark
and lepton sectors originates  from
the complexity of the CG coefficients of $T^{\pr}$.
This was possible by fixing the phases
of the flavon vevs using the method of the so-called
``discrete vacuum alignment'',
which was advocated in \cite{Antusch:2011sx}
and used in a variety of other
models with discrete flavour symmetries
\cite{Antusch:2013wn}.
The value of the angle $\theta_{13}$
was generated by charged lepton corrections to
the TBM mixing using non-standard GUT relations
\cite{Marzocca:2011dh,Antusch:2011qg,Antusch:2009gu}.

  After the publication of  \cite{Meroni:2012ty} it was realised
in \cite{Holthausen:2012dk,Feruglio:2012cw} that the requirement
of CP invariance in the context of theories with discrete
flavour symmetries, imposed before the breaking of the discrete
symmetry leading  to CP nonconservation and generation of
the masses of the matter fields of the theory,
requires the introduction of the so-called
``generalised CP transformations'' of the
matter fields charged under the discrete symmetry.
The explicit form of the generalised CP transformations
is dictated by the type of the discrete symmetry.
It was noticed in \cite{Holthausen:2012dk}, in particular,
that due to a subtle intimate relation between CP symmetry and
certain discrete family symmetries, like the one associated with
the group $T^{\pr}$, it can happen that the CP symmetry
does not enforce the Yukawa type couplings, which generate
the matter field mass matrices after the symmetry breaking,
to be real but to have certain discrete phases predicted by
the family symmetry in combination with the
generalised CP transformations. In the
$SU(5)\times T^{\pr}$ model proposed in \cite{Meroni:2012ty},
these phases, in principle, can change or modify completely
the pattern of CP violation obtained by
exploiting the complexity of some of the  $T^{\pr}$
CG coefficients.

  In the present article we address the problem
of the relation between the $T^{\pr}$ symmetry and
the CP symmetry  in models of lepton flavour.
After some general remarks about the connection
between the  $T^{\pr}$ and  CP symmetries in
Section~\ref{Sec:TprandCP}, we present in Section~\ref{Sec:Model}
a  fully consistent and explicit model of
lepton flavour with a $T^{\pr}$ family symmetry and geometrical
CP violation. We show that the model reproduces correctly
the charged lepton masses, all leptonic mixing angles and
neutrino mass squared differences and predicts
the values of the leptonic CP violating phases and the neutrino mass
spectrum. We show also that this model indeed exhibits
geometrical CP violation. We clarify how the CP symmetry
is broken in the model
by using the explicit form of the constructed
flavon vacuum alignment sector;
without the knowledge of the flavon potential
it is impossible to make conclusions about
the origin of CP symmetry breaking in
flavour models with $T^{\pr}$ symmetry.
In the Appendix we give some technical details about
the group $T^{\pr}$ and present a ``UV completion'' of
the model, which is necessary in order to
to select correctly certain $T^{\pr}$
contractions in the relevant  effective operators.

\section{\texorpdfstring{$\boldsymbol
{T^{\prime}}$ Symmetry and Generalised CP Transformations}
{T prime and generalised CP}}
\label{Sec:TprandCP}

In this Section we would like to clarify the role of a generalised CP
transformation  combined with the non-Abelian discrete symmetry
group $T^{\prime}$. Let $G_f= T^{\prime}\rtimes H_{\text{CP}}$ be
the symmetry group acting in the lepton sector such that both
$T^{\prime}$ and $H_{\text{CP}}$ act on the lepton flavour space.
Motivated by this study we will present in the next section a model
where $G_f$ is broken such that all lepton mixing angles and
physical CP phases of the PMNS mixing matrix can be predicted in
terms of two mixing angles and two phases. The breaking of $G_f$
will be achieved through non zero vacuum expectation values (vevs) of
some scalar fields, the so-called flavons.

\subsection{The consistency conditions}

The discrete non-Abelian family symmetry group $T^{\prime}$ is the
double covering of the tetrahedral group $A_4$ and its complete
description in terms of generators, elements and representations is
given in Appendix \ref{App:groupTp}. An interesting feature of this
group is the fact that it is the smallest group that admits 1-, 2-,
and 3-dimensional representations and for which the three
representations can be related by  the multiplication rule $\db
\otimes \db = \tp \oplus \si$\footnote{
 The only other
24-element group that has representation of the same dimensions is
the octahedral group $O$ (which is isomorphic to $S_4$). In this
case, however, the product of two doublet reps does not contain a
triplet \cite{Aranda:2000tm}.}. $T^{\pr}$ has seven different
irreducible representations: the 1- and 3-dimensional
representations $\1$, $\1^{\pr}$, $\1^{\sec}$, $\3$ are not
faithful, i.e., not injective, while the doublet representations
$\2$, $\2^{\pr}$ and $\2^{\sec}$ are faithful. One interesting
feature of the $T^{\pr}$ group is related to the tensor products
involving the 2-dimensional representation since the CG
coefficients are complex.

We define now the transformation of a field $\phi(x)$ under the
group $T^{\pr}$ and $H_{\text{CP}}$ respectively as:
\begin{equation}
\phi(x) \rightarrow \rho_r(g) \phi(x), \qquad \phi(x)
\rightarrow X_r \phi^*(x^{\pr}),
\label{eq:fieldsCP}
\end{equation}
where $\rho_r(g)$ is an irreducible representation $r$ of the group
element $g \in T^{\pr}$, $x^{\pr} \equiv (x^0,-\vec{x})$ and $X_r$
is the unitary matrix representing the generalised CP transformation.
In order to
introduce consistently the CP transformation for the family symmetry
group $T^{\pr}$, the matrix $X_r$ should satisfy the consistency
conditions \cite{Holthausen:2012dk,Feruglio:2012cw,Ding:2013bpa}:
\begin{equation}
X_r \rho_r^*(g) X_r^{-1} = \rho_r(g^{\pr}) \,,\quad g,g^{\pr}
\in T^{\pr}\,. \label{consistency}
\end{equation}

Following the discussion given in
\cite{Holthausen:2012dk,Feruglio:2012cw,Ding:2013bpa} it is important to remark that
the consistency condition corresponds to a similarity transformation
between the representation $\rho_r^*$ and
$\rho \circ \text{CP}$. 
Since the structure of the group is preserved and an element $g \in
T^{\pr}$ is always mapped into an element $g^{\pr} \in T^{\pr}$,
this map defines an automorphism of the group. In general $g$ and
$g^{\pr}$  might belong to different conjugacy classes: in this case
the map defines an outer automorphism \footnote{For details
concerning the group of outer and inner automorphisms,
$\text{Out}(G)$ and $\text{Inn}(G)$, see
\cite{Holthausen:2012dk, Ding:2013bpa}.}.

It is worth noticing that the matrices $X_r$ are defined up to
an arbitrary global  phase.
Indeed, without loss of generality, for each matrix  $X_r$, one can
define different phases $\theta_r$ for different irreducible
representations and moreover one can define $X_r$ up to a group
transformation (change of basis): in
fact the consistency conditions in Eq.~(\ref{consistency})  are
invariant under  $X_r \rightarrow \text{e}^{\ci \theta_r} X_r$ and $X_r
\rightarrow \rho_r(\tilde g) X_r$ with $\tilde g \in T^{\pr}$.

 It proves convenient to use the  freedom associated with the
arbitrary phases $\theta_r$  to define  the  generalised CP
transformation for which the vev alignments of the flavon fields can
be chosen to be all real. We will show later on that the phases
$\theta_r$ are not physical and therefore the results we present are
independent from the specific values we assume.  In the context of
the $T^{\pr}$ group this choice however helps us to extract a real
flavon vev structure which  is a distinctive feature of some models
proposed in the literature 
where the origin of the physical CP violation arising in the lepton
sector is tightly related to the combination of real vevs, complex
CGs \footnote{This idea  was pioneered in \cite{Chen:2007afa}.} and
eventual phases arising from the requirement of invariance of the
superpotential under the generalised CP transformation.

Before going into details of the computations, let us comment  that
in the analysis presented in \cite{Holthausen:2012dk} related to the
group $T^{\pr}$,  the CP transformations are defined as  elements of
the outer automorphism group and  are derived up to inner
automorphisms of $T^{\pr}$ (up to conjugacy transformations). In the
present work we will consider instead  all the possible
transformations including the inner automorphism group and
we will discuss all the convenient CP transformations which
can be used to clarify the role of a generalised CP symmetry in the
context of the group $T^{\pr}$.

\subsection{Transformation properties under generalised CP}

We give now  all the possible equivalent choices of generalised CP
transformations for any irreducible representation of $T^{\pr}$.

The group $T^{\pr}$ is defined by the group generators $T$ and $S$,
then from the consistency conditions in Eq.~\eqref{consistency} it
is sufficient to require that
\begin{equation}
X_r \rho_r^*(S) X_r^{-1} = \rho_r(\hat S)\,, \qquad X_r
\rho_r^*(T) X_r^{-1} = \rho_r(\hat T)\,.
\end{equation}

It is easy  to show that the CP transformation leaves invariant the
order of the element of the group $g$ meaning  that denoting $n(g)$
the order of $g$, we have $n(g) = n(g^{\pr})$. Since the element $S$
has order four and the element $T$ has order three we have $\hat S
\in 6\,C^4$ and $\hat T \in 4^{\pr} C^3$ \cite{Grimus:2011fk}. The
latter result is derived using the action of CP on the
one-dimensional representations, i.e.~$\rho_{1,1^{\pr},1^{\sec}}(\hat T) =
\rho^*_{1,1^{\pr},1^{\sec}}(T)$ which can be satisfied only if
$\hat T \in 4^{\pr} C^3$.

The conjugacy classes $6\,C^4$ and $4^{\pr} C^3$ contain
the group elements
\begin{equation}
\begin{split}
\hat S \in 6\,C^4 & = \left \{S,S^3,T S T^2,T^2 S T, S^2 T S T^2, S^2 T^2 S T \right \} \,, \\
\hat T \in 4^{\pr} C^3 & = \left \{ T^2, S^2 T S T, S^2 T^2 S, S^3 T^2 \right \} \, .\\
\end{split}
\end{equation}

We recall that we have the freedom to choose arbitrary phases
$\theta_r$, so for instance in the case of $X_{1}$, $X_{1^{\pr}}$
and $X_{1^{\sec}}$ we are allowed to write the most general CP
transformations for the three inequivalent singlets of $T^{\pr}$ as
\begin{equation}
\mathbf{1} \rightarrow  \text{e}^{\ci \theta_{1}} \mathbf{1}^* \;, \quad
\mathbf{1^{\prime}} \rightarrow \text{e}^{\ci \theta_{1^{\pr}}} \mathbf{1^{\prime}}^* \;, \quad
\mathbf{1^{\prime \prime}} \rightarrow \text{e}^{\ci \theta_{1^{\sec}}} \mathbf{1^{\prime \prime}}^* \,.
\end{equation}

Differently from the case of the $A_4$ family symmetry discussed in
\cite{Ding:2013bpa} in which one can  show that the generalised CP
transformation can be represented as a group transformation,
in the case of $T^{\pr}$ we will show that this is true only for the
singlet and the triplet representations. For the doublets
the action of the CP transformation cannot be written as an action
of a group element (i.e. $\nexists \,\, g \in T^{\prime}$ such that
$X_r = \rho_r(g)$ for $r = \2, \2^{\pr}, \2^{\sec}$).

We give a list of all the possible forms of $X_r$, which can be in
general different for each representation: the CP transformations on
the singlets, $X_{\1,\1^{\pr},\1^{\sec}}$, are complex phases,
as mentioned above while
the CP transformations on the doublets, $X_{\2,\2^{\pr},\2^{\sec}}$,
and the  triplets $X_{\3}$, are given respectively in Table
\ref{CPtriplets} and \ref{CPdoublets}. We stress that all the
possible forms of $X_r$ are defined up to a phase, which can be in
general different for each representation.
Each CP transformation we found generates a $Z_2$
symmetry.

\begin{table}
\centering
\resizebox{\textwidth}{!}{
\begin{tabular}{lcccc}
\toprule
 $g,g^{\pr}$ & $X_\3 = \rho_3(g) = \rho_3(g^{\pr}) $  & $T \rightarrow \hat T$ & $S \rightarrow \hat S$ \\
\midrule
$T$, $S^2 T$ &
$\left (
\begin{array}{ccc}
 1 & 0 & 0 \\
 0 & \omega & 0 \\
 0 & 0 & \omega^2 \\
\end{array}
\right)$
& $T^2$ & $S^3$ \\
\midrule
$T^2$, $S^2 T^2$ &
$\left (
\begin{array}{ccc}
 1 & 0 & 0 \\
 0 & \omega^2 & 0 \\
 0 & 0 & \omega \\
\end{array}
\right)$
& $T^2$ & $S^2 T S T^2 $ \\
\midrule
$E$, $S^2$ &
$\left (
\begin{array}{ccc}
 1 & 0 & 0 \\
 0 & 1 & 0 \\
 0 & 0 & 1 \\
\end{array}
\right)$
& $T^2$ & $S^2 T^2 S T $ \\
\midrule
$T S$, $S^2 T S$ &
$\left(
\begin{array}{ccc}
 -1/3 \, & 2/3 \, \omega & 2/3 \, \omega^2 \\
 2/3 \,  & -1/3 \, \omega  & 2/3 \,  \omega^2 \\
 2/3 \, & 2/3 \,  \omega& -1/3 \,  \omega^2 \\
\end{array}
\right)$
& $S^2 T^2 S$ & $S$ \\
\midrule
$T S T^2$, $S^2 T S T^2$ &
$\left(
\begin{array}{ccc}
 -1/3 \, & 2/3 \,  & 2/3 \,  \\
 2/3 \,  & -1/3 \,  & 2/3 \,  \\
 2/3 \, & 2/3 \, & -1/3 \,  \\
\end{array}
\right)$
& $S^2 T^2 S$ & $T^2 S T$ \\
\midrule
$S^2 T S T$, $T S T$ &
$\left(
\begin{array}{ccc}
 -1/3 \, & 2/3 \, \omega^2 & 2/3 \, \omega \\
 2/3 \,  & -1/3 \, \omega^2 & 2/3 \, \omega \\
 2/3 \, & 2/3 \, \omega^2& -1/3 \, \omega \\
\end{array}
\right)$
& $S^2 T^2 S$ & $S^2 T S T^2$ \\
\midrule
$ST$, $S^3 T$ &
$\left(
\begin{array}{ccc}
 -1/3 \, & 2/3 \, \omega^2 & 2/3 \, \omega \\
 2/3 \, \omega^2 & -1/3 \, \omega & 2/3 \, \\
 2/3 \, \omega & 2/3 \, & -1/3 \, \omega^2 \\
\end{array}
\right)$
& $S^2 T S T$ & $S^3$\\
\midrule
$S^3 T^2$, $S T^2$ &
$\left(
\begin{array}{ccc}
 -1/3 \, & 2/3 \,  & 2/3 \,  \\
 2/3 \, \omega^2 & -1/3 \,  \omega^2 & 2/3 \,  \omega^2 \\
 2/3 \, \omega & 2/3 \,  \omega & -1/3 \, \omega \\
\end{array}
\right)$
& $S^2 T S T$ & $T S T^2$\\
\midrule
$S$, $S^3$ &
$\left(
\begin{array}{ccc}
 -1/3 \, & 2/3 \, \omega & 2/3 \, \omega^2 \\
 2/3 \, \omega^2 & -1/3 \,  & 2/3 \,  \omega \\
 2/3 \, \omega & 2/3 \,  \omega^2 & -1/3 \,  \\
\end{array}
\right)$
& $S^2 T S T$ & $T^2 S T$\\
\midrule
$S^3 T S $, $S T S$ &
$\frac{1}{9} \left(
\begin{array}{ccc}
 4 \omega ^2+ 4 \omega +1 & - 2 \omega ^2 - 2 \omega + 4 &
   - 2 \omega ^2 - 2 \omega + 4  \\
 - 2 \omega ^2 + 4 \omega -  2 & 4 \omega ^2 + \omega + 4 & - 2
   \omega ^2 + 4 \omega - 2 \\
 4 \omega ^2 - 2 \omega - 2  & 4 \omega ^2 - 2 \omega  - 2 &
   \omega ^2 + 4 \omega + 4 \\
\end{array}
\right)$
& $S^3 T^2$ & $S$\\
\midrule
$S^2 T^2 S $, $T^2 S$ &
$\left(
\begin{array}{ccc}
 -1/3 \, & 2/3 \, \omega & 2/3 \, \omega^2 \\
 2/3 \, \omega & -1/3 \, \omega^2 & 2/3 \,  \\
 2/3 \, \omega^2 & 2/3 \, & -1/3 \, \omega \\
\end{array}
\right)$
& $S^3 T^2$ & $T S T^2$\\
\midrule
$T^2 S T$, $S^2 T^2 S T$ &
$\left(
\begin{array}{ccc}
 -1/3 \, & 2/3 \, \omega^2 & 2/3 \, \omega \\
 2/3 \, \omega & -1/3 \,  & 2/3 \, \omega^2 \\
 2/3 \, \omega^2 & 2/3 \, \omega & -1/3 \, \\
\end{array}
\right)$
& $S^3 T^2$ & $S^2 T^2 S T$\\
\bottomrule
\end{tabular}
}
\caption{The generalised CP transformation for the triplet representation of the group $T^{\pr}$ derived using the consistency conditions. We have defined $\omega = \text{e}^{\ci 2 \pi/3}$.
\label{CPtriplets}}
\end{table}

\begin{table}
\centering \resizebox{15cm}{!}{
\begin{tabular}{lccclcc}
\cmidrule[\heavyrulewidth]{1-3}  \cmidrule[\heavyrulewidth]{5-7}
 $X_\2$, $X_{\2^{\pr}}$, $X_{\2^{\sec}}$  & $T \rightarrow \hat T$ & $S \rightarrow \hat S$ &  & $X_\2$, $X_{\2^{\pr}}$, $X_{\2^{\sec}}$  & $T \rightarrow \hat T$ & $S \rightarrow \hat S$\\
\cmidrule{1-3}  \cmidrule{5-7}
$\left (
\begin{array}{cc}
 \bar p^5 & 0 \\
 0 & p^5 \\
\end{array}
\right)$ & $T^2$ & $S^3$
&& $\sqrt{\frac{2}{3}} \left (
\begin{array}{cc}
 \bar p^5/\sqrt{2} & 1 \\
 1 &  \bar p^7/\sqrt{2} \\
\end{array}
\right)$
& $S^2 T S T$ & $S^3$\\
\cmidrule{1-3}  \cmidrule{5-7}
$\left (
\begin{array}{cc}
 \bar p& 0 \\
 0 & p \\
\end{array}
\right)$ & $T^2$ & $S^2 T S T^2 $
&& $\frac{1}{\sqrt{3}}\left (
\begin{array}{cc}
 p^5& \sqrt{2} q\\
 \sqrt{2} q^5 & \bar p^5 \\
\end{array}
\right)$
& $S^2 T S T$ & $T S T^2$\\
\cmidrule{1-3}  \cmidrule{5-7}
$\left (
\begin{array}{cc}
e^{\ci \pi/4} & 0 \\
 0 & \text{e}^{-i \pi/4} \\
\end{array}
\right)$ & $T^2$ & $S^2 T^2 S T $
&& $\left (
\begin{array}{cc}
e^{\ci \pi/4} & \sqrt{2} q^5 \\
 \sqrt{2} q & \text{e}^{-i \pi/4} \\
\end{array}
\right)$
& $S^2 T S T$ & $T^2 S T$\\
\cmidrule{1-3}  \cmidrule{5-7}
$\left (
\begin{array}{cc}
p & \sqrt{2} \bar q^5 \\
 \sqrt{2} \bar q & \bar p \\
\end{array}
\right)$ & $S^2 T^2 S$ & $S$
&& $\frac{1}{\sqrt{3}}\left (
\begin{array}{cc}
p & \sqrt{2} \bar q \\
 \sqrt{2} \bar q^5 & \bar p \\
\end{array}
\right)$
& $S^3 T^2$ & $S$\\
\cmidrule{1-3}  \cmidrule{5-7}
$ \frac{1}{\sqrt{3}} \left (
\begin{array}{cc}
e^{-i \pi/4} & - i \sqrt{2} \\
 - i \sqrt{2} & \text{e}^{\ci \pi/4} \\
\end{array}
\right)$ & $S^2 T^2 S$ & $T^2 S T$
&& $\frac{1}{\sqrt{3}} \left (
\begin{array}{cc}
\bar p & \sqrt{2} \\
 \sqrt{2} & p \\
\end{array}
\right)$
& $S^3 T^2$ & $T S T^2$\\
\cmidrule{1-3}  \cmidrule{5-7}
$ \sqrt{\frac{2}{3}} \left (
\begin{array}{cc}
q^4 / \sqrt{2} & \bar p \\
 p^5 & \bar q / \sqrt{2} \\
\end{array}
\right)$ & $S^2 T^2 S$ & $S^2 T S T^2$
&& $ \sqrt{\frac{2}{3}} \left
(
\begin{array}{cc}
e^{\ci 5 \pi/8} / \sqrt{2} & \text{e}^{-i \pi/24} \\
 \text{e}^{-i 7 \pi/24}& \text{e}^{-i 7 \pi/8} / \sqrt{2}  \\
\end{array}
\right)$
& $S^3 T^2$ & $S^2 T^2 S T$\\
\cmidrule[\heavyrulewidth]{1-3}  \cmidrule[\heavyrulewidth]{5-7}
\end{tabular}
} \caption{The generalised CP transformation for the doublet
representation of the group $T^{\pr}$ derived using the consistency
conditions. We have defined $\omega = \text{e}^{\ci 2 \pi/3}$, $p =
\text{e}^{\ci \pi/12}$, $q = \text{e}^{\ci \pi/6}$ and note that
$\omega \bar p^5 = \text{e}^{\ci \pi /4}$. \label{CPdoublets}}
\end{table}

The generalised CP transformation $H_{\text{CP}}$, acting on the
lepton flavour space is given by, see also \cite{Holthausen:2012dk},
\begin{equation}
u : \begin{cases}
T \rightarrow T^2\;,\\
S \rightarrow S^2 T^2 S T \;.\\
\end{cases}
\label{eq:CPu}
\end{equation}
This definition of the CP symmetry is particularly
convenient because it acts on the 3- and 1- dimensional
representations trivially. This particular transformation however is
related to any other possible CP transformation by a group
transformation.

In other words, different choices of CP are
related to each other by inner automorphisms of the group i.e. the
CP transformations listed in Tables~\ref{CPtriplets} and
\ref{CPdoublets} are related to each other through a conjugation
with a group element. For example, another possible CP transformation
would be
\begin{equation}
 v : \begin{cases}
T \rightarrow T^2\;,\\
S \rightarrow S^3\;,\\
\end{cases}
\end{equation}
which is related to $u$ via $u =
conj(T^2) \, \circ \, v$. Indeed
\be
\begin{split}
& S \xmapsto{v} S^3 \xmapsto{conj(T^2)}  T^2 S^3 (T^2)^{-1}  = S^2 T^2 S T \;,\\
& T \xmapsto{v} T^2 \xmapsto{conj(T^2)} T^2 T^2 (T^2)^{-1}  = T^2 \;.\\
\end{split}
\ee

Without loss
of generality we choose as CP transformation the one defined through
Eq.~\eqref{eq:CPu} and from Eq.~\eqref{eq:fieldsCP} using the
results of Table~\ref{CPtriplets} and Table~\ref{CPdoublets}
we can write the representation of the CP transformation acting on
the fields as
\begin{gather}
 \mathbf{1} \rightarrow  \text{e}^{\ci \theta_{1}} \mathbf{1}^* \;,\quad
  \mathbf{1^{\prime}} \rightarrow \text{e}^{\ci \theta_{1^{\pr}}} \mathbf{1^{\prime}}^* \;, \quad
  \mathbf{1^{\prime \prime}} \rightarrow \text{e}^{\ci \theta_{1^{\sec}}} \mathbf{1^{\prime \prime}}^* \;, \quad
  \mathbf{3} \rightarrow \text{e}^{\ci \theta_{3}} \mathbf{3}^* \;, \label{CPgeneral}\\
 \mathbf{2} \rightarrow \text{e}^{\ci \theta_{2}}  \begin{pmatrix} \omega \bar p^{5} & 0 \\ 0 & \bar \omega p^5 \end{pmatrix} \mathbf{2}^*\;, \quad
  \mathbf{2^{\prime}} \rightarrow \text{e}^{\ci \theta_{2^{\pr}}} \begin{pmatrix} \omega \bar p^{5} & 0 \\ 0 & \bar \omega p^5 \end{pmatrix} \mathbf{2^{\prime}}^*\;, \quad
  \mathbf{2^{\sec}} \rightarrow \text{e}^{\ci \theta_{2^{\sec}}} \begin{pmatrix} \omega \bar p^{5} & 0 \\ 0 & \bar \omega p^5 \end{pmatrix} \mathbf{2^{\sec}}^* \;, \nonumber
\end{gather}
where $\omega = \text{e}^{\ci 2 \pi /3}$, $p = \text{e}^{\ci \pi / 12}$
and $\omega \bar p^5 = \text{e}^{\ci \pi/4}$.
Notice that we did not specify the values of the phases $\theta_r$.
Further we can check that the CP symmetry transformation chosen
generates a $Z_2$ symmetry group. Indeed it is easy to show
that $u ^ 2 = E$, therefore the multiplication table of the group
$H_{\text{CP}} = \{ E,\,u\}$ is obviously equal to the
multiplication table of a $Z_2$ group, from which we can
write $H_{\text{CP}}  \cong Z_2$.

Since we want to have real flavon vevs  -- following
the setup given in \cite{Meroni:2012ty} --  it turns out to be
convenient to select the CP transformations with
$\theta_1 = \theta_{1^{\pr}} = \theta_{1^{\sec}} = \theta_3 = 0$
and $\theta_{2^{\sec}} = -\theta_{2^{\pr}} = \pi/4$
\footnote{Since in our model later on we do not have fields
in a $\2$ representation of $T^{\prime}$ the phase $\theta_2$
is irrelevant in our further discussion and we do not fix its
value. A possible convenient choice might be $\theta_{2} = 0$
which makes the mass term of a two-dimensional representation real.}.
With this choice the phases of the couplings of renormalisable 
operators is fixed up to a sign by the CP symmetry. In fact,
supposing one has a renormalisable operator of the form $\lambda
\mathcal{O} = \lambda (A \times B \times C)$ where $\lambda$ is the
coupling constant and $A$, $B$, $C$ represent the fields, then the
generalised CP phase of the operator is defined as $\beta \equiv
\text{CP}[\mathcal{O}]/\mathcal{O}^*$. The phase of $\lambda$
is hence given by the equation $\lambda = \beta \lambda^*$ which is solved
by
\begin{equation}
\begin{cases}
\arg(\lambda) = \arg(\beta)/2 \,\,\, \mbox{or} \,\,\,
\arg(\beta)/2 - \pi \hspace{1cm} \quad \mbox{if} \,\,\,
\arg(\beta) > 0 \;,\\
\arg(\lambda) = \arg(\beta)/2 \,\,\, \mbox{or} \,\,\,
\arg(\beta)/2 + \pi \hspace{1cm} \quad \mbox{if} \,\,\,
\arg(\beta) \leq 0 \;.\\
\end{cases}
\end{equation}
In Table~\ref{tab:GenCP} we give a list of the
phases of $\lambda$ for all renormalisable operators
without fixing the $\theta_r$ and with the above
choice for $\theta_r$ in Table~\ref{tab:GenCPgen}.

\begin{table}
\centering
\begin{tabular}{lc}
\toprule
$\lambda \mathcal{O} = \lambda (A \times B \times C)$ & $\beta \equiv \text{CP}[\mathcal{O}]/\mathcal{O}^*$  \\
\midrule
($\mathbf{2} \times \mathbf{2})_{\mathbf{1}}
\times \mathbf{1}$  & $\text{e}^{\ci (\theta_1 + 2 \theta_2)}$ \\
$(\mathbf{2^{\prime}} \times
\mathbf{2^{\prime \prime}})_{\mathbf{1}} \times
\mathbf{1}$ & $\text{e}^{\ci (\theta_1 +
\theta_{2^{\prime}} + \theta_{2^{\sec}} )}$  \\
($\mathbf{2^{\prime}} \times
\mathbf{2^{\prime}})_{\mathbf{1^{\prime \prime}}}
 \times \mathbf{1^{\prime}}$  &
 $\text{e}^{\ci (\theta_{1^{\pr}} + 2 \theta_{2^{\pr}})}$   \\
$(\mathbf{2} \times
\mathbf{2^{\prime \prime}})_{\mathbf{1^{\prime \prime}}}
\times \mathbf{1^{\prime}}$ &
$\text{e}^{\ci (\theta_{1^{\pr}} +  \theta_2 + \theta_{2^{\sec}} )}$  \\
($\mathbf{2^{\prime \prime}} \times
\mathbf{2^{\prime \prime}})_{\mathbf{1^{\prime}}}
\times \mathbf{1^{\prime \prime}}$  &
$\text{e}^{\ci (\theta_{1^{\sec}} + 2 \theta_{2^{\sec}})}$  \\
$(\mathbf{2} \times \mathbf{2^{\prime}})_{\mathbf{1^{\prime}}}
\times \mathbf{1^{\prime \prime}}$ &
$\text{e}^{\ci (\theta_{1^{\sec}} + \theta_2 +  \theta_{2^{\pr}})}$  \\
\midrule
$\left[ (\mathbf{2} \times \mathbf{2})_{\mathbf{3}} \times
\mathbf{3} \right]_{\mathbf{1}}$  &
$-\ci$ $\text{e}^{\ci ( 2 \theta_2 + \theta_3 ) }$ \\
$\left[ (\mathbf{2^{\prime}} \times
\mathbf{2^{\prime \prime}})_{\mathbf{3}} \times \mathbf{3}
\right]_{\mathbf{1}}$ & $-\ci$
$\text{e}^{\ci (\theta_{2^{\prime}} + \theta_{2^{\sec}} + \theta_3  )}$ \\
$\left[(\mathbf{2^{\prime}} \times
\mathbf{2^{\prime}})_{\mathbf{3}}
 \times \mathbf{3} \right]_{\mathbf{1}}$  &
 $-\ci$ $\text{e}^{\ci (2 \theta_{2^{\pr}} + \theta_3 )}$   \\
$\left[ (\mathbf{2} \times
\mathbf{2^{\prime \prime}})_{\mathbf{3}}
\times \mathbf{3} \right]_{\mathbf{1}} $ & $-\ci$
$\text{e}^{\ci (\theta_2 + \theta_{2^{\sec}} + \theta_3 )}$  \\
$\left[(\mathbf{2^{\prime \prime}} \times
\mathbf{2^{\prime \prime}})_{\mathbf{3}}
\times \mathbf{3}  \right]_{\mathbf{1}}$  &
$-\ci$ $\text{e}^{\ci (2 \theta_{2^{\sec}} + \theta_3 )}$   \\
$\left[ (\mathbf{2} \times
\mathbf{2^{\prime}})_{\mathbf{3}} \times
\mathbf{3} \right]_{\mathbf{1}} $ & $-\ci$
 $\text{e}^{\ci (\theta_2 +  \theta_{2^{\pr}} + \theta_3 )}$  \\
\midrule
$\left[ (\mathbf{3} \times
\mathbf{3})_{\mathbf{1}} \times \mathbf{1}
\right]_{\mathbf{1}} $ & $\text{e}^{\ci (\theta_1 + 2 \theta_3) }$   \\
$\left[ (\mathbf{3} \times
 \mathbf{3})_{\mathbf{1^{\pr}}} \times
 \mathbf{1^{\sec}} \right]_{\mathbf{1}} $ &
 $\text{e}^{\ci (\theta_{1^{\sec}} + 2 \theta_3 )}$   \\
$\left[ (\mathbf{3} \times
\mathbf{3})_{\mathbf{1^{\sec}}}
\times \mathbf{1^{\pr}} \right]_{\mathbf{1}} $ &
$\text{e}^{\ci (\theta_{1^{\pr}} + 2 \theta_3 )}$   \\
$\left[ (\mathbf{3} \times
\mathbf{3})_{\mathbf{3_S \,,\, 3_A}}
\times \mathbf{3} \right]_{\mathbf{1}} $ & $\text{e}^{3 i \theta_3}$  \\
\midrule
$\left[ (\mathbf{1} \times \mathbf{1})_{\mathbf{1}}
\times \mathbf{1} \right]_{\mathbf{1}} $ &
$\text{e}^{\ci (3 \theta_{1})}$   \\
$\left[ (\mathbf{1^{\pr}}
\times \mathbf{1^{\pr}})_{\mathbf{1^{\sec}}}
\times \mathbf{1^{\pr}} \right]_{\mathbf{1}} $ &
$\text{e}^{\ci (3 \theta_{1^{\pr}})}$   \\
$\left[ (\mathbf{1^{\sec}}
 \times \mathbf{1^{\sec}})_{\mathbf{1^{\pr}}}
\times \mathbf{1^{\sec}} \right]_{\mathbf{1}} $ &
$\text{e}^{\ci (3 \theta_{1^{\sec}})}$   \\
$\left[ (\mathbf{1^{\pr}} \times
\mathbf{1^{\sec}})_{\mathbf{1}}
\times \mathbf{1} \right]_{\mathbf{1}} $ &
$\text{e}^{\ci ( \theta_1 + \theta_{1^{\pr}} +
\theta_{1^{\sec}})}$   \\
\bottomrule
\end{tabular}
\caption{
List of operators, which form a singlet
and constraints on the
phase of the coupling $\lambda$ from
the invariance under the generalised CP
transformations Eq.~\eqref{CPgeneral}.
\label{tab:GenCP}}
\end{table}

\begin{table}
\centering
\begin{tabular}{lcc}
\toprule
$\lambda \mathcal{O} = \lambda (A \times B \times C)$ & $\beta \equiv \text{CP}[\mathcal{O}]/\mathcal{O}^*$  & $\arg(\lambda)$ \\
\midrule
$(\mathbf{2^{\prime}} \times \mathbf{2^{\prime \prime}})_{\mathbf{1}} \times \mathbf{1}$ & $1$ & $0$, $\pi$ \\
($\mathbf{2^{\prime}} \times \mathbf{2^{\prime}})_{\mathbf{1^{\prime \prime}}} \times \mathbf{1^{\prime}}$  & $-\ci$ & $\pm \bar \Omega$\\
($\mathbf{2^{\prime \prime}} \times \mathbf{2^{\prime \prime}})_{\mathbf{1^{\prime}}} \times \mathbf{1^{\prime \prime}}$  &$\ci$  & $\pm \Omega$\\
\midrule
$\left[\2^{\pr} \times (\2^{\sec} \times \3)_{\2^{\sec}} \right]_{\1} $ & $1$ & $0$, $\pi$ \\
$\left[\2^{\sec} \times (\2^{\pr} \times \3)_{\2^{\pr}} \right]_{\1} $ & $1$ & $0$, $\pi$ \\
$\left[ (\mathbf{2^{\prime}} \times \mathbf{2^{\prime \prime}})_{\mathbf{3}} \times \mathbf{3} \right]_{\mathbf{1}}$ & $-\ci$ & $\pm \bar \Omega$\\
$\left[(\mathbf{2^{\prime}} \times \mathbf{2^{\prime}})_{\mathbf{3}} \times \mathbf{3} \right]_{\mathbf{1}}$  & $-1$   & $\pm \pi/2$ \\
$\left[(\mathbf{2^{\prime \prime}} \times \mathbf{2^{\prime \prime}})_{\mathbf{3}} \times \mathbf{3}  \right]_{\mathbf{1}}$  & $1$   & $0$, $\pi$ \\
\midrule
$\left[ (\mathbf{3} \times \mathbf{3})_{\mathbf{1}} \times \mathbf{1} \right]_{\mathbf{1}} $ & $1$  & $0$, $\pi$ \\
$\left[ (\mathbf{3} \times \mathbf{3})_{\mathbf{1^{\pr}}} \times \mathbf{1^{\sec}} \right]_{\mathbf{1}} $ & $1$  & $0$, $\pi$ \\
$\left[ (\mathbf{3} \times \mathbf{3})_{\mathbf{1^{\sec}}} \times \mathbf{1^{\pr}} \right]_{\mathbf{1}} $ & $1$  & $0$, $\pi$ \\
$\left[ (\mathbf{3} \times \mathbf{3})_{\mathbf{3_S \,,\, 3_A}} \times \mathbf{3} \right]_{\mathbf{1}} $ & $1$ & $0$, $\pi$ \\
\midrule
$\left[ (\mathbf{1} \times \mathbf{1})_{\mathbf{1}} \times \mathbf{1} \right]_{\mathbf{1}} $ & $1$  & $0$, $\pi$ \\
$\left[ (\mathbf{1^{\pr}} \times \mathbf{1^{\pr}})_{\mathbf{1^{\sec}}} \times \mathbf{1^{\pr}} \right]_{\mathbf{1}} $ & $1$   & $0$, $\pi$ \\
$\left[ (\mathbf{1^{\sec}} \times \mathbf{1^{\sec}})_{\mathbf{1^{\pr}}} \times \mathbf{1^{\sec}} \right]_{\mathbf{1}} $ & $1$ & $0$, $\pi$ \\
$\left[ (\mathbf{1^{\pr}} \times \mathbf{1^{\sec}})_{\mathbf{1}} \times \mathbf{1} \right]_{\mathbf{1}} $ & $1$ & $0$, $\pi$ \\
\bottomrule
\end{tabular}
\caption{
List of some operators, for which it is possible construct a singlet,
and constraints on the phase of the coupling $\lambda$ from
invariance under the generalised CP transformations
Eq.~\eqref{CPgeneral} with the choice
$\theta_1 = \theta_{1^{\pr}} =  \theta_{1^{\sec}} =
\theta_3 = 0$ and $\theta_{2^{\sec}} =
-\theta_{2^{\pr}} = \pi/4$.
We have omitted the transformations which include
the $\2$ representation of $T^{\pr}$ because
they do not appear in our model, but they can be
read off from Table~\ref{tab:GenCP} after choosing
$\theta_2$.
\label{tab:GenCPgen}}
\end{table}

Under the choice we made, the CP transformation acting on the fields
using the above choice for the $\theta_r$ reads
\begin{equation}
\begin{split}
 \mathbf{1} \rightarrow \mathbf{1}^* \;, \quad
  \mathbf{1^{\prime}} \rightarrow  \mathbf{1^{\prime}}^* \;&, \quad
  \mathbf{1^{\prime \prime}} \rightarrow \mathbf{1^{\prime \prime}}^* \;,\quad
  \mathbf{3} \rightarrow  \mathbf{3}^* \;, \\
\mathbf{2^{\prime}} \rightarrow \begin{pmatrix} 1 & 0 \\ 0 & -\ci \end{pmatrix} \mathbf{2^{\prime}}^* \;&, \quad
\mathbf{2^{\prime \prime}} \rightarrow \begin{pmatrix} \ci & 0 \\ 0 & 1 \end{pmatrix} \mathbf{2^{\prime \prime}}^* \,, \\
\end{split}
\label{CPgeneralFinal}
\end{equation}
where we have again skipped the $\2$
representation because we will not need it later on.

\subsection{Conditions to violate physical CP}
\label{Sec:CPviolConditions}

In this section we try to clarify the
origin of the phases entering the Lagrangian
after $T^{\pr}$ breaking which are then
responsible for physical CP violation.
We will use
the choice of the $\theta_r$ discussed in the previous
section, i.e. $\theta_1 = \theta_{1^{\pr}} =
\theta_{1^{\sec}} = \theta_3 = 0$ and
$\theta_{2^{\sec}} = -\theta_{2^{\pr}} = \pi/4$.

We already know that the singlets and triplets do not
introduce CP violation, see also \cite{Holthausen:2012dk}. Therefore
we only want to consider the doublets. Suppose we couple the doublet
flavons $\psi_i$ to an operator $\mathcal{O}_r$ containing matter
fields and transforming in the representation $r$ of $T^{\pr}$. This
means that the superpotential contains the operator
\begin{equation}
 \mathcal{W} \supset \mathcal{O}_r \Phi_{\bar r} \,\text{, where } \Phi_{\bar r} = \left( \prod_i \psi_i \right)_{\bar r} \;.
\end{equation}
In order to obtain a singlet, the flavons (the doublets) have to be
contracted to the representation $\bar r$ which is the complex
conjugate representation of $r$.

If the operator $\mathcal{O}_r$ by itself conserves physical CP -- by
which we mean that it does not introduce any complex phases into the
Lagrangian including the associated coupling constant -- the
only possible source of CP violation is coming from the doublet vevs
and the complex CG factors appearing in the contraction with the operator
and the doublets. For illustrative purpose we want to discuss this
explicitly if we have two doublets $\psi^{\pr} \sim \2^{\pr}$ and
$\psi^{\sec} \sim \2^{\sec}$.

For $r = \mathbf{1}$ there is only one possible combination using
only $\psi^{\pr}$ and $\psi^{\sec}$ which is $\left( \psi^{\pr}
\otimes \psi^{\sec} \right)_{\1}$.
Using the tensor products of $T^{\pr}$ ---see for
example\cite{Meroni:2012ty}--- we find that the combination is real
if the vevs fulfill the following conditions
\begin{equation}
\psi^{\pr} =
\left( \begin{array}{c} X_1 \, \text{e}^{\ci \alpha} \\ X_2 \, \text{e}^{\ci \beta} \\ \end{array} \right) \;,
\quad
\psi^{\sec} =
\left( \begin{array}{c} Y_1 \, \text{e}^{-\ci \beta} \\ Y_2 \, \text{e}^{-\ci\alpha} \\ \end{array} \right) \;,
\end{equation}
with $X_1$, $X_2$, $Y_1$, $Y_2$, $\alpha$ and $\beta$ real
parameters. For $r = \mathbf{1}^{\pr}$ and $r = \mathbf{1}^{\sec}$
the only possible contractions $\left( \psi^{\pr} \otimes \psi^{\pr}
\right)_{\1^{\sec}}$ and $\left( \psi^{\sec} \otimes \psi^{\sec}
\right)_{\1^{\pr}}$ vanish due to the antisymmetry  of the
contraction.

For $r = \mathbf{1}$ there are three possible contractions.
Either a flavon with itself or both flavons together.
\begin{itemize}
\item  For the selfcontractions
$\left( \psi^{\pr} \otimes \psi^{\pr} \right)_{\3}$
and $\left( \psi^{\sec} \otimes \psi^{\sec} \right)_{\3}$
the Lagrangian will not contain a phase if the flavon
fields $\psi^{\pr}$ and $\psi^{\sec}$ have
the following structure
\begin{equation}
\psi^{\pr} =
\left( \begin{array}{c} X_1 \, \text{e}^{\ci \alpha} \\ X_2 \, \text{e}^{-\ci( \alpha + \pi/4)} \\ \end{array} \right) \;,
\quad
\psi^{\sec} =
\left( \begin{array}{c} Y_1 \, \text{e}^{-\ci (\beta - \pi/4)} \\ Y_2 \, \text{e}^{\ci \beta} \\ \end{array} \right) \;,
\end{equation}
with $X_1$, $X_2$, $Y_1$, $Y_2$ being
real and $\alpha$, $\beta = 0, \, \pm \pi/2, \, \pi$.

\item The contraction $\left( \psi^{\pr} \otimes \psi^{\sec} \right)_{\3}$
does not introduce phases if $\psi^{\pr}$, $\psi^{\sec}$ have the
following structure:
\begin{equation}
\psi^{\pr} \sim
\left( \begin{array}{c} X_1 \, \text{e}^{\ci ( \beta - \pi/4) } \\ X_2 \, \text{e}^{\ci ( \alpha + \pi/4) } \\ \end{array} \right) \;,
\quad
\psi^{\sec} \sim
\left( \begin{array}{c} Y_1 \, \text{e}^{-i \beta} \\ Y_2 \, \text{e}^{-i\alpha} \\ \end{array} \right) \;,
\label{Eq:condvev}
\end{equation}
with $X_1$, $X_2$, $Y_1$, $Y_2$ real and
$\beta - \alpha = -\pi/4$.

\end{itemize}

The previous results allow us to
distinguish in a particular model
the alignments
which can introduce phases
with a specific superpotential in the
Yukawa sector. For example,
if we consider a model in which one
entry of the Yukawa matrix is filled
by a term of the form
$\left( \psi^{\pr} \otimes \psi^{\sec} \right)_{\3}$
and another entry by
$\left( \psi^{\pr} \otimes \psi^{\sec} \right)_{\1}$
we see that we cannot fulfill both conditions
simultaneously if the doublet vevs do not vanish.
That means we would expect CP violation
if both of these contractions are present
in a given model.

Later on in our model we will have real doublet
alignments with
\begin{equation}
 \psi^{\pr} \sim \begin{pmatrix} 1 \\ 0 \end{pmatrix} \text{ and } \psi^{\sec} \sim \begin{pmatrix} 0 \\ 1 \end{pmatrix}  \;.
\end{equation}
These alignments would conserve CP for sure only
if the model contains only the contractions
$\left( \psi^{\pr} \otimes \psi^{\pr} \right)_{\3}$,
$\left( \psi^{\sec} \otimes \psi^{\sec} \right)_{\3}$
and
$\left( \psi^{\pr} \otimes \psi^{\sec} \right)_{\1}$.
Adding the contraction
$\left( \psi^{\pr} \otimes \psi^{\sec} \right)_{\3}$
would add a phase to the Yukawa matrix resulting
possibly in physical CP violation.

\section{The Model}
\label{Sec:Model}

In this section we discuss a supersymmetric model of lepton flavour
based on $T^{\pr}$ as a family symmetry.  Because it considers only
the lepton sector we can consider it as a toy model.
The generalised CP symmetry will be broken in a geometrical way as
we will discuss later on and we can fit all the available data of
masses and mixing in the lepton sector.

The gauge symmetry of the model is the Standard Model gauge
group $G_{\text{SM}} = SU(3)_c \times SU(2)_L \times U(1)_Y$.
The discrete symmetries of the model are
$T^{\pr} \rtimes H_{\text{CP}} \times Z_8 \times Z_4^2 \times Z_3^2 \times Z_2$,
where the $Z_n$ factors are the shaping symmetries of the
superpotential required to forbid unwanted operators.

There are a few comments about this symmetry in order. First of all,
the symmetry seems to be rather large but in fact compared to the
first works on $T^{\pr}$ with geometrical CP violation
\cite{Chen:2007afa, Chen:2009gf} we have only added a factor of
$Z_8 \times Z_2$ but included the full flavon vacuum alignment
and messenger sector. This symmetry is also much smaller than
the shaping symmetry we have used before in \cite{Meroni:2012ty}.

One might wonder where this symmetry originates from and it might be
embedded into (gauged) continuous symmetries or might be a remnant
of the compactification of extra-dimensions. But a discussion of such
an embedding goes clearly beyond the scope of this work where we just
want to discuss the connection of a $T^{\pr}$ family symmetry with CP
and illustrate it by a toy model which is nevertheless in full
agreement with experimental data.

In this section we will only discuss the effective operators
generated after integrating out the heavy messenger fields. The full
renormalisable superpotential including the messenger fields is given in
Appendix~\ref{App:Mess}.

\subsection{The Flavon Sector}

\begin{table}
\centering
\begin{tabular}{lccccccccc}
\toprule
 & $G_{\text{SM}}$ & $T'$ & $U(1)_R$ & $Z_8$ & $Z_4$ & $Z_4$ & $Z_3$ & $Z_3$ & $Z_2$ \\
 \midrule
$\phi$ & $(\mathbf{1},\mathbf{1},0)$ & $\mathbf{3}$  & 0 & 2 & 0 & 0 & 1 & 0 & 1\\
$\tilde{\phi}$  & $(\mathbf{1},\mathbf{1},0)$ & $\mathbf{3}$  & 0 & 2 & 0 & 2 & 0 & 1 & 0\\
$\hat{\phi}$  & $(\mathbf{1},\mathbf{1},0)$ & $\mathbf{3}$  & 0 & 5 & 0 & 3 & 2 & 0 & 0\\
$\xi$ & $(\mathbf{1},\mathbf{1},0)$ & $\mathbf{3}$  & 0 & 0 & 2 & 2 & 0 & 0 & 0\\
 \midrule
$\psi'$  & $(\mathbf{1},\mathbf{1},0)$ & $\mathbf{2}'$ & 0 & 3 & 2 & 3 & 2 & 0 & 0\\
$\psi''$ & $(\mathbf{1},\mathbf{1},0)$ & $\mathbf{2}''$  & 0 & 7 & 2 & 1 & 2 & 0 & 1\\
$\tilde{\psi}''$  & $(\mathbf{1},\mathbf{1},0)$ & $\mathbf{2}''$  & 0 & 1 & 2 & 3 & 0 & 1 & 1\\
 \midrule
$\zeta$ & $(\mathbf{1},\mathbf{1},0)$ & $\mathbf{1}$  & 0 & 5 & 0 & 1 & 2 & 0 & 1\\
$\zeta'$  & $(\mathbf{1},\mathbf{1},0)$ & $\mathbf{1}'$ & 0 & 4 & 0 & 2 & 0 & 0 & 1\\
$\tilde{\zeta}'$ & $(\mathbf{1},\mathbf{1},0)$ & $\mathbf{1}'$ & 0 & 2 & 0 & 2 & 0 & 1 & 0\\
$\zeta''$ & $(\mathbf{1},\mathbf{1},0)$ & $\mathbf{1}''$  & 0 & 2 & 0 & 0 & 1 & 0 & 1\\
$\tilde{\zeta}''$  & $(\mathbf{1},\mathbf{1},0)$ & $\mathbf{1}''$  & 0 & 0 & 0 & 0 & 0 & 1 & 0\\
$\rho$ & $(\mathbf{1},\mathbf{1},0)$ & $\mathbf{1}$  & 0 & 0 & 2 & 2 & 0 & 0 & 0\\
$\tilde \rho$ & $(\mathbf{1},\mathbf{1},0)$ & $\mathbf{1}$  & 0 & 0 & 2 & 2 & 0 & 0 & 0\\
\midrule
$\epsilon_1$  & $(\mathbf{1},\mathbf{1},0)$ & $\mathbf{1}$  & 0 & 4 \
& 1 & 0 & 0 & 0 & 0\\
$\epsilon_2$  & $(\mathbf{1},\mathbf{1},0)$ & $\mathbf{1}$  & 0 & 4 \
& 2 & 2 & 0 & 0 & 1\\
$\epsilon_3$  & $(\mathbf{1},\mathbf{1},0)$ & $\mathbf{1}$  & 0 & 4 \
& 2 & 0 & 0 & 0 & 0\\
$\epsilon_4$  & $(\mathbf{1},\mathbf{1},0)$ & $\mathbf{1}$  & 0 & 0 \
& 0 & 0 & 1 & 1 & 0\\
$\epsilon_5$  & $(\mathbf{1},\mathbf{1},0)$ & $\mathbf{1}$  & 0 & 0 \
& 0 & 0 & 2 & 2 & 0\\
\bottomrule
\end{tabular}
\caption{
List of the flavon fields and their transformation properties. We also list
here the auxiliary flavon fields $\epsilon_i$, $i = 1,\ldots,5$, which are needed to fix
the phases of the vevs of the other flavon fields.
\label{tab:FlavonFields}}
\end{table}

We will start the discussion of the model with the flavon
sector which is self-contained. How the flavons couple
to the matter sector will be discussed afterwards.

The model contains 14 flavon fields in 1-, 2- and 3-dimensional
representations of $T^{\pr}$ and 5 auxiliary flavons in 1-dimensional
representations. Before we will discuss the
superpotential which fixes the directions and phases of the flavon
vevs we will first define them. We have four flavons in the
3-dimensional representation of $T^{\pr}$ pointing in the directions
\begin{equation}
\langle \phi \rangle = \left( \begin{array}{c} 0 \\ 0 \\ 1\\ \end{array} \right) \phi_0  \;, \quad
\langle \tilde \phi \rangle = \left( \begin{array}{c} 0 \\ 1 \\ 0\\ \end{array} \right) \tilde \phi_0 \; , \quad
\langle \hat \phi \rangle = \left( \begin{array}{c} 1 \\ 0 \\ 0\\ \end{array} \right) \hat \phi_0  \; , \quad
\langle \xi \rangle = \left( \begin{array}{c} 1 \\ 1 \\ 1\\ \end{array} \right) \xi_0 \;.
\label{eq:tripletsVEVs}
\end{equation}
The first three flavons will be used in the charged lepton sector
and the fourth one couples only to the neutrino sector. These flavon
vevs, like all the other flavon vevs, are real.

Further we introduce three doublets of $T^{\pr}$:
$\psi^{\pr} \sim \2^{\pr}$, $\psi^{\sec} \sim \2^{\sec}$
and $\tilde \psi^{\sec} \sim \2^{\sec}$. We
recall that the doublets are the only representations of the family
group $T^{\pr}$ which introduce phases, due to the complexity
of the Clebsh-Gordan coefficients.
For the doublets we will find the alignments
\begin{equation}
\langle \psi^{\pr} \rangle = \left( \begin{array}{c} 1 \\ 0 \\ \end{array} \right) \psi^{\pr}_0 \;, \quad
\langle \psi^{\sec} \rangle = \left( \begin{array}{c} 0 \\ 1 \\ \end{array} \right) \psi^{\sec}_0 \;, \quad
\langle \tilde \psi^{\sec} \rangle = \left( \begin{array}{c} 0 \\ 1 \\ \end{array} \right) \tilde \psi^{\sec}_0 \;.
\label{eq:doubletsVEVs}
\end{equation}

And finally, we introduce 7 flavon fields in one-dimensional
representations of the family group. In particular, we have
(the primes indicate the types of singlet)
\begin{equation}
\langle \zeta^{\pr} \rangle =  \zeta^{\pr}_0 \;, \quad   \langle \zeta^{\sec} \rangle =  \zeta^{\sec}_0 \;,\quad
\langle \tilde \zeta^{\pr} \rangle = \tilde  \zeta^{\pr}_0 \;, \quad  \langle \tilde \zeta^{\sec} \rangle =  \tilde \zeta^{\sec}_0 \;, \quad
\langle \zeta \rangle =  \zeta_0 \;, \quad
\langle \rho \rangle =  \rho_0\;,  \quad \langle \tilde \rho \rangle =  \tilde \rho_0 \;.
\label{eq:singletsVEVs}
\end{equation}
The $\rho$ and $\tilde \rho$ couple only to the
neutrino sector while the other one-dimensional
flavons couple only to the charged lepton sector.
Also the five auxiliary flavons $\epsilon_i$,
$i = 1, \ldots, 5$ get real vevs which we
do not label here explicitly.

The flavon quantum numbers are summarized in
Table \ref{tab:FlavonFields}. In this table we have also
included the five auxiliary flavon fields $\epsilon_i$
which are only needed to fix the phases of the other flavon
vevs and all acquire real vevs by themselves.

\begin{table}
\centering
\begin{tabular}{lccccccccc}
\toprule
 & $G_{\text{SM}}$ & $T'$ & $U(1)_R$ & $Z_8$ & $Z_4$ & $Z_4$ & $Z_3$ & $Z_3$ & $Z_2$ \\
 \midrule
$D_{\phi}$  & $(\mathbf{1},\mathbf{1},0)$ & $\mathbf{3}$  & 2 & 0 & 2 & 0 & 1 & 0 & 0\\
$\tilde{D}_{\phi}$ & $(\mathbf{1},\mathbf{1},0)$ & $\mathbf{3}$  & 2 & 0 & 3 & 0 & 0 & 1 & 0\\
$\hat{D}_{\phi}$ & $(\mathbf{1},\mathbf{1},0)$ & $\mathbf{3}$  & 2 & 6 & 0 & 2 & 1 & 2 & 0\\
$D_{\psi}$  & $(\mathbf{1},\mathbf{1},0)$ & $\mathbf{3}$  & 2 & 2 & 0 & 2 & 2 & 0 & 0\\
$\bar D_{\psi}$  & $(\mathbf{1},\mathbf{1},0)$ & $\mathbf{3}$  & 2 & 2 & 0 & 2 & 2 & 0 & 0\\
$\tilde{D}_{\psi}$ & $(\mathbf{1},\mathbf{1},0)$ & $\mathbf{3}$  & 2 & 6 & 0 & 2 & 0 & 1 & 0\\
$D_{\xi}$  & $(\mathbf{1},\mathbf{1},0)$ & $\mathbf{3}$  & 2 & 4 & 2 & 2 & 0 & 0 & 1\\
\midrule
$S_{\zeta}$  & $(\mathbf{1},\mathbf{1},0)$ & $\mathbf{1}$  & 2 & 6 & 0 & 2 & 2 & 0 & 0\\
$S_{\zeta}'$ & $(\mathbf{1},\mathbf{1},0)$ & $\mathbf{1}'$ & 2 & 0 & 0 & 0 & 0 & 0 & 0\\
$\tilde{S}'_{\zeta}$  & $(\mathbf{1},\mathbf{1},0)$ & $\mathbf{1}'$ & 2 & 4 & 0 & 0 & 0 & 1 & 0\\
$P$ & $(\mathbf{1},\mathbf{1},0)$ & $\mathbf{1}$  & 2 & 0 & 0 & 0 & 0 & 0 & 0\\
\bottomrule
\end{tabular}
\caption{
List of the driving fields and their $T^{\pr}$ transformation properties.
The field $P$ stands for the fields $\tilde S_{\zeta}$, $S_{\xi}$, $S_{\rho}$
and $S_{\varepsilon_i}$, with $i=1,\ldots,5$, which all have the same
quantum numbers.
\label{tab:drivFields}}
\end{table}

We discuss now the superpotential in the flavon sector
which ``aligns'' the flavon vevs.  We will use so-called
$F$-term alignment where the vevs are determined
from the $F$-term conditions of the driving fields. The
driving fields are listed with their
quantum numbers in Table~\ref{tab:drivFields}, where
we have indicated for simplicity
$P = \tilde S_{\zeta}$, $S_{\xi}$, $S_{\rho}$ and
$S_{\varepsilon_i}$, with $i = 1,\ldots,5$, because they
have all the same quantum numbers under the whole
symmetry group.

The fields labeled as $P$ play a crucial role in fixing
the phases of the flavon vevs. They are fixed by the
discrete vacuum alignment method as it was first
proposed in \cite{Antusch:2011sx}. Having a flavon $\epsilon$
(for the moment we assume it is a singlet under the
family symmetry) charged under a $Z_n$
symmetry the superpotential will contain a term
\begin{equation}
\mathcal{W} \supset P \left( \frac{\epsilon^n}{\Lambda^{n-2}} \mp M^2 \right) \;.
\label{eq:W-strucutre}
\end{equation}
Remember that the $P$ fields are total singlets. Due to CP
symmetry in this simple example all
parameters and couplings are real. The $F$-term equation for $P$
reads
\begin{equation} \label{eq:flavonpotentialZn}
|F_P|^2 =  \left| \frac{\epsilon^n}{\Lambda^{n-2}} \mp M^2 \right|^2 = 0 .
\end{equation}
which gives for the phase of the flavon vev
\begin{equation}\label{eq:phaseswithZn}
\arg(\langle \epsilon \rangle) =   \left\{ \begin{array}{ll}
\frac{2 \pi}{n}q \;,\quad q = 1, \dots , n & \mbox{\vphantom{$\frac{f}{f}$} for ``$-$'' in Eq.~(\ref{eq:flavonpotentialZn}),}\\
\frac{2 \pi}{n} q +\frac{\pi}{n} \;,\quad q = 1, \dots , n & \mbox{\vphantom{$\frac{f}{f}$} for ``$+$'' in Eq.~(\ref{eq:flavonpotentialZn}).}
\end{array}
\right.
\end{equation}
This method will be used to fix the phases of the singlet and triplet
flavon vevs (including the $\epsilon_i$). Note that we have to introduce
for every phase we fix in this way a $P$ field and only after
a suitable choice of basis for this fields we end up with the simple structure
we show later, see also the appendix of \cite{Antusch:2011sx}.
For the directions of the triplets we use standard expressions,
cf.~also the previous paper \cite{Meroni:2012ty}.

For the doublets, nevertheless, we use here a different method.
Take for example the term  $D_{\psi} \,   \left[ \left( \psi^{\sec} \right)^2 - \phi \, \zeta^{\pr} \right] $.
The $F$-term equations read
\begin{align}
|F_{D_{\psi_1}}| &=  (\psi_2^{\sec})^2 - \phi_3 \, \zeta^{\pr} = 0 \;,\\
|F_{D_{\psi_1}}| &= \ci (\psi_1^{\sec})^2  -\phi_2 \, \zeta^{\pr} = 0 \;,\\
|F_{D_{\psi_1}}| &= (1 - \ci) \psi_1^{\sec}\,\psi_2^{\sec} - \phi_1 \, \zeta^{\pr} = 0\;.
\end{align}
Note the phases coming from the complex CG coefficients of $T^{\pr}$.
Plugging in the (real) vevs of $\phi$ and $\zeta^{\pr}$ it turns
out that only the second component of $\psi^{\sec}$ does not vanish
and is indeed real as well.

The full superpotential for the flavon vacuum alignment reads
\be
\begin{split}
\mathcal{W}_f & =
     \frac{D_{\phi} \, \varepsilon_3}{\Lambda} \, \left[ \phi^2 - \phi \,  \zeta^{\sec} \right]
 + \frac{\tilde D_{\phi} \, \varepsilon_1}{\Lambda} \, \left[ \tilde \phi^2 - \tilde \phi \, \tilde \zeta^{\pr} \right]
 + \frac{ \hat D_{\phi}}{\Lambda} \left[ \varepsilon_4  \, \hat \phi \, \hat \phi + \varepsilon_5 \, \tilde \phi \, \tilde \zeta^{\sec} + \frac{\varepsilon_4^2 \, \tilde \zeta^{\sec} \, \tilde \phi}{\Lambda} \right] \\
 &+ D_{\psi} \,   \left[ \left( \psi^{\sec} \right)^2 - \phi \, \zeta^{\pr} \right]
 + \bar D_{\psi} \, \left( \ci \psi^{\pr} \, \psi^{\pr} + \phi \, \zeta^{\pr} \right)
 + \tilde D_{\psi} \,   \left[ \left( \tilde \psi^{\sec} \right)^2 - \tilde \phi \, \tilde \zeta^{\sec} - \frac{\varepsilon_4^2 \, \hat \phi \, \hat \phi}{\Lambda^2} \right] \\
& + S^{\pr}_{\zeta} \left( \zeta^{\pr} \, \zeta^{\pr} - \xi\,\xi - \frac{\varepsilon_2^2 \, \xi^2}{\Lambda^2} \right)
 + \tilde S^{\pr}_{\zeta} \left( \tilde \zeta^{\pr} \, \tilde \zeta^{\pr} - \tilde \phi \, \tilde \phi \right)
 + \frac{\tilde S_{\zeta}}{\Lambda} \left[\left(\tilde \zeta^{\sec} \right)^3-  M^3_{\tilde \zeta^{\sec}} \right]  \\
& + S_{\zeta} \left( \zeta \, \zeta + \hat \phi \, \hat \phi \right)
 + \frac{S_{\varepsilon_1}}{\Lambda^2} \left( \varepsilon_1^4 - M_{\varepsilon_1}^4 \right)
 + S_{\varepsilon_4} \left( \frac{\varepsilon_4^3}{\Lambda} - M_{\varepsilon_4}^2 \right) \\
& + S_{\varepsilon_3} \left(  \varepsilon_2^2 - M_{\varepsilon_2}^2 \right)  + S_{\varepsilon_3} \left(  \varepsilon_3^2 - M_{\varepsilon_3}^2 \right)  + S_{\varepsilon_5} \left( \varepsilon_4  \, \varepsilon_5 - M_{\varepsilon_5}^2 \right) \\
&+ \frac{D_{\xi} \, \varepsilon_2}{\Lambda} \left( \xi^2  + \xi \rho  + \xi \tilde \rho  \right) +  S_{\xi} \left( \xi^2 - M^2_{\xi} \right) + S_{\rho} \left( \rho^2 + \tilde \rho^2 - M_{\rho}^2 \right)\;.
\end{split}
\ee
We will not go through all the details and discuss each $F$-term
condition but this potential is minimized by  the vacuum structure
as in Eqs.~\eqref{eq:tripletsVEVs}, \eqref{eq:doubletsVEVs} and
\eqref{eq:singletsVEVs}. Finally, we want to remark
that the $F$-term equations do not fix the phase of the field
$\zeta$. However, the phase of this field will turn out to be
unphysical because it can be canceled out through
an unphysical unitary transformation of the
right-handed charged lepton fields
as we will show later explicitly.

\subsection{The Matter Sector}

Since we have discussed now the symmetry breaking
flavon fields we will now proceed with the discussion
on how these fields couple to the matter sector and
generate the Yukawa couplings and right-handed Majorana
neutrino masses.

\begin{table}
\centering
\begin{tabular}{ccccccc}
\toprule
 & $L$ & $\bar{E}$ & $\bar{E}_3$ & $N_R$ & $H_d$ & $H_u$ \\ \midrule
$SU(2)_L$ & $\mathbf{2}$ & $\mathbf{1}$ & $\mathbf{1}$ & $\mathbf{1}$ & $\mathbf{2}$ & $\mathbf{2}$ \\
$U(1)_Y$ & -1 & 2 & 2 & 0 & -1 & 1 \\ \midrule
$T^{\pr}$ & $\mathbf{3}$ & $\mathbf{2}''$ & $\mathbf{1}''$ & $\mathbf{3}$ & $\mathbf{1}$ & $\mathbf{1}$ \\
$U(1)_R$ & 1 & 1 & 1 & 1 & 0 & 0 \\
$Z_8$ & 5 & 7 & 2 & 4 & 7 & 7 \\
$Z_4$ & 1 & 3 & 1 & 3 & 2 & 2 \\
$Z_4$ & 1 & 2 & 1 & 3 & 2 & 2 \\
$Z_3$ & 0 & 0 & 2 & 0 & 0 & 0 \\
$Z_3$ & 2 & 0 & 0 & 0 & 1 & 1 \\
$Z_2$ & 0 & 1 & 1 & 0 & 0 & 0 \\
\bottomrule
\end{tabular}
\caption{
List of the matter and Higgs fields of the model and their
transformation properties under $T^{\pr}$, $U(1)_R$ and the
shaping symmetries. We also give the quantum numbers under
$SU(2)_L \times U(1)_Y$. All fields are singlets of $SU(3)_C$.
\label{tab:MatterFields}}
\end{table}

The model contains three generations of lepton fields, the
left-handed $SU(2)_L$ doublets are organized in a triplet
representation of $T^{\pr}$, the first two families of right-handed
charged lepton fields are organized in a two dimensional
representation, $\2^{\sec}$, and the third family sits in a
$\1^{\sec}$. There are two Higgs doublets as usual in supersymmetric
models. They are both singlets, $\1$ under $T^{\pr}$. The model
includes three heavy right-handed Majorana neutrino fields $N$,
which are organized in a triplet. The light active neutrino masses
are generated through the type~I~seesaw mechanism~\cite{seesaw}. At
leading order  tri-bimaximal mixing (TBM) is predicted in the the
neutrino sector which is corrected by the charged lepton sector
allowing a realistic fit of the measured parameters of the PMNS
mixing matrix.
The quantum numbers of the matter
fields are summarized in Table
\ref{tab:MatterFields}.

In this work we use the right-left convention for the
Yukawa matrices
\begin{equation}
-\mathcal{L} \supset \left({\rm Y}_e \right)_{ij} \, \bar
e_{R\,i}
\, e_{L\,j}
 H_d + \text{ H.c.} \;,
\end{equation}
i.e. there exists a unitary matrix $U_e$ which diagonalizes the
product ${{\rm Y}_e^{\dag}} \, {{\rm Y}_e}$ and contributes to the
physical PMNS mixing matrix.

\subsubsection{The Charged Lepton Sector}

The Yukawa matrix ${{\rm Y}_e}$ is generated after the
flavons acquire their vevs and $T^{\pr}$ is broken.
The effective superpotential describing the couplings
of the matter sector to the flavon sector is given by
\begin{align}
\mathcal{W}_{{\rm Y}_{e}} & =   \frac{y_{33}^{(e)}}{\Lambda} \left( \bar E_3 \, H_d \right)_{\1^{\sec}}\left( L \, \phi \right)_{\1^{\pr}}  +
\frac{y_{32}^{(e)}}{\Lambda^2} \,  \left( \bar E_3 \, H_d \right)_{\1^{\sec}} \left( L \, \hat \phi \right)_{\1^{\pr}} \, \zeta+
\frac{\hat y_{32}^{(e)}}{\Lambda^2} \, \bar \Omega \,  \left( \bar E_3 \, H_d \right)_{\1^{\sec}} \left[ L \left( \psi^{\pr} \, \psi^{\sec} \right)_{\3} \right]_{\1^{\pr}} \nonumber\\
 & + \frac{y_{22}^{(e)}}{\Lambda^2} \left( \bar E \, \psi^{\pr} \right)_{\1}  H_d   \left( L \, \phi \right)_{\1}  +
  \frac{y_{21}^{(e)}}{\Lambda^3} \, \bar \Omega  \, \left (  \bar E \, \psi^{\pr} \right)_{\1} H_d \left[ \left ( \psi^{\pr} \, \psi^{\sec} \right)_{\3} \, L \right]_{\1}  \nonumber\\
& + \frac{y_{11}^{(e)}}{\Lambda^3} \, \Omega \, \left ( \bar E \, \tilde \psi^{\sec} \right)_{\1^{\pr}} H_d \,\tilde \zeta^{\pr} \left ( L \, \tilde \phi \right)_{\1^{\pr}} \;, \label{eq:Yde}
\end{align}
where $\Lambda$ denotes a generic messenger scale. Note the
explicit phase factors $\Omega = (1 +\ci)/\sqrt{2}$ and
$\bar \Omega = (1 - \ci)/\sqrt{2}$ appearing
in some of the operators. They are determined by the
invariance under the generalised CP transformations
and they can be evaluated from Table \ref{tab:GenCPgen}.
We also give here explicitly the contraction of $T^{\pr}$
as indices at the brackets. These contractions
are determined by the messenger sector which will be
discussed in Appendix~\ref{App:Mess}.

After plugging in the flavon vevs from Eqs.~\eqref{eq:tripletsVEVs}-\eqref{eq:singletsVEVs}
we find for the structure of the Yukawa matrix
${\rm Y}_e$
\begin{equation}
{\rm Y}_e = \left( \begin{array}{ccc}
\Omega \, a & 0 & 0 \\
 \ci b &  c & 0 \\
 0 & d  + \ci k & e \\
\end{array}
\right) \equiv
 \left( \begin{array}{ccc}
\Omega \, a & 0 & 0 \\
 \ci b &  c & 0 \\
 0 & \rho \, \text{e}^{\ci \eta}  & e \\
\end{array}
\right) \;,
\label{eq:Ye}
\end{equation}
where we define
$\rho = \sqrt{d^2 + k^2}$ and $\eta = \arg \left( d + \ci k \right)$.

The parameters $a$, $b$, $c$, $d$, $e$, $k$ depend
on the unfixed phase of the vev of $\zeta$, $\zeta_0$,
which can be explicitly factorized as 
\begin{equation}
{\rm Y}_e =
\begin{pmatrix}
\Omega \, \bar a(\zeta_0) & 0 & 0 \\
 \ci \, \zeta_0^{3} \, \overline b   & \zeta_0^{3} \, \overline c & 0 \\
 0 & \zeta_0^2 ( \overline d   + \ci \overline k) & \zeta_0^2 \, \overline e \\
\end{pmatrix} =
\begin{pmatrix}
\text{e}^{\ci \arg(\Omega \, \bar a(\zeta_0))}  & 0 & 0 \\ 0 & \zeta_0^3 & 0 \\ 0 & 0 & \zeta_0^2
\end{pmatrix}
\begin{pmatrix}
| \Omega \, \bar a(\zeta_0)  |  & 0 & 0 \\
 \ci \, \overline b   & \overline c & 0 \\
 0 & \overline d   + \ci \overline k & \overline e \\
\end{pmatrix}
\;,
\end{equation}
from which it is clear that an eventual phase of
$\zeta_0$ drops out in the physical
combination ${\rm Y}_e^\dagger {\rm Y}_e$
and we can choose the parameters in the Yukawa
matrix to be real. 

We remind that there are in principle three possible sources of
complex phases which can lead to physical CP violation: complex
vevs, complex couplings whose phases are determined by the
invariance under the generalised CP symmetry and complex CG
coefficients. In our model all vevs are real due to our flavon alignment
and the convenient choice of the $\theta_r$ phases.

Then the (physical) phases in
${\rm Y}_e$ are completely induced by the complex couplings and
complex CG coefficients. In fact the insights we have gained before in
Section~\ref{Sec:CPviolConditions} can be used here.
The phase in the 1-1 element is unphysical (it drops out
in the combination ${{\rm Y}_e^{\dag}} \, {{\rm Y}_e}$. So the
physical CP violation is to leading order given by the phases
of the ratios $({{\rm Y}_e})_{21}/({{\rm Y}_e})_{22}$ and
$({{\rm Y}_e})_{32}/({{\rm Y}_e})_{33}$. Let us study for illustration
the second ratio which has two components, one with a
non-trivial relative phase and one without.
The real ratio $d/e$ is coming from the operators with the
coefficients $y_{32}^{(e)}$ and $y_{33}^{(e)}$ and from the
viewpoint of $T^{\pr} \rtimes H_{\text{CP}}$ there is not really
any difference between the two because we have only added a singlet
which cannot break CP in our setup as we said before.

For the second ratio $\ci k/e$ this is different. Using the notation
from Section~\ref{Sec:CPviolConditions} we have
$\mathcal{O}_{\3} \Phi_{\3} = \left( \bar E_3 \, H_d \right)_{\1^{\sec}}\left( L \, \Phi_{\3} \right)_{\1^{\pr}}$.
If $\Phi_{\3} = \phi$ (the operator with $y_{32}^{(e)}$) we cannot have a phase because $\phi$
is a triplet flavon. For $\Phi_{\3} = (\psi^{\pr} \psi^{\sec})_{\3}$
(the operator with $\hat y_{32}^{(e)}$)
we can check if condition \eqref{Eq:condvev} is fulfilled
which is not the case because both vevs are real, while
the condition demands a relative phase difference between
the vevs of $\pi/4$. This demonstrates the usefulness of the
conditions given in Section~\ref{Sec:CPviolConditions} in
understanding the origin of physical CP violation in this setup.

\subsubsection{The Neutrino Sector}

The neutrino sector is constructed using a  superpotential
similar to that used in \cite{Meroni:2012ty}:
the light neutrino masses are generated through the type~I~see-saw
mechanism, i.e. introducing right-handed heavy Majorana states which are
accommodated in a triplet under $T^{\pr}$.  We have the effective
superpotential
\begin{equation}
 \mathcal{W}_{{\rm Y}_{\nu}}= \lambda_1 N\, N \,\xi+ N\,N\,(\lambda_2 \rho + \lambda_3 \tilde \rho)  + \frac{y_\nu}{\Lambda}(N  L)_1 (H_u \rho)_1 +  \frac{\tilde y_\nu}{\Lambda}(N  L)_1 (H_u\tilde \rho)_1 \;.
\end{equation}
The Dirac and the Majorana mass matrices obtained from this superpotential are identical to those
described in \cite{Meroni:2012ty} and we quote them here for completeness
\begin{equation}
 M_R = \begin{pmatrix} 2 Z+X & -Z & -Z \\ -Z & 2 Z & -Z+X \\ -Z & - Z+X  & 2 Z \end{pmatrix} \;,\quad
 M_D = \begin{pmatrix} 1 & 0 & 0 \\ 0 & 0 & 1 \\ 0 & 1  & 0 \end{pmatrix}
\frac{\rho^{\prime}}{\Lambda}\; ,
\label{eq:numasses}
\end{equation}
%
where $X,Z$ and $\rho^{\prime}$ are real parameters
which can be written explicitly as
\begin{equation}
 \begin{split}
  X &= \frac{\lambda_2}{\sqrt{3}} \rho_0 +  \frac{\lambda_3}{\sqrt{3}}  \tilde{\rho}_0 \;, \;\; Z = \frac{\lambda_1}{\sqrt{18}} \xi_0  \; \text{ and }  \rho' = \frac{y_\nu}{\sqrt{3}} \rho_0 v_u +  \frac{\tilde{y}_\nu}{\sqrt{3}}  \tilde{\rho}_0 v_u \;.
 \end{split}
\end{equation}
The right-handed neutrino mass matrix $M_R$ is diagonalised
by the TBM matrix \cite{TBM}
\begin{equation}
\TBM =
\begin{pmatrix}
\sqrt{2/3} & \sqrt{1/3} & 0 \\
-\sqrt{1/6} & \sqrt{1/3}  &  -\sqrt{1/2}\\
-\sqrt{1/6} &  \sqrt{1/3}  & \sqrt{1/2}
\end{pmatrix} \;,
\end{equation}
such that the heavy RH neutrino masses read:
\begin{equation}
\begin{split}
\label{eq:MR} \TBM^T\, M_R\,\TBM &= D_{N} = \diag(3Z + X,X,3Z - X) \\&=
\diag(M_1 \,\text{e}^{\ci \phi_1}, M_2 \, \text{e}^{\ci \phi_2}, \, M_3 \, \text{e}^{\ci \phi_3})\;,~M_{1,2,3} > 0\,.
\end{split}
\end{equation}
Since $X$ and $Z$ are real parameters, the phases
$\phi_1$, $\phi_2$ and $\phi_3$ take values 0 or $\pi$.
A light neutrino Majorana mass term is generated
after electroweak symmetry breaking
via the type I see-saw mechanism:
\begin{equation}\label{mnuLO}
M_\nu  \;=\;- M_D^T\, M_R^{-1}\,M_D =
U^*_{\nu}\diag\left(m_1,m_2,m_3\right)U^\dagger_{\nu}\,,
\end{equation}
where
\begin{equation}
U_{\nu} = \ci \, \TBM \,\diag\left(\text{e}^{\ci \phi_1/2},\text{e}^{\ci \phi_2/2},
\text{e}^{\ci \phi_3/2}\right) \equiv \ci\, \TBM \,\Phi_\nu\,,~~
\Phi_\nu\equiv \diag\left(\text{e}^{\ci \phi_1/2},\text{e}^{\ci \phi_2/2},\text{e}^{\ci \phi_3/2}\right)\,,
\label{Unu}
\end{equation}
%
and $m_{1,2,3} > 0$ are the light neutrino masses,
\begin{equation}
 m_i = \left(\frac{\rho^{\prime}}{\Lambda}\right )^2\,
\frac{1}{M_i}\,,\,\,\,i=1,2,3\,.
\label{numasses}
\end{equation}
%
The phase factor $\ci$ in Eq.\ (\ref{Unu})
corresponds to an unphysical
phase and we will drop it in what follows.
Note also that one of the phases $\phi_k$, say $\phi_1$,
is physically irrelevant since it can be considered
as a common phase of the neutrino mixing matrix.
In the following we will always set $\phi_1=0$.
This corresponds to the choice $(X + 3Z) > 0$.

\subsection{\texorpdfstring{Comments about the $\boldsymbol{\theta_r}$}{Comments about the theta r}}

At this point we want to comment 
on the role of the phases $\theta_r$ appearing in the definition of
the CP transformation in Eq.~\eqref{CPgeneralFinal}. These phases
are arbitrary and hence they should not contribute to physical
observables. This means, for instance, that these arbitrary phases
must not appear in the Yukawa matrices after $T^{\pr}$ is broken.
However it is not enough to look at the Yukawa couplings alone but
one also has to study the flavon vacuum alignment sector. We want to
show next a simple example for which, as expected, these phases
turn out to be unphysical.

In order to show this we consider as example $({\rm Y}_e)_{22}$
and $({\rm Y}_e)_{21}$ respectively generated by the following operators:
\begin{equation}
({\rm Y}_e)_{22} \sim\left ( \bar E \, \psi^{\pr} \right)_{\1} \left( L \, \phi \right)_{\1} \zeta \, H_d \;, \quad
({\rm Y}_e)_{21} \sim\left( \bar E \, \psi^{\pr} \right)_{\1} \left( L \, \phi \right)_{\1^{\sec}} \zeta^{\pr} \, H_d \;. \label{eq:tetaop}
\end{equation}
The fields together with their charges have been defined before in
Table~\ref{tab:FlavonFields}. We will now be more explicit and
consider all the possible phases arising in each of the given
operators under the CP transformation of Eq.~\eqref{CPgeneral} where the $\theta_r$
were included explicitly. For each flavon vev in the operators we will
denote the arising phase with a bar correspondingly,
i.e. for the vev of the  flavon $\phi$ we will have
$\phi \rightarrow \text{e}^{\ci\bar \phi} \phi_0$ where
$\phi_0$ is the modulus of the vev.
Then using the transformations in Eq.~\eqref{CPgeneral} and
Table~\ref{CPgeneral} we get
\begin{equation}
\arg \left(({\rm Y}_e)_{22}-({\rm Y}_e)_{21}\right) = \bar\zeta-\bar\zeta^{\pr} + (\theta_{\1}-\theta_{\1^{\pr}})/2 \;.
\end{equation}
The vevs of the flavons $\zeta$ and $\zeta^{\pr}$ are determined
at leading order by
\begin{equation}
S_{\zeta} \,   \left[ \zeta^2 - (\hat\phi\, \hat\phi)_{\1} \right] \text{ and }
S_{ \zeta^{\pr}} \, \left[ (\zeta^{\pr})^2 - (\xi \, \xi)_{\1^{\sec}}  \right].
\end{equation}
where $S_{\zeta, \zeta^{\pr}}$ are two of the so-called ``driving
fields'' which in this case are singlet of type $\1$ and $\1^{\pr}$
under $T^{\pr}$. From the $F$-term equations one gets
that $\bar\zeta = \bar{\hat{\phi}} + (\theta_{\3}-\theta_{\1})/2$ and
$\bar\zeta^{\pr}=\bar\xi + (\theta_{\3}-\theta_{\1^{\pr}})/2$ and
thus in the physical phase difference
\begin{equation}
\arg \left(({\rm Y}_e)_{22}-({\rm Y}_e)_{21}\right)= \bar{\hat{\phi}}-\bar \xi \;.
\end{equation}
the phases $\theta_{\1}$, $\theta_{\1^{\pr}}$ and $\theta_{\3}$ cancel out.

This shows how the $\theta_{r}$ cancel out in a complete model
and become unphysical. Including them only in one sector, for
instance, in the Yukawa sector they might appear to be physical
and only after considering also the flavon alignment sector it can
be shown that they are unphysical which is nevertheless quite
cumbersome in a realistic model due to the many fields and
couplings involved.

\subsection{Geometrical CP violation and residual symmetries}

In this section we want to provide a better understanding
of the quality of symmetry breaking our model exhibits.
To be more precise we will argue that our model
breaks CP in a geometrical fashion and then we will
discuss the residual symmetries of the mass matrices.

Geometrical CP violation was first defined in \cite{Branco:1983tn}
and there it is tightly related to the so-called ``calculable
phases'' which are phases of flavon vevs which do not
depend on the parameters of the potential but only on the
geometry of the potential. This applies also
to our model. All complex phases are determined in the
end by the (discrete) symmetry group of our model.
In particular the symmetries $T' \rtimes H_{\text{CP}}$
and the $Z_n$ factors play a crucial role here. For the
singlets and triplets in fact the $Z_n$ symmetries (in
combination with CP) make the
phases calculable using the discrete vacuum alignment
technique \cite{Antusch:2011sx}. For the doublets then
the symmetry $T' \rtimes H_{\text{CP}}$ enters via fixing
the phases of the couplings and fixing relative phases
between different components of the multiplets.
In particular, all flavon vevs are left invariant
under the generalised CP symmetry and hence protected
by it.
However the calculable phases are necessary but not sufficient
for geometrical CP violation. For this we have to see if
CP is broken or not.

For this we will have a look at the residual symmetries
of the mass matrices after $T^{\pr} \rtimes H_{\text{CP}}$
is broken. First of all, we observe that the vev structure
mentioned in Section~\ref{Sec:Model} gives a breaking pattern which
is different in the neutrino and in the charged
lepton sector, i.e. the residual groups $G_{\nu}$
and $G_e$ are different.

In the charged lepton
sector the group $T^{\pr}$ is fully broken by
the singlet, doublet and triplet vevs.
If it exists, the residual group in
the charged lepton sector is defined
through the elements which leave invariant
the flavon vevs and satisfy
 \begin{equation}
 \begin{split}
  \rho^{\dag}(g_{e_i}) \, {\rm Y}_e^{\dag} \, {\rm Y}_e \, \rho(g_{e_i})
   &=  {\rm Y}_e^{\dag} \, {\rm Y}_e \text{ with $g_{e_i} \in G_e < T^{\pr} \rtimes H_{\text{CP}}$} \;,\\
  X_e^{\dag}  \, \left( {\rm Y}_e^{\dag} \, {\rm Y}_e \right) \, X_e
   &= \left( {\rm Y}_e^{\dag} \, {\rm Y}_e \right)^{\ast} \mbox{ with $X_e \in G_e < T^{\pr} \rtimes H_{\text{CP}}$} \;.\\
 \end{split}
 \end{equation}
The first condition is the ordinary condition
to study residual symmetries while the second
one is relevant only for models with spontaneous
CP violation. 

In our model this conditions are not satisfied for any
$\rho(g)$ or $X_e$. 
Hence there is no residual symmetry group in the charged
lepton sector and even more CP is broken spontaneously.
Together with the fact that all our phases are determined
by symmetries (up to signs and discrete choices) we have
demonstrated now that our model exhibits geometrical CP
violation.

In the neutrino sector we can write similar relations
that take into account the symmetrical structure of the
Majorana mass matrix, and in particular as before the
residual symmetry is defined through the elements
which leave invariant the flavon vevs and satisfy
 \begin{equation}
 \begin{split}
 \rho^{T}(g_{\nu_i}) \, M_{\nu} \, \rho(g_{\nu_i})
  &= M_{\nu} \text{ with $g_{\nu_i} \in G_\nu < T^{\pr} \rtimes H_{\text{CP}}$}\;,\\
 X_{\nu}^{T}  \, M_{\nu}  \, X_{\nu}
  &= M_{\nu}^{\ast} \text{ with $X_{\nu} \in G_{\nu} < T^{\pr} \rtimes H_{\text{CP}}$} \;.
\label{Eq:nu}
 \end{split}
\end{equation}
In our model $M_{\nu}$ is a real matrix and therefore
$\rho(g_{\nu_i})$ and $X_{\nu}$ are  defined through 
the same conditions.
Defining $O$ as the orthogonal matrix which diagonalizes 
the real symmetric matrix $M_{\nu}$ 
we find from Eq.~\eqref{Eq:nu}
\be
(O \, X_{\nu}^T \, O^T) \, M^{\text{diag}}_{\nu} \, (O \, X_{\nu} \, O^T) = M^{\text{diag}}_{\nu}
\ee
and hence the matrix $D = O \, X_{\nu} \, O^T$ has to be of the form
\be
D = \begin{pmatrix}
(-1)^p & 0 & 0 \\
0 & (-1)^q & 0 \\
0 & 0 & (-1)^{p+q} \\
\end{pmatrix} \;.
\ee

The same argument can be applied to the matrix $\rho(g_{\nu_i})$,
because the matrices $X_{\nu}$, $\rho(g_{\nu_i})$
and $M_{\nu}$ are simultaneously diagonalisable by
the same orthogonal matrix $O$.
It is easy to find that
\begin{equation}
 O = \begin{pmatrix}
1/\sqrt{3} & 1/\sqrt{3} & 1/\sqrt{3} \\
0 & -1/\sqrt{2} & 1/\sqrt{2} \\
-\sqrt{2/3} & 1/\sqrt{6} & 1/\sqrt{6} \\
\end{pmatrix} \;,
\end{equation}
which is expected since $M_{\nu}$ is diagonalized by
$U_{\text{TBM}}$ and $O$ is just a permutation of
$U_{\text{TBM}}^T$ which corresponds to a permutation
of the eigenvalues.

The residual symmetry coming from $T^{\pr}$ is generated only by
\begin{equation}
 T \, S \, T^2 = O^T \cdot
\begin{pmatrix}
1 & 0 & 0 \\
0 & -1 & 0 \\
0 & 0 & -1\\
\end{pmatrix}
\cdot O \;,
\end{equation}
which also leaves invariant the vev structure.
This symmetry is a $Z_2$ symmetry.
In summary the residual symmetry in the neutrino sector
is a Klein group $K_4 \cong Z_2 \times Z_2$,
in which one $Z_2$ comes from $H_\text{CP}$ and the other
one from $T^{\pr}$.
$H_\text{CP}$ is conserved because in the neutrino
sector $X_{\nu}$ can be chosen as  the identity matrix
and $M_{\nu}$ is real.

Combining the two we find
\begin{equation}
G_f \equiv T^{\pr} \rtimes H_{\text{CP}} \equiv T^{\pr} \rtimes Z_2
\longrightarrow
\begin{cases}
G_e & = \emptyset \;, \\
G_{\nu} & = K_4 \;,\\
\end{cases}
\end{equation}
so that  $T^{\pr} \rtimes H_{\text{CP}}$ is completely broken
and there is no residual symmetry left.

\subsection{Predictions}
\subsubsection{Absolute Neutrino Mass Scale}

Before we consider the mixing angles and phases in the PMNS matrix
we first will discuss the neutrino spectra predicted by the model.
We get the same results as in \cite{Meroni:2012ty}
because our neutrino mass matrix has exactly the same
structure. The forms of the  Dirac and Majorana
mass terms given in Eq.~\eqref{eq:numasses}
imply that in the model considered by us both
light neutrino mass spectra with normal ordering (NO) and with
inverted ordering (IO) are
allowed (see also \cite{Meroni:2012ty}).
In total three different spectra for the light active neutrinos
are possible. They correspond to the different choices of the values
of the phases $\phi_i$ in Eq.~\eqref{Unu}).
More specifically, the cases $\phi_1 = \phi_2 = \phi_3 = 0$
and $\phi_1 = \phi_2 = 0$ and $\phi_3 = \pi$
correspond to NO spectra of the type A and B, respectively.
For $\phi_1 = \phi_2 = 0$ and $\phi_3 = \pi$
also  IO spectrum is possible.
The neutrino masses in cases of the three 
spectra are given by:
\begin{eqnarray}
{\rm NO~spectrum~A}:~~~(m_1,m_2,m_3) = 
(4.43, 9.75,48.73)\cdot 10^{-3}~{\rm eV}\,,\\ [0.30cm]
{\rm NO~spectrum~B}:~~~(m_1,m_2,m_3) = 
(5.87, 10.48, 48.88)\cdot 10^{-3}~{\rm eV}\,,\\ [0.30cm]
{\rm IO~spectrum}:~~~(m_1,m_2,m_3) = 
(51.53, 52.26, 17.34)\cdot 10^{-3}~{\rm eV}\,,
\end{eqnarray}
%
where we have used the best fit values of 
$\Delta m^2_{21}$ and $|\Delta m^2_{31(32)}|$ 
given in Table 1. Employing the 3$\sigma$ allowed ranges 
of values of the two neutrino mass squared differences 
quoted in Table 1, we find the intervals in which 
$m_{1,2,3}$ can vary:

\begin{itemize}
\item NO spectrum A:\\
$m_1 \in [4.23, 4.66] \cdot 10^{-3}$ eV,
$m_2 \in [9.23, 10.17] \cdot 10^{-3}$ eV,
$m_3 \in [4.28, 5.56] \cdot 10^{-2}$ eV;
\item NO spectrum B:\\
$m_1 \in [5.56, 6.20] \cdot 10^{-3}$ eV,
$m_2 \in [9.83, 11.20] \cdot 10^{-3}$ eV,
$m_3 \in [4.23, 5.74] \cdot 10^{-2}$ eV;
\item IO spectrum:\\
$m_1 \in [4.57, 5.87] \cdot 10^{-2}$ eV,
$m_2 \in [4.63, 5.96] \cdot 10^{-2}$ eV,
$m_3 \in [1.53, 1.98] \cdot 10^{-2}$ eV.
\end{itemize}

Correspondingly, we get for the sum of the neutrino masses:
\begin{eqnarray}
\label{summjNOA}
{\rm NO~A}:~~\sum_{j=1}^{3} m_j = 6.29 \times 10^{-2}~{\rm eV}\,,~~
5.63 \times 10^{-2} \leq \sum_{j=1}^{3} m_j \leq 7.04 \times 10^{-2}~{\rm eV}\,,~~\\ [0.30cm]
\label{summjNOB}
{\rm NO~B}:~~
~~\sum_{j=1}^{3} m_j = 6.52 \times 10^{-2}~{\rm eV}\,,~~
\label{summjIO}
5.77 \times 10^{-2} \leq \sum_{j=1}^{3} m_j \leq 7.48 \times 10^{-2}~{\rm eV}\,,~~\\ [0.30cm]
{\rm IO}:~~\sum_{j=1}^{3} m_j = 12.11 \times 10^{-2}~{\rm eV}\,,~~
10.73 \times 10^{-2} \leq \sum_{j=1}^{3} m_j \leq 13.81 \times 10^{-2}~{\rm eV}\,,~~
\end{eqnarray}
%
where we have given the predictions using the 
best fit values and the $3\sigma$ intervals of the 
allowed values of $m_1$, $m_2$ and $m_3$ quoted above.

\subsubsection{The Mixing Angles and Dirac CPV Phase} 

  We will derive next expressions for the mixing
angles and the CPV phases in the standard parametrisation
of the PMNS matrix in terms of the parameters
of the model. The expression for the
charged lepton mass matrix ${\rm Y}_{e}$
given in Eq.~\eqref{eq:Ye} contains altogether
seven parameters: five real parameters and two phases, one of which
is equal to $\pi/2$. Three (combinations of) parameters are
determined by the three charged lepton masses.
The remaining two real parameters and two phases are
related to two angles and two phases in the matrix
$U_{e}$ which diagonalises the product
${\rm Y}_e^{\dag} \,{\rm Y}_e$ and enters
into the expression of the PMNS matrix:
$U_{\text{PMNS}}=  U_{e}^\dagger U_{\nu}$, where
$U_{\nu}$ is of TBM form (see Eq.~(3.19)), while
$U_{e}\propto R_{23}R_{12}$,
$R_{23}$ and $R_{12}$ are orthogonal matrices
describing rotations in the 2-3 and 1-2 planes,
respectively. It proves convenient to adopt
for the matrices $U_e$ and $U_{\nu}$
the notation used in \cite{Marzocca:2013cr}:
\begin{equation}
\begin{cases}
U_e = \Psi_e \, R^{-1}_{23} \left( \theta^e_{23} \right)
R^{-1}_{12} \left( \theta^e_{12} \right) \\
U_{\nu} = R_{23} \left( \theta^{\nu}_{23} \right)
R_{12} \left( \theta^{\nu}_{12} \right) \Phi_{\nu} \\
\end{cases}
\label{UeUnu}
\end{equation}
%
where  $\Psi_e =
{\rm diag} \left(1,\text{e}^{\ci \psi_e}, \text{e}^{\ci \omega_e } \right)$,
$\theta^{\nu}_{23} =  -\pi/4$,
$\theta^{\nu}_{12} = \sin^{-1}(1/\sqrt{3})$,
$\Phi_\nu $ is a diagonal phase
matrix  defined in Eq.~\eqref{Unu},
and
%
\begin{equation}
R_{12}\left( \theta^e_{12} \right) = \begin{pmatrix}
\cos \theta^e_{12} & \sin \theta^e_{12} & 0\\
- \sin \theta^e_{12} & \cos \theta^e_{12} & 0\\
0 & 0 & 1 \end{pmatrix} \;,
\quad
R_{23}\left( \theta^e_{23} \right) = \begin{pmatrix}
1 & 0 & 0\\
0 & \cos \theta^e_{23} & \sin \theta^e_{23} \\
0 & - \sin \theta^e_{23}  & \cos \theta^e_{23} \\
\end{pmatrix} \;.
\label{R1223}
\end{equation}
%

Using the expression for the
charged lepton mass  matrix ${\rm Y}_{e}$
given in Eq.~\eqref{eq:Ye} and 
comparing the right and the left sides of the equation
\begin{equation}
{\rm Y}_e^{\dag} \, {\rm Y}_e = U_e {\rm
diag}\left(m_e^2,m_{\mu}^2,m_{\tau}^2\right) U_e^{\dag}\,,
\label{YeUe}
\end{equation}
%
we find that $m_{e}^2 = a^2$, $m_{\mu}^2 = c^2$ 
and $m_{\tau}^2 = e^2$.
For $U_e$ given in Eq. (\ref{UeUnu})
this equality  
holds only under the condition that  $\sin \theta^e_{12}$
and $\sin \theta^e_{23}$ are sufficiently small. 
Using the leading terms in powers of  
the small parameters $\sin \theta^e_{12}$
and $\sin \theta^e_{23}$ 
we get the approximate relations:\\
\begin{equation}
\sin \theta^e_{12} \, \text{e}^{\ci \psi_e} \simeq
\pm \ci \left| \frac{b}{c} \right| \;, \quad
\sin \theta^e_{23} \, \text{e}^{\ci (\psi_e - \omega_e)} =
\frac{e \, \rho}{c^2 - e^2}  \, \text{e}^{ \ci \eta} \simeq
\left | \frac{\rho}{e}  \right |  \, \text{e}^{ \ci \xi_e } \;,
\end{equation}
%
where $\psi_e = \pm  \pi/2$,
$\xi_e = \psi_e - \omega_e$, $\xi_e \in [0, 2 \pi]$,
$\theta^e_{12} \simeq |b/c|$ and $\theta^e_{23} \simeq |\rho/e|$.

In the discussion that follows $\theta^e_{12}$, $\theta^e_{23}$,
$\psi_e$ and $\omega_e$ are treated as arbitrary angles and phases,
i.e., no assumption about their magnitude is made.

   The lepton mixing we obtain in the model
we have constructed, including the Dirac CPV phase but not
the Majorana CPV phases, was investigated in detail on general
phenomenological grounds in ref. \cite{Marzocca:2013cr}
and we will use the results obtained in \cite{Marzocca:2013cr}.
The three angles $\theta_{12}$, $\theta_{23}$ and $\theta_{13}$
and the Dirac and Majorana CPV phases $\delta$ and $\beta_1$
and $\beta_2$ (see Eqs. (1.1) - (1.3)),
of the PMNS mixing matrix
$U_{\text{PMNS}}=  U_{e}^\dagger U_{\nu}
= R_{12}( \theta^e_{12})R_{23}( \theta^e_{23})  \Psi^{*}_e
 R_{23}(\theta^{\nu}_{23}) R_{12}( \theta^{\nu}_{12})\; \Phi_{\nu} $,
can be expressed as functions of the two real angles,
$\theta^e_{12}$ and $\theta^e_{23}$,
and the two phases, $\psi_e$ and $\omega_e$
present in $U_{e}$.  However, as was shown in
\cite{Marzocca:2013cr}, the three angles
$\theta_{12}$, $\theta_{23}$ and $\theta_{13}$
and the Dirac phase $\delta$ are expressed
in terms of the angle
$\theta^e_{12}$, an angle $\hat\theta_{23}$
and just one phase $\phi$, where
\begin{equation}
\sin^2\hat\theta_{23} =
\frac{1}{2}\,\left ( 1 - 2\sin\theta^e_{23}\cos \theta^e_{23}\cos (\omega_e -\psi_e) \right )\,,
\label{th23hat}
\end{equation}
%
and the phase  $\phi = \phi(\theta^e_{23},\omega_e,\psi_e)$.
Indeed, it is not difficult to show that
(see the Appendix in \cite{Marzocca:2013cr})
\begin{equation}
R_{23}( \theta^e_{23})\, \Psi^{*}_e \, R_{23}(\theta^{\nu}_{23}) =
P\, \Phi\, R_{23}(\hat\theta_{23})\tilde{Q}\,.
\end{equation}
%
Here
$P={\rm diag}(1,1, \text{e}^{-\,\ci \alpha})$,
$\Phi = {\rm diag}(1,\text{e}^{\ci \phi},1)$ and
$\tilde{Q} = {\rm diag} \left(1,1, \text{e}^{\ci \beta} \right)$,
where
\begin{equation}
\alpha = \gamma + \psi_e + \omega_e  \,,~~~~
\beta = \gamma - \phi\,,
\label{alphabeta}
\end{equation}
%
and
\begin{equation}
\gamma = \arg \left (\,-\text{e}^{ -\ci \psi_e}\cos \theta^e_{23}
+ \text{e}^{-\ci \omega_e}\sin\theta^e_{23}\right)\,,~~
\phi= \arg \left (\text{e}^{ -\ci \psi_e}\cos \theta^e_{23}
+ \text{e}^{-\ci \omega_e}\sin\theta^e_{23}\right)\,.
\label{gammaphi}
\end{equation}
%
The phase $\alpha$ is unphysical (it can be absorbed in
the $\tau$ lepton field). The phase $\beta$ contributes
to the matrix of physical Majorana phases,
which now is equal to $\overline{Q} = \tilde{Q} \, \Phi_{\nu}$.
The PMNS matrix takes the form:
\begin{equation}
U_{\text{PMNS}}=
R_{12}(\theta^e_{12})\,\Phi(\phi)\, R_{23}(\hat\theta_{23})\,
R_{12}(\theta^{\nu}_{12})\,\overline{Q}\,,
\label{UPMNSthhat}
\end{equation}
%
where $\theta^{\nu}_{12} = \sin^{-1}(1/\sqrt{3})$.
Thus, the four observables $\theta_{12}$, $\theta_{23}$, $\theta_{13}$
and  $\delta$ are functions of three parameters
$\theta^e_{12}$, $\hat\theta_{23}$ and $\phi$.
As a consequence, the Dirac phase $\delta$ can be expressed
as a function of the three PMNS angles
$\theta_{12}$, $\theta_{23}$ and $\theta_{13}$,
leading to a new ``sum rule''
relating $\delta$ and $\theta_{12}$, $\theta_{23}$ and $\theta_{13}$
\cite{Marzocca:2013cr}. Using the measured values of
$\theta_{12}$, $\theta_{23}$ and $\theta_{13}$,
the authors of \cite{Marzocca:2013cr}
obtained  predictions for the values of $\delta$ and of
the rephasing invariant
$J_{\text{CP}} = {\rm Im}(U_{e1}^* U_{\mu 3}^* U_{e3} U_{\mu 1})$,
which controls the magnitude of CP violating effects
in neutrino oscillations \cite{PKSP3nu88},
as well as for the $2\sigma$ and $3\sigma$
ranges of allowed values of $\sin\theta_{12}$,
$\sin\theta_{23}$ and $\sin\theta_{13}$.
These predictions are valid also in the model
under discussion.

 To be more specific, using Eq.~(\ref{UPMNSthhat})
we get for the angles $\theta_{12}$, $\theta_{23}$ and $\theta_{13}$
of the standard parametrisation of $U_{\text{PMNS}}$
\cite{Marzocca:2013cr}:
\be
\begin{split}
    \sin \theta_{13} &= \left| U_{e3} \right| =
\sin \theta^e_{12} \sin \hat{\theta}_{23}, \\
\sin^2 \theta_{23} &=
\frac{\left| U_{\mu3} \right|^2}{1- \left| U_{e 3} \right|^2 } =
\frac{\sin^2\hat\theta_{23} - \sin^2 \theta_{13}}
{1 - \sin^2\theta_{13}}\,,~~~
\cos^2 \theta_{23} = \frac{\cos^2\hat\theta_{23}}
{1 - \sin^2\theta_{13}}\,, \\
\sin^2 \theta_{12} &=
\frac{\left| U_{e2} \right|^2}{1- \left| U_{e3} \right|^2 } =
\frac{1}{3}\left (2  +
\frac{ \sqrt{2}\,\sin2\theta_{23}\sin\theta_{13}\cos\phi - \sin^2\theta_{23}}
{ 1 - \cos^2\theta_{23}\cos^2\theta_{13} } \right )\,,
\label{eq:chlep_corrections}
\end{split}
\ee
%
where the first relation
$\sin \theta_{13}=\sin \theta^e_{12} \sin \hat{\theta}_{23}$
was used in order to obtain the expressions for
$\sin^2 \theta_{23}$ and $\sin^2 \theta_{12}$.
Clearly, the angle $\hat\theta_{23}$ differs little from
the atmospheric neutrino mixing angle $\theta_{23}$.
For $\sin^2\theta_{13} = 0.024$
and $\sin^2\theta_{23} \cong 0.39$
we have $\sin \theta^e_{12}\cong 0.2$.
Comparing the imaginary and real parts of
$U_{e1}^* U_{\mu 3}^* U_{e3} U_{\mu 1}$, obtained
using Eq.~(\ref{UPMNSthhat})
and the standard parametrisation of $U_{\text{PMNS}}$,
one gets the following relation between the phase
$\phi$ and the Dirac phase $\delta$
\cite{Marzocca:2013cr}:
\begin{align}
\label{sindNOsqrt3}
\sin\delta =&\; -\, \frac{2\sqrt{2}}{3}\,\frac{\sin\phi}{\sin2\theta_{12}}\,,
\\[0.30cm]
\cos\delta =&\; \frac{2\sqrt{2}}{3\sin2\theta_{12}}\,\cos\phi\,
\left (-1 + \frac{2\sin^2\theta_{23}}
{\sin^2\theta_{23}\cos^2\theta_{13} + \sin^2\theta_{13}}\,\right )
\nonumber \\[0.30cm]
 &\;+ \frac{1}{3\sin2\theta_{12}}\,
\frac{\sin2\theta_{23}\, \sin\theta_{13}}
{\sin^2\theta_{23}\cos^2\theta_{13} + \sin^2\theta_{13}}\,.
\label{cosdNOsqrt3}
\end{align}
%
The results quoted above, including those
for $\sin\delta$ and $\cos\delta$, are exact.
As can be shown, in particular, we have:
$\sin^2\delta + \cos^2\delta = 1$.

  Equation  (\ref{eq:chlep_corrections}) allows to express
$\cos\phi$ in terms of
$\theta_{12}$, $\theta_{23}$ and $\theta_{13}$,
%
and substituting the result thus obtained for $\cos\phi$
in Eqs. (\ref{sindNOsqrt3}) and (\ref{cosdNOsqrt3}),
one can get expressions for $\sin\delta$ and $\cos\delta$
in terms of $\theta_{12}$, $\theta_{23}$ and $\theta_{13}$.
We give below the result for $\cos\delta$
\cite{Marzocca:2013cr}:
\begin{equation}
\cos\delta =  \frac{\tan\theta_{23}}{3\sin2\theta_{12}\sin\theta_{13}}\,
\left [ 1 + \left (3\sin^2\theta_{12} - 2 \right )\,
 \left (1 - \cot^2\theta_{23}\,\sin^2\theta_{13}\right )\right ]\,.
\label{2cosdNOsqrt3}
\end{equation}
%
For the best fit values of $\sin^2\theta_{12}$,
$\sin^2\theta_{23}$ and $\sin\theta_{13}$,
one finds in the case of NO and IO spectra 
\footnote{
Due to the slight difference between 
the best fit values of 
$\sin^2\theta_{23}$ and $\sin\theta_{13}$
in the cases of NO and IO spectra (see Table 1), 
the values we obtain for $\cos\delta$ 
in the two cases differ somewhat.
However, this difference is 
equal to $10^{-4}$ in absolute value and we 
will neglect it in what follows.}
(see also \cite{Marzocca:2013cr}):
\begin{equation}
\cos\delta \cong -\,0.069\,,~~\sin\delta = \pm 0.998\,.
\end{equation}
%
These values correspond to 
\begin{equation}
\delta = 93.98^{\circ}\,~~{\rm or}\,
~~\delta = 266.02^{\circ}\,. 
\label{delta}
\end{equation}
%

 Thus, our model predicts $\delta \simeq \pi/2$ or $3\pi/2$.
The fact that the value of the Dirac CPV phase
$\delta$ is determined (up to an ambiguity of
the sign of $\sin\delta$) by the values of the
three mixing angles  $\theta_{12}$, $\theta_{23}$
and $\theta_{13}$ of the PMNS matrix,
(\ref{2cosdNOsqrt3}),
is the most striking prediction of the
model considered.
For the best fit values of $\theta_{12}$,
$\theta_{23}$ and $\theta_{13}$ we get
$\delta \cong \pi/2$ or $3\pi/2$.
These result implies also that in the model
under discussion, the $J_{\text{CP}}$ factor,
which determines the magnitude
of CP violation in neutrino oscillations,
is also a function of the three angles
$\theta_{12}$, $\theta_{23}$ and $\theta_{13}$
of the PMNS matrix:
\begin{equation}
J_{\text{CP}} = J_{\text{CP}}(\theta_{12},\theta_{23},\theta_{13},
\delta(\theta_{12},\theta_{23},\theta_{13})) =
J_{\text{CP}}(\theta_{12},\theta_{23},\theta_{13})\,.
\label{JCPNO}
\end{equation}
%
This allows to obtain predictions for the range of
possible values of $J_{\text{CP}}$ using the current data on
$\sin^2\theta_{12}$, $\sin^2\theta_{23}$
and $\sin^2\theta_{13}$. For the best fit values of these parameters
(see Table 1) we find: $J_{\text{CP}} \simeq \pm 0.034$.

The quoted results on $\delta$ and
$J_{\text{CP}}$ were obtained first on the basis of a phenomenological
analysis in \cite{Marzocca:2013cr}.
Here they are obtained for the first time within
a selfconsistent model of lepton flavour based on the
$T^{\prime}$ family symmetry.

 In \cite{Marzocca:2013cr} the authors performed a detailed
statistical analysis which permitted to determine
the ranges of allowed values of
$\sin^2\theta_{12}$, $\sin^2\theta_{23}$,
$\sin\theta_{13}$, $\delta$ and
$J_{\text{CP}}$ at a given confidence level.
We quote below some of the results obtained in
\cite{Marzocca:2013cr}, which are valid also in the
model constructed by us.

 Most importantly, the CP conserving values of $\delta =0;\pi;2\pi$
are excluded with respect to the best fit CP violating
values  $\delta \cong \pi/2;3\pi/2$ at more than $4\sigma$.
Correspondingly,  $J_{\text{CP}} = 0$ is also excluded with respect
to the best-fit values $J_{\text{CP}} \simeq (-0.034)$ and
$J_{\text{CP}}\simeq 0.034$ at more than $4 \sigma$.
Further, the $3\sigma$ allowed ranges of values
of both $\delta$ and $J_{\text{CP}}$ form rather
narrow intervals. In the case of
the best fit value $\delta \cong 3\pi/2$, for instance,
we have in the cases of NO and IO spectra:
\begin{eqnarray}
\label{eq:JCPSONHTBMpi2}
{\rm NO: } \quad J_{\text{CP}} \cong -0.034\,,& 0.028 \ltap J_{\text{CP}} \ltap 0.039 \,,
&{\rm or}\\
\label{eq:JCPSONHTBM3pi2}
&  -0.039 \ltap J_{\text{CP}} \ltap -0.028\,,&\\
\label{eq:JCPSOIHTBMpi2}
{\rm IO: } \quad J_{\text{CP}} \cong -0.034\,,&  0.027 \ltap J_{\text{CP}} \ltap 0.039 \,,
&{\rm or}\\
&  -0.039 \ltap J_{\text{CP}} \ltap -0.026 \,,&
\label{eq:JCPSOIHTBM3pi2}
\end{eqnarray}
%
where we have quoted the best fit value of $J_{\text{CP}}$
as well. The positive values are related
to the $\chi^2$ minimum at $\delta = \pi/2$.

\begin{figure}
  \begin{center}
 \subfigure
  {\includegraphics[width=7cm]{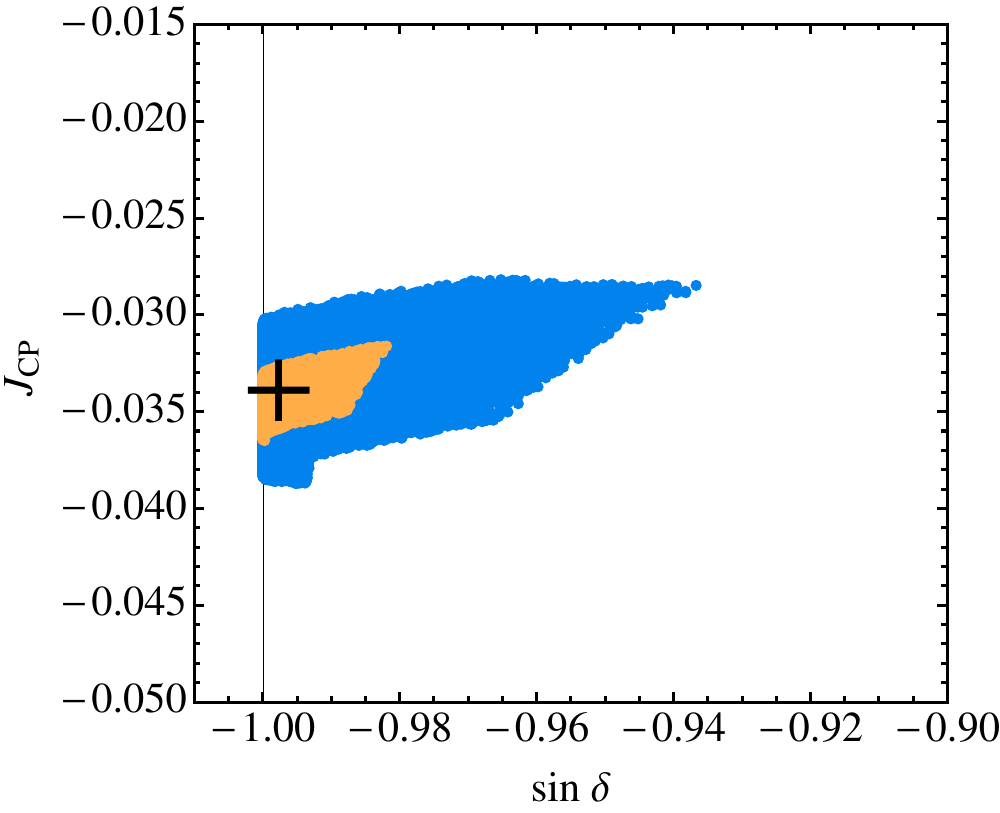}}
 \vspace{5mm}
 \subfigure
    {\includegraphics[width=7cm]{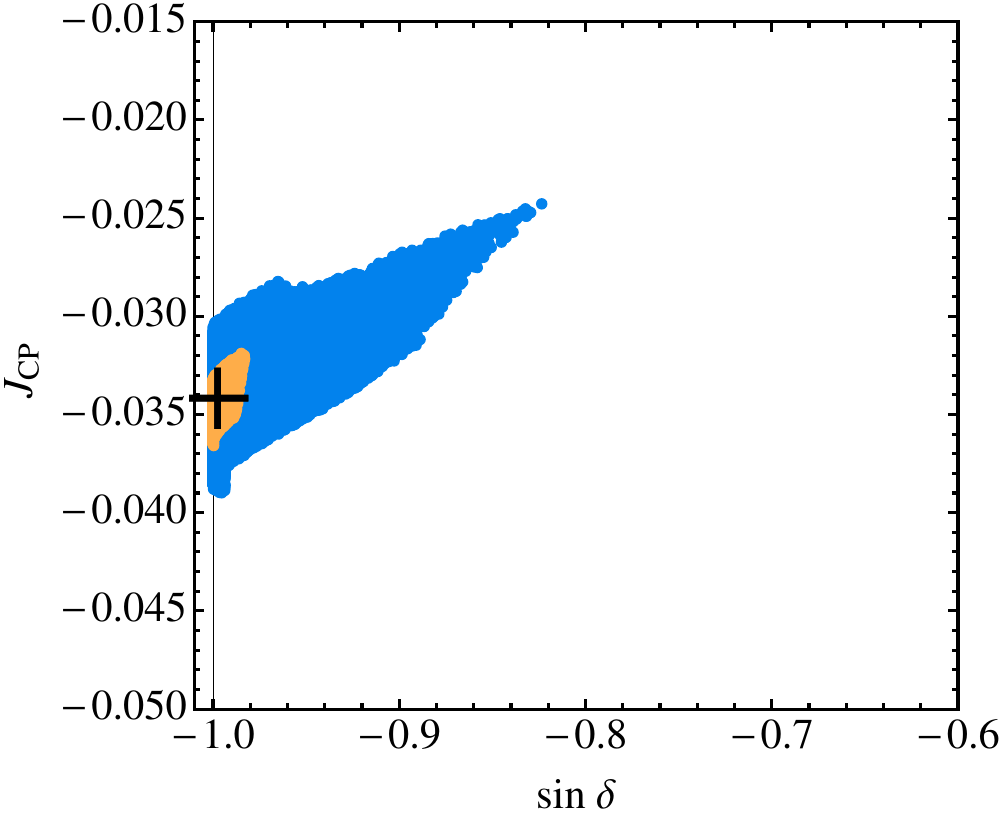}}
    \subfigure
 {\includegraphics[width=7cm]{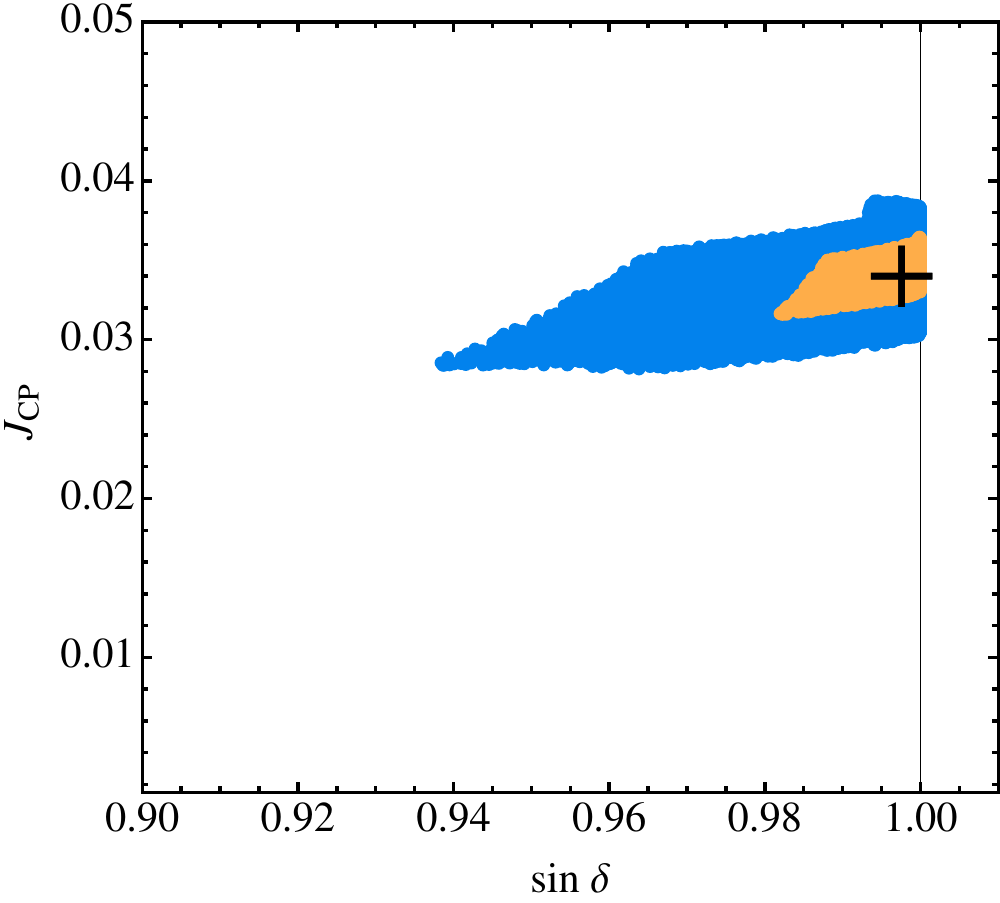}}
 \vspace{5mm}
 \subfigure
    {\includegraphics[width=7cm]{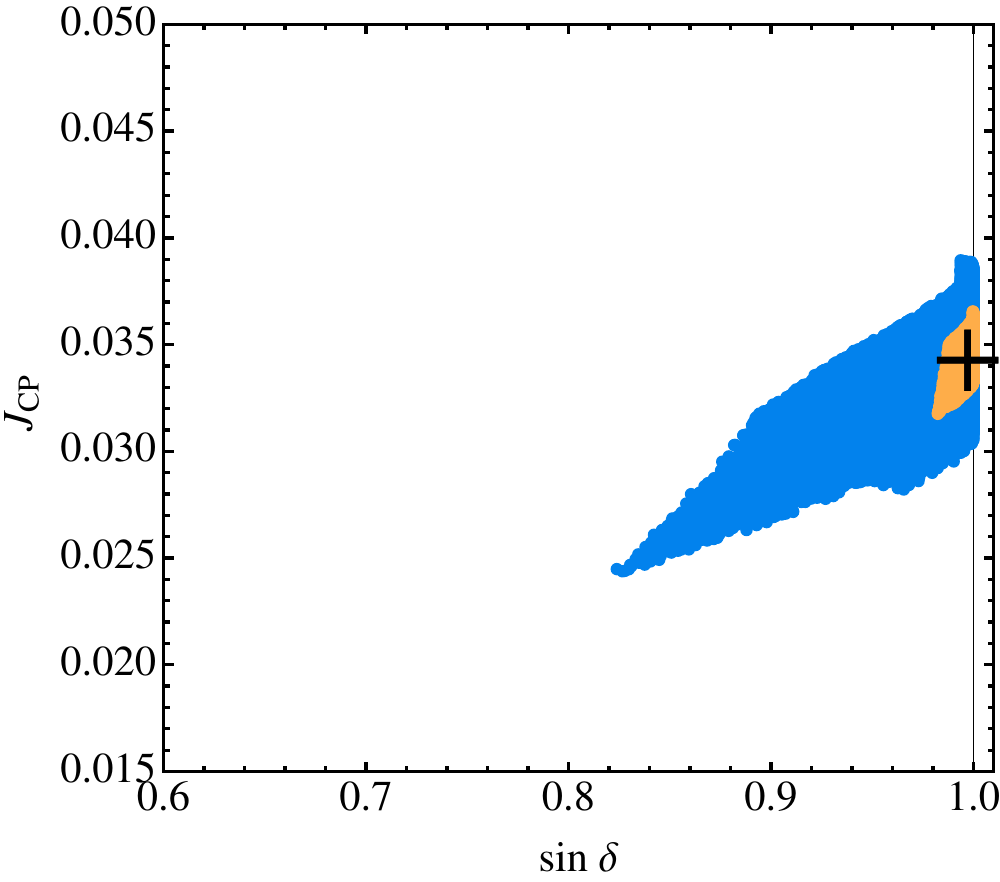}}
     \end{center}
\vspace{-1.0cm}
\caption{\label{fig:FigJCPNI}
Possible values of $\sin\delta$ and
$J_{\text{CP}}$, obtained by using   
the 1$\sigma$ (light brown areas) and 2$\sigma$ ranges 
(blue + light brown areas) of allowed 
values of the mixing angles  
$\theta_{12}$, $\theta_{23}$ and $\theta_{13}$
for NO spectrum (left panels) and
IO spectrum (right panels), and for 
$\sin \delta <0 $ 
(upper panels) and $\sin \delta > 0$ 
(lower panels). The predictions for the best fit 
values of $\theta_{12}$, $\theta_{23}$ and $\theta_{13}$, 
corresponding to $\delta = 266.02^{\circ}$ 
($\sin \delta <0 $) and $\delta = 93.98^{\circ}$ 
($\sin \delta > 0$), are indicated with crosses. 
See text for further details.
}
\end{figure}

The preceding results and discussion 
are illustrated qualitatively in Fig. \ref{fig:FigJCPNI}, 
where we show  the correlation between
the value of $\sin\delta$ and $J_{\text{CP}}$ for the 1$\sigma$ and
2$\sigma$ ranges of allowed values of 
$\sin^2\theta_{12}$, $\sin^2\theta_{23}$ and $\sin^2\theta_{13}$, 
which were taken from Table 1.
The figure was produced assuming flat distribution of the values of 
$\sin^2\theta_{12}$, $\sin^2\theta_{23}$ and $\sin^2\theta_{13}$  
in the quoted intervals around the corresponding best fit 
values. As can be seen from  Fig. \ref{fig:FigJCPNI}, the 
predicted values of both  $\sin\delta$ and $J_{\text{CP}}$ thus 
obtained form rather narrow intervals 
\footnote{The 2$\sigma$ ranges of allowed values of 
$\sin\delta$ and $J_{\text{CP}}$  shown in Fig. \ref{fig:FigJCPNI} 
match approximately the 3$\sigma$ ranges of allowed values of 
$\sin\delta$ and $J_{\text{CP}}$ obtained in 
\cite{Marzocca:2013cr} by performing a more 
rigorous statistical ($\chi^2$) analysis.
}.
%

As it follows from Table 1, the angle $\theta_{23}$ is determined 
using the current neutrino oscillation data
with largest uncertainty. We give next the values of the 
Dirac phase $\delta$ for two values of $\sin^2\theta_{23}$ 
from its $3\sigma$ allowed range, 
$\sin^2\theta_{23} =0.50~{\rm and}~0.60$, and for the best fit 
values of  $\sin^2\theta_{12}$  and $\sin^2\theta_{13}$: 
\begin{eqnarray}
\label{deltas2th2305}
\sin^2\theta_{23} = 0.50:~~\cos\delta = -\,0.123\,,~~
\delta = 97.09^{\circ}~~{\rm or}~~262.91^{\circ}\,;\\[0.30cm]
\label{deltas2th2306}
\sin^2\theta_{23} = 0.60:~~\cos\delta = -\,0.176\,,~~
\delta = 100.12^{\circ}~~{\rm or}~~259.88^{\circ}\,.
\end{eqnarray}
%
These results show that $|\sin\delta|$, which determines the 
magnitude of the CP violation effects in neutrino oscillations,
exhibits very weak dependence on the value of 
$\sin^2\theta_{23}$: for any value of $\sin^2\theta_{23}$ 
from the interval $0.39 \leq \sin^2\theta_{23} \leq 0.60$ 
we get  $|\sin\delta| \geq 0.98$.

  The predictions of the model for $\delta$ and $J_{\text{CP}}$ will be
tested in the experiments searching for CP violation in neutrino
oscillations, which will provide information on the value of the
Dirac phase $\delta$.\\

\subsubsection{The Majorana CPV Phases} 

 Using the expressions for the
angles $\theta^e_{12}$ and $\hat\theta_{23}$ and 
for $\cos\phi$ in terms of
 $\sin\theta_{13}$, $\sin\theta_{12}$ and $\sin\theta_{23}$
and the best fit values of 
 $\sin\theta_{13}$, $\sin\theta_{12}$ and $\sin\theta_{23}$,
we can calculate the numerical form of $U_{\rm PMNS}$ from which  we
can extract the values of the physical CPV Majorana phases. 
We follow the procedure described in \cite{AMEMSPSU5T}.
Obviously, there are two such forms of $U_{\rm PMNS}$
corresponding to the two possible values of $\delta$. 
In the case of 
$\delta =  266.02^{\circ}$  and $\phi \simeq 102.55^{\circ}$ we find: 
\be
 U_{\rm PMNS}  = \left(
\begin{array}{ccc}
 0.822 \; \text{e}^{- \ci 7.47^{\circ}} & 0.547  \; \text{e}^{\ci 16.04^{\circ}} & 0.155  \; \text{e}^{\ci 102.55^{\circ}} \\
 0.436 \; \text{e}^{- \ci 104.08^{\circ}} & 0.658 \; \text{e}^{\ci 114.67^{\circ}} & 0.614 \; \text{e}^{\ci 102.55^{\circ}} \\
 0.365 & -0.517 & 0.774 \\
\end{array}
\right) \; \overline{Q} \;.\\
\ee
%
Recasting this expression in the form of the standard parametrisation of 
$U_{\rm PMNS}$  we get:
\be 
U_{\rm PMNS} = P \; \left(
\begin{array}{ccc}
 0.822  & 0.547  & 0.155  \; \text{e}^{\ci 93.98^{\circ}} \\
 0.436 \; \text{e}^{ \ci 169.41^{\circ}} & 0.658 \; \text{e}^{\ci 4.65^{\circ}} & 0.614 \\
 0.365 \; \text{e}^{\ci 16.04 }& 0.517 \; \text{e}^{\ci 172.53} & 0.774 \\
\end{array}
\right) \; Q_2 \; \overline{Q} \;, \\
\ee
%
where 
$P = {\rm diag}(\text{e}^{\ci(16.04 - 7.47)^{\circ}},\text{e}^{\ci 102.55^{\circ}},1)$,
$Q_2 =  {\rm diag}(\text{e}^{-\ci 16.04^{\circ}},\text{e}^{\ci 7.47^{\circ}},1)$ 
and  $\text{e}^{\ci\,93.98^{\circ}} =  \text{e}^{-\,\ci(360-93.98)^{\circ}} = 
 \text{e}^{-\ci\, 266.02^{\circ}}$.\\
Similarly, in the case of  
$\delta = 93.98$ and $\phi \simeq 257.45^{\circ}$ we obtain:
\be 
U_{\rm PMNS}  = \left(
\begin{array}{ccc}
 0.822 \; \text{e}^{ \ci 7.47^{\circ}} & 0.547  \; \text{e}^{-\ci 16.04^{\circ}} & 0.155  \; \text{e}^{-\ci 102.55^{\circ}} \\
 0.436 \; \text{e}^{\ci 104.08^{\circ}} & 0.658 \; \text{e}^{-\ci 114.67^{\circ}} & 0.614 \; \text{e}^{-\ci 102.55^{\circ}} \\
 0.365 & -0.517 & 0.774 \\
\end{array}
\right) \; \overline{Q} \;.\\
\ee
%
Extracting again phases in diagonal matrices 
on the right hand and left hand sides to get the standard 
parametrisation of $U_{\rm PMNS}$ we find:
\be 
U_{\rm PMNS} = \tilde P \; \left(
\begin{array}{ccc}
 0.822  & 0.547  & 0.155  \; \text{e}^{-\ci 93.98^{\circ}} \\
 0.436 \; \text{e}^{ -\ci 169.41^{\circ}} & 0.658 \; \text{e}^{-\ci 4.65^{\circ}} & 0.614 \\
 0.365 \; \text{e}^{-\ci 16.04 }& 0.517 \; \text{e}^{-\ci 172.53} & 0.774 \\
\end{array}
\right) \; \tilde Q_2 \; \overline{Q}  \;,\\
\ee 
%
where $P = {\rm diag}(\text{e}^{\ci(-16.04 + 7.47)^{\circ}},\text{e}^{-\ci
102.55^{\circ}},1)$ and $\tilde Q_2 =  {\rm diag}(\text{e}^{ \ci
16.04^{\circ}},\text{e}^{- \ci 7.47^{\circ}},1)$.
The phases in the matrices $P$ and $\tilde P$ 
can be absorbed by the charged lepton fields and are unphysical.  
In contrast, the phases in the matrices $Q_2$ and $\tilde Q_2$ 
contribute to the physical Majorana phases.  
We can finally write the Majorana phase matrix in
the parametrization given in (\ref{eq:UPMNS}) ($\phi_1 =0$):
\be  
-\frac{\beta_1}{2} =  \mp  16.04^{\circ} -\beta -\frac{\phi_3}{2}\,,~~
\sin \delta = \mp\, 0.976\,, 
\ee
\be  -\frac{\beta_2}{2} =  
\pm   7.47^{\circ} -\beta
-\frac{\phi_3-\phi_2}{2}\,,~~
\sin \delta = \mp\, 0.976\,. 
\ee
%

In order to calculate the phase $\beta = \gamma - \phi$ we have 
to find the value of $\gamma$. It follows from Eqs. (\ref{th23hat}) 
and (\ref{gammaphi}) that
\ba
\label{cosgamma}
\cos\gamma = \frac{\sin\theta^e_{23}\cos\omega_e}
{\sqrt{2} \sin\hat\theta_{23}}\,, \;\;\; 
\label{singamma}
\sin\gamma = \frac{\pm \cos\theta^e_{23} - \sin\theta^e_{23}\sin\omega_e}
{\sqrt{2} \sin\hat\theta_{23}}\,,\\[0.30cm]
\label{cosphi}
\cos\phi = \frac{\sin\theta^e_{23}\cos\omega_e}
{\sqrt{2} \cos\hat\theta_{23}}\,, \;\;\;
\label{sinphi}
\sin\phi = \frac{\mp \cos\theta^e_{23} - \sin\theta^e_{23}\sin\omega_e}
{\sqrt{2} \cos\hat\theta_{23}}\,, \,
\ea
%
where we used the fact that $\psi_e = \pm \pi/2$.
These equations imply the following relations: 
\ba 
\label{cosgammaphi} 
\cos\gamma = \cos\phi \, \frac{\cos\hat\theta_{23}}
{\sin\hat\theta_{23}}\,,\hspace{4.35cm} \\[0.30cm]
\sin\phi \cos\hat\theta_{23} + \sin\gamma \sin\hat\theta_{23} = 
-\, \sqrt{2} \sin\theta^e_{23}\sin\omega_e\,.
\label{singammaphi}
\ea 
%
It is clear from Eq. (\ref{cosgammaphi}) 
that the value of  $\cos\gamma$ can be determined 
knowing the values of $\cos\phi$ and $\sin\hat\theta_{23}$,
independently of the values of $\theta^e_{23}$ and $\omega_e$.
This, obviously, allows to find also $|\sin \gamma|$, 
but not the sign of $\sin\gamma$.
In the case of  $\sin\theta^e_{23}\sin\omega_e \ll 1$ of interest, 
Eq. (\ref{singammaphi}) allows to correlate the sign of 
 $\sin \gamma$ with the sign of $\sin\phi$ 
and thus to determine $\gamma$ for a given $\phi$:
we have $\sin\gamma < 0$ if $\sin\phi >0$, 
and $\sin\gamma > 0$ for $\sin\phi < 0$.
Thus, for $\phi =  102.5530^{\circ}$ 
(corresponding to $\delta = 266.02^{\circ}$) we 
find $\gamma = -\,105.4118$ and
$\beta = \gamma - \phi = -\,207.9648^{\circ} = 
-\,(180 + 27.9648)^{\circ}$, 
while for $\phi = -\,102.5530^{\circ}$ 
(corresponding to $\delta = 93.98^{\circ}$)
we obtain $\gamma = +\,105.4118$ and 
$\beta = + 207.96^{\circ} = + \,(180 + 27.96)^{\circ}$.

The results thus derived allow us to calculate
numerically the Majorana CPV phases.
For the best fit values of the neutrino 
mixing angles we get:
\be
\begin{split}
& \beta_1 = (23.84 + 360 +\phi_3)^{\circ}  \;,\; \beta_2 = (70.88 + 360 - \phi_2 + \phi_3)^{\circ}~ 
\mbox{for $\phi = -102.55^{\circ}$ ($\delta = 93.98^{\circ}$)}\;; ~~~\\
\end{split}
\ee
\be
\begin{split}
& \beta_1 = (-23.84 - 360 + \phi_3)^{\circ} = 
(-23.84 + 360 + \phi_3)^{\circ}\;, \\
& \beta_2 = (-70.88 -360 -\phi_2 + \phi_3)^{\circ}  
= (-70.88 + 360 -\phi_2 + \phi_3)^{\circ}~ 
\mbox{for $\phi = 102.55^{\circ}$ ($\delta = 266.02^{\circ}$)}\;, \\
\end{split}
\ee
%
where we have used the fact that $\beta_{1(2)}$ 
and $\beta_{1(2)} + 4\pi$ lead to the same physical 
results. In the cases of the three types of 
neutrino mass spectrum allowed by the model, which are 
characterised, in particular, by specific values of the 
$\phi_2$  and $\phi_3$ we find:
\begin{itemize}
\item NO A spectrum, i.e., $\phi_2 = \phi_3 = 0$:
\be
\begin{split}
& \beta_1 = (23.84 + 360)^{\circ}  \;,\; \beta_2 = (70.88 + 360)^{\circ}~ 
\mbox{for $\phi = -102.55^{\circ}$ ($\delta = 93.98^{\circ}$)}\;, \\
& 
\beta_1 = (-23.84 + 360)^{\circ} \;,\; \beta_2 = (-70.88 + 360)^{\circ}~
\mbox{for $\phi = 102.55^{\circ}$ ($\delta = 266.02^{\circ}$)}\;; \\
\end{split}
\ee
%
\item  NO B spectrum, i.e., $\phi_2 = 0$ and $\phi_3 = \pi$:
\be
\begin{split}
& \beta_1 = (23.84 + 540)^{\circ}  \;,\; \beta_2 = (70.88 + 540)^{\circ} 
\mbox{for $\phi = -102.55^{\circ}$ ($\delta = 93.98^{\circ}$)} \;, \\
& 
\beta_1 = (-23.84 + 540)^{\circ} \;,\; \beta_2 = (-70.88 + 540)^{\circ} 
\mbox{for $\phi = 102.55^{\circ}$ ($\delta = 266.02^{\circ}$)} \;; \\
\end{split}
\ee
%
\item IO spectrum, $\phi_3 = 0$ and $\phi_2 = \pi$:
\be
\begin{split}
& \beta_1 = (23.84 + 360)^{\circ}  \;,\; \beta_2 = (70.88 + 180)^{\circ}~ 
\mbox{for $\phi = -102.55^{\circ}$ ($\delta = 93.98^{\circ}$)} \;, \\
& 
\beta_1 = (-23.84 + 360)^{\circ} \;,\; \beta_2 = (-70.88 + 180)^{\circ}~ 
\mbox{for $\phi = 102.55^{\circ}$ ($\delta = 266.02^{\circ}$)} \;, \\
\end{split}
\ee
%
\end{itemize}
where again we have used the fact that 
$\beta_{1(2)}$ and $\beta_{1(2)} + 4\pi$ are 
physically indistinguishable.

\subsubsection{The Neutrinoless Double Beta Decay Effective Majorana Mass} 

Knowing the values of the neutrino masses and 
the Majorana and Dirac CPV phases 
we can  derive predictions for the neutrinoless double beta
(\betabeta-) decay effective Majorana mass $\meff$ (see, e.g., \cite{BiPet87}).
Since $\meff$ depends only on the cosines of the CPV phases, 
we get the same result for $\phi = +\, 102.55^{\circ}$
($\delta = 266.02^{\circ}$) and  $\phi = -\, 102.55^{\circ}$
($\delta = 93.98^{\circ}$).

Thus, for $\phi = \pm 102.55^{\circ}$, using the 
best fit values of the neutrino mixing angles, we 
obtain:
\begin{eqnarray}
\meff = 4.88 \times 10^{-3} \, { \rm eV}\,,~~{\rm NO~A~spectrum}\,; \label{MeffNOA}\\[0.30cm]
\meff = 7.30 \times 10^{-3} \, { \rm eV}\,,~~{\rm NO~B~spectrum}\,;\\[0.30cm]
\meff = 26.34 \times 10^{-3} \, { \rm eV}\,,~~{\rm IO~spectrum} \label{MeffIO}\;.
\end{eqnarray}
\begin{figure}
\begin{center}
\includegraphics[width=9cm]{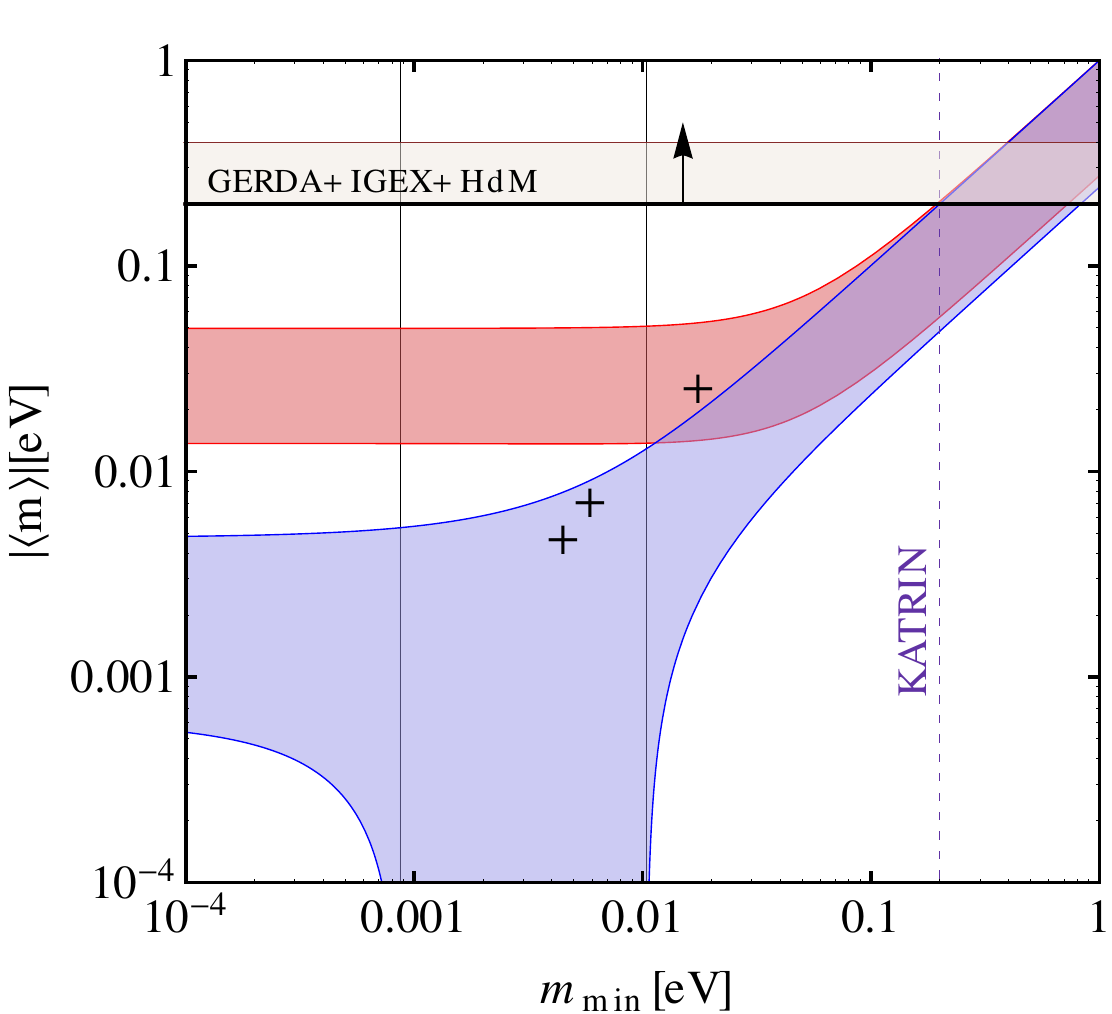}
\end{center}
\caption{\label{fig:FigMeff}
The 3$\sigma$ allowed regions of values of
the effective Majorana mass $\meff$ as functions 
of the lightest neutrino mass $m_{min}$ for 
the NO (blue area) and  IO (red area) 
neutrino mass spectra. The regions are obtained 
by using the experimentally determined values of the 
neutrino oscillation parameters 
(including the $1\sigma$ uncertainties) 
quoted in Table 1. The black crosses 
correspond to the predictions of the model 
constructed in the present article, Eqs. (\ref{MeffNOA}) - (\ref{MeffIO}).
The horizontal band indicates the upper bound
$\meff  \sim0.2-0.4$ eV obtained using the 90 \% C.L.
limit on the half-life of $^{76}$Ge
reported in \cite{Agostini:2013mba}.
The  dotted line represents
the prospective upper limit from
the $\beta$-decay experiment
KATRIN \cite{MainzKATRIN}.
}
\end{figure}

In Fig. \ref{fig:FigMeff} we show the 
general phenomenologically allowed 3$\sigma$ 
range of values of $\meff$ for the NO (blue area) 
and IO (red area) neutrino mass spectra as a function of the 
lightest neutrino mass. 
The values of $\meff$ quoted above and 
corresponding to the three types 
of neutrino mass spectrum (NO A, NO B and IO), 
predicted by the model 
constructed in the present article, 
are indicated with black crosses. 
The vertical lines in Fig. \ref{fig:FigMeff}
correspond to 
$m_{min} = 8.6 \times 10^{-4}\,{\rm eV}$ and 
$1.0 \times 10^{-2}\, {\rm eV}$; 
for a given value of $m_{min}$ from the interval 
determined by these two values, 
[$8.6 \times 10^{-4},1.0 \times 10^{-2}$] eV,
one can have $\meff = 0$ for 
specific values of the Majorana CPV phases.

%
\subsubsection{Limiting Cases}
%
%
Finally, there are two interesting 
limiting forms of the charged lepton 
Yukawa coupling (mass) matrix ${\rm Y}_e$: 
they correspond to i) $k = 0$, i.e., $\eta = 0$ or $\pi$, and 
ii) $d = 0$, i.e., $\eta = \pm \pi/2$.
In the case of $k=0$, the TBM prediction for $\theta_{12}$ 
does not depend on $\theta_{23}^e$ anymore; if  $d = 0$,
even $\theta_{23}$ itself does not depend on $\theta_{23}^e$ anymore.
Up to next-to-leading order we find:
\begin{equation}
 \begin{split}
 \text{i) }  & \sin^2 \theta_{12}= \frac{1}{3} + \frac{1}{3} \sin^2\theta_{13}\approx \frac{1}{3} \text{ for } k=0, \eta = 0,\pi \;,\\
 \text{ii) } & \sin^2 \theta_{23}= \frac{1}{2} - \frac{1}{2} \sin^2\theta_{13}\approx \frac{1}{2} \text{ for } d=0, \eta = \pm\pi/2 \;, \\
 \end{split}
\end{equation}
%
where we have written the corrections in terms of $\theta_{13}$.
Both cases could be realised by choosing a certain set of messengers.
If we remove the messenger pair 
$\Sigma_{2^{\pr}}^A$, $\bar \Sigma_{2^{\sec}}^A$,
our model would correspond to the case i), 
while if we remove the messenger pair
$\Sigma_{1^{\sec}}^C$, $\bar \Sigma_{1^{\pr}}^C$, the model 
would correspond to the case ii). 
The model we have constructed, which includes both messenger
pairs, gives a somewhat better description of the current data 
on the neutrino mixing angles. 
This brief discussion shows how important 
the messenger sector can be for getting 
meaningful predictions.

\section{Summary and Conclusions}

 In this work we have analyzed  the presence of a generalised CP
symmetry, $H_{\text{CP}}$, combined with the non-Abelian discrete group
$T^{\prime}$ in the lepton flavour space, i.e. the possibility of
the existence of a symmetry group $G_f= T^{\prime}\rtimes H_{\text{CP}}$
acting among the three generations of charged leptons and neutrinos.
The phenomenological implications of the breaking of such a symmetry
group both in the charged lepton and neutrino sectors are thus
explored especially in connection with the CP violation appearing in
the leptonic mixing matrix, $U_{\text{PMNS}}$.

First of all we have derived in Section \ref{Sec:TprandCP} all the
possible generalised CP transformations for all the representations
of the $T^{\prime}$ group i.e. we found all possible outer
automorphisms of the group $T^{\prime}$ following the consistency
conditions given in \cite{Holthausen:2012dk,Feruglio:2012cw,Ding:2013bpa}.
We have
chosen as generalised CP symmetry the transformation $u:(T,
S)\rightarrow (T^2, S^2T^2 S\,T)$
which corresponds
to a $Z_2$ symmetry 
and it is defined up to an
inner automorphism.
The transformation $u$ is particularly convenient since, in
the basis chosen for the generators $S$ and $T$, for the 1 and
3-dimensional representations it is trivially defined as the
identity up to a global unphysical phase $\theta_r$ where the index
$r$ refers to the representation. More importantly we found that,
given this specific generalised CP symmetry combined with
$T^{\prime}$, it is possible to fix the vevs of the flavon fields to
real values in such a way that no complex phases, and thus no physical
CP violation, stem from the vevs themselves.

Moreover, for a list of possible renormalisable operators, namely
$\lambda \mathcal{O} = \lambda (A \times B \times C)$ where
$\lambda$ is the coupling constant and $A$, $B$, $C$ are fields, we
derived the constraints on the phase  $\lambda$ under the assumption
of invariance under the generalised CP transformation. This list of
possible operators can be used to construct a CP-conserving
renormalisable superpotential for the flavon sector and therefore
can be used in order to show that real vev structures can be
achieved.

Motivated by this preliminary study we constructed in Section~\ref{Sec:Model}
a supersymmetric flavour model able to describe the
observed patterns and mixing for three generations of charged lepton
fields and the three light active neutrinos.

We have constructed an effective
superpotential with operators up to mass dimension six giving the
charged lepton and neutrino Yukawa couplings and the Majorana
mass term for the RH neutrinos.
Naturally small neutrino masses are generated by
the type I see-saw mechanism.
At leading order, the mixing in the neutrino sector
is described by the tri-bimaximal
mixing, which is then perturbed by additional contributions coming
from the charged lepton sector. The latter are responsible for the
compatibility of the predictions on the mixing angles with the
experimental values and, in particular, with the non-zero value of the
reactor mixing angle $\theta_{13}$.

Similarly to what was found in \cite{Meroni:2012ty}, we find
that both types of neutrino mass spectrum - with normal 
ordering (NO) and inverted ordering (IO) -  
are possible within the model and that 
the NO spectrum can be of two varieties, A and B. They differ 
by the value of the lightest neutrino mass.
Only one spectrum of the IO type is compatible with the model.
For each of the three neutrino mass spectra,   
NO A, NO B and IO, the
absolute scale of neutrino masses is predicted
with relatively small uncertainty.
This allows us to predict the value of the sum of the 
neutrino masses for the three spectra.
The Dirac phase $\delta$ is predicted to be approximately
$\delta \cong \pi/2$ or $3\pi/2$. More concretely, 
for the best fit values of the neutrino mixing angles 
quoted in Table 1 we get
$\delta = 93.98^{\circ}$ or $\delta = 266.02^{\circ}$. 
The deviations of $\delta$ from the values 
$90^{\circ}$ and $270^{\circ}$ are
correlated with the deviation of atmospheric 
neutrino mixing angle $\theta_{23}$ from $\pi/4$.
Thus, the CP violating effects in neutrino oscillations
are predicted to be nearly maximal (given the values of the
neutrino mixing angles) and experimentally observable.
The values of the Majorana CPV phases are also predicted 
by the model. This allows us to predict 
the neutrinoless double beta decay effective Majorana mass 
in each of the three cases of neutrino mass spectrum 
allowed by the model, NO A, NO B and IO.
The predictions of the model can be tested
in ongoing and future planned
i) accelerators experiments searching for CP violation in 
neutrino oscillations (T2K, NO$\nu$A, etc.), 
ii) experiments aiming to determine the absolute 
neutrino mass scale, and 
iii) experiments searching for 
neutrinoless double beta decay.

It is important to comment  that in this model the physical CP
violation emerging in the PMNS mixing matrix stems only from the
charged lepton sector. Indeed, in the neutrino sector the Majorana
mass matrix and the Dirac Yukawa couplings are real and the CP
violation is caused by the complex CP violating
phases arising in the charged lepton sector.
The presence of the latter is a consequence of
the requirement of invariance of the theory
under the generalised CP symmetry at the
fundamental level and
of the complex CGs of the $T^{\pr}$ group.

We also found that the residual group in the charged lepton sector
is trivial i.e. $G_e = \emptyset$ and 
since the phases of the flavon vevs are
completely independent of the coupling constants of the flavon
superpotential, the CP symmetry is broken geometrically 
(according to the definition of ``geometrical CP violation'' 
given in \cite{Branco:1983tn}).
In the neutrino sector,  the residual subgroup
is instead a Klein group, $G_\nu=K_4 = Z_2 \times Z_2$ with
one $Z_2$ coming from the generalised CP symmetry
$H_{\text{CP}}$.

 Concluding,  we have  shown that the spontaneous breaking of a
symmetry group $G_f= T^{\prime}\rtimes H_{\text{CP}}$ in the leptonic
sector through a real flavon vev structure is possible and, at the
same time, CP violation in the leptonic sector can take place.
In this scenario the appearance of the CP violating phases
in the PMNS mixing matrix can be traced to two factors:
i) the requirement of invariance of the Lagrangian of the theory
under $H_{\text{CP}}$ at the fundamental level, and
ii) the complex CGs of the $ T^{\prime}$ group.
The model we have constructed allows for two neutrino mass spectra
with normal ordering (NO) and one with inverted ordering (IO).
For each of the three spectra the
absolute scale of neutrino masses is predicted
with relatively small uncertainty.
The value of the Dirac CP violation (CPV) phase
$\delta$ in the lepton mixing
matrix is predicted to be $\delta \cong \pi/2~{\rm or}~ 3\pi/2$.
Thus, the CP violating effects in neutrino oscillations
are predicted to be nearly maximal and 
experimentally observable.
We present also predictions
for the sum of the neutrino masses, 
for the Majorana CPV phases and
for the effective Majorana mass
in neutrinoless double beta decay.
The predictions of the model can be tested
in a variety of ongoing and future planned
neutrino experiments.

\section*{Note added}

After the submission of our article to the arXiv,
an update of the global fits to the neutrino oscillation
data appeared \cite{Capozzi:2013csa}. The results reported in
\cite{Capozzi:2013csa} are in agreement with the predictions of
our model. More specifically, the authors of
\cite{Capozzi:2013csa} find  that the best fit value of
$\delta$ is $\delta \approx 3 \pi/2$,
which is one of the two possible values predicted by
in our model. Similar results on $\delta$ were
obtained in the global analysis of the neutrino oscillation data
performed in \cite{TAUP}.

\section*{Acknowledgements}

The work of M.~Spinrath was partially supported by the ERC Advanced Grant
no.~267985 ``DaMESyFla'', by the EU Marie Curie ITN ``UNILHC''
(PITN-GA-2009-237920). A. Meroni acknowledges MIUR (Italy)
for financial support under the program Futuro in Ricerca 2010 (RBFR10O36O).
This work was supported in part also by the European Union FP7 
ITN INVISIBLES (Marie Curie Actions, PITN-GA-2011-289442-INVISIBLES),
by the INFN program on ``Astroparticle Physics''(A.M., S.T.P.) and by
the World Premier International Research Center
Initiative (WPI Initiative), MEXT, Japan  (S.T.P.).

\appendix

\section{\texorpdfstring{Technicalities about $\boldsymbol{T^{\pr}}$}{Technicalities about T prime}}

\label{App:groupTp}

The group $T^{\pr}$ is the double covering group of $A_4$ and it is defined through the algebraic relations:
\be
S^2 = R \quad R^2 = T ^3 = (S T)^3 = E \quad RT = TR \,.
\ee
The number of the unitary irreducible representations of a discrete group is equal to the number of the conjugacy classes. For $T^{\prime}$ they are seven, which are classified given the elements $T$, $S$,  because $R \equiv S^2$, we summarize them as
\begin{equation}
\begin{split}
& 1\,C^1:  \left \{E\right \} \;, \quad 1^{\pr}\,C^2:  \left \{ S^2 \right \} \\
& 4 \,C^3:  \left \{ T, S^3 T S, S T, T S \right \} \;, \quad 4^{\pr} \, C^3: \left \{ T^2, S^2 T S T, S^2 T^2 S, S^3 T^2\right \} \\
& 4^{\sec} \, C^6: \left \{ S^2 T, S T S, S^3 T, S^2 T S \right \} \;, \quad 4^{\pr \pr \pr} \, C^6: \left \{ S^2 T^2, T S T, T^2 S, S T^2 \right \} \\
& 6 \,C^4: \left \{  S, S^3, T S T^2, T^2 S T, S^2 T S T^2, S^2 T^2 S T \right \}
\end{split}
\end{equation}
The representations of $T^{\prime}$ can be expressed as
\beq
\begin{split}
& \mathbf{1}: T = 1 \;, \; R = 1 \;, \; S = 1 \,;\\
& \mathbf{1^{\prime}}: T = \omega  \;, \; R = 1  \;, \; S = 1 \,;\\
& \mathbf{1^{\prime \prime}}: T = \omega^2  \;, \; R = 1  \;, \; S = 1 \,;\\
& \mathbf{2}:
T = \left(
\begin{array}{cc}
 \omega ^2 & 0 \\
 0 & \omega  \\
\end{array}
\right)  \;, \;
 R = \left(
\begin{array}{cc}
 -1 & 0 \\
 0 & -1 \\
\end{array}
\right)  \;, \;
 S = \left(
\begin{array}{cc}
 -\frac{i}{\sqrt{3}} & -\sqrt{\frac{2}{3}} p \\
 \sqrt{\frac{2}{3}} \bar p & \frac{i}{\sqrt{3}} \\
\end{array}
\right); \\
& \mathbf{2^{\prime}}:
T = \left(
\begin{array}{cc}
 \omega^3 & 0 \\
 0 & \omega^2  \\
\end{array}
\right)  \;, \;
 R = \left(
\begin{array}{cc}
 -1 & 0 \\
 0 & -1 \\
\end{array}
\right)  \;, \;
 S = \left(
\begin{array}{cc}
 -\frac{i}{\sqrt{3}} & -\sqrt{\frac{2}{3}} p \\
 \sqrt{\frac{2}{3}} \bar p & \frac{i}{\sqrt{3}} \\
 \end{array}
\right); \\
& \mathbf{2^{\prime \prime}}:
T = \left(
\begin{array}{cc}
 \omega & 0 \\
 0 & 1  \\
\end{array}
\right)  \;, \;
 R = \left(
\begin{array}{cc}
 -1 & 0 \\
 0 & -1 \\
\end{array}
\right)  \;, \;
 S = \left(
\begin{array}{cc}
 -\frac{i}{\sqrt{3}} & -\sqrt{\frac{2}{3}} p \\
 \sqrt{\frac{2}{3}} \bar p & \frac{i}{\sqrt{3}} \\
 \end{array}
\right); \\
& \mathbf{3}:
T = \left(
\begin{array}{ccc}
 1 & 0 & 0 \\
 0 & \omega  & 0 \\
 0 & 0 & \omega ^2 \\
\end{array}
\right) \;, \;
 R = \left(
\begin{array}{ccc}
 1 & 0 & 0 \\
 0 & 1 & 0 \\
 0 & 0 & 1 \\
\end{array}
\right)  \;, \;
 S =\left(
\begin{array}{ccc}
 -\frac{1}{3} & \frac{2 \omega }{3} & \frac{2 \omega ^2}{3} \\
 \frac{2 \omega ^2}{3} & -\frac{1}{3} & \frac{2 \omega }{3} \\
 \frac{2 \omega }{3} & \frac{2 \omega ^2}{3} & -\frac{1}{3} \\
\end{array}
\right). \\
\end{split}
\eeq
We use the definition of the representation of $T^{\pr}$ given in \cite{Feruglio:2007uu} in which $\omega$ and $p$ are fixed to be respectively $\omega = \text{e}^{\frac{2 \ci \pi }{3}}$ and $p = \text{e}^{\frac{\ci \pi }{12}}$.
Finally $T^{\pr}$ has $n = 13$ subgroups excluding the whole group:
\begin{itemize}
\item Trivial subgroup
\\
$\mathbb{E} = \left \{ E \right \}$;
\item $Z_2$ subgroup
\\
$Z^{S^2}_2 = \left \{ E, S^2 \right \}$;
\item $Z_3$ subgroups
\\
$Z^{T}_3 = \left \{ E, T, T^2 \right \}$, $Z^{S^3 T S}_3 = \left \{ E, S^3 T S, S^2 T S T \right \}$,
$Z^{ST}_3 = \left \{ E, S T, S^2 T^2 S \right \}$, $Z^{T S}_3 = \left \{ E, T S, S^3 T^2 \right \}$;
\item $Z_4$ subgroups
\\
$Z^{S}_4 = \left \{ E, S, S^2, S^3 \right \}$, $Z^{T S T^2}_4 = \left \{ E, T S T^2, S^2, S^2 T S T^2 \right \}$,
$Z^{T^2 S T}_4 = \left \{ E, T^2 S T, S^2, S^2 T^2 S T \right \}$;
\item $Z_6$ subgroups
\\
$Z^{S^2 T}_6 = \left \{ E, S^2 T, T^2, S^2, T, S^2 T^2 \right \}$,
$Z^{S T S}_6 = \left \{ E, S T S, S^2 T S T, S^2, S^3 T S, T S T \right \}$,
\\
$Z^{S^3 T}_6 = \left \{ E, S^3 T, S^2 T^2 S , S^2, S T, T^2 S \right \}$,
$Z^{S^2 T S}_6 = \left \{ E, S^2 T S , S^3 T^2, S^2, T S, S T^2 \right \}$.
\end{itemize}

A complete table of the CGs coefficients can be found in \cite{Meroni:2012ty}.

\section{Messenger Sector}
\label{App:Mess}

\begin{table}
\centering
\begin{tabular}{lcccccccccc}
\toprule
 & $SU(2)$ & $U(1)_Y$ & $T'$ & $U(1)_R$ & $Z_8$ & $Z_4$ & $Z_4$ & $Z_3$ & $Z_3$ & $Z_2$ \\
 \midrule
$\Xi_1$ , $\bar{\Xi}_1$  &  $\mathbf{2}$ ,$\mathbf{2}$ &  1,1  &  $\mathbf{1}$ ,$\mathbf{1}$ &  0,2  &  7,1  &  0,0  &  0,0  &  0,0  &  1,2  &  0,0\\
\midrule
$\Sigma_1^A$,$\bar{\Sigma}_1^A$ &  $\mathbf{2}$ ,$\mathbf{2}$ &  1,1  &  $\mathbf{1}$ ,$\mathbf{1}$ &  1,1  &  1,7  &  3,1  &  3,1  &  2,1  &  1,2  &  1,1\\
$\Sigma_1^B$,$\bar{\Sigma}_1^B$ &  $\mathbf{1}$ ,$\mathbf{1}$ &  2,2  &  $\mathbf{1}$ ,$\mathbf{1}$ &  1,1  &  2,6  &  1,3  &  1,3  &  2,1  &  0,0  &  1,1\\
$\Sigma_{1'}^A$ ,$\bar{\Sigma}_{1''}^A$ &  $\mathbf{2}$ ,$\mathbf{2}$ &  1,1  &  $\mathbf{1}'$,$\mathbf{1}''$ &  1,1  &  7,1  &  3,1  &  3,1  &  0,0  &  2,1  &  0,0\\
$\Sigma_{1'}^B$ ,$\bar{\Sigma}_{1''}^B$ &  $\mathbf{1}$ ,$\mathbf{1}$ &  2,2  &  $\mathbf{1}'$,$\mathbf{1}''$ &  1,1  &  0,0  &  1,3  &  1,3  &  0,0  &  1,2  &  0,0\\
$\Sigma_{1''}^A$,$\bar{\Sigma}_{1'}^A$  &  $\mathbf{2}$ ,$\mathbf{2}$ &  1,1  &  $\mathbf{1}''$ ,$\mathbf{1}'$  &  1,1  &  1,7  &  3,1  &  3,1  &  2,1  &  1,2  &  1,1\\
$\Sigma_{1''}^B$,$\bar{\Sigma}_{1'}^B$  &  $\mathbf{2}$ ,$\mathbf{2}$ &  1,1  &  $\mathbf{1}''$ ,$\mathbf{1}'$  &  1,1  &  1,7  &  3,1  &  1,3  &  0,0  &  0,0  &  0,0\\
$\Sigma_{1''}^C$,$\bar{\Sigma}_{1'}^C$  &  $\mathbf{2}$ ,$\mathbf{2}$ &  1,1  &  $\mathbf{1}''$ ,$\mathbf{1}'$  &  1,1  &  6,2  &  3,1  &  0,0  &  1,2  &  1,2  &  0,0\\
$\Sigma_{2'}^A$ ,$\bar{\Sigma}_{2''}^A$ &  $\mathbf{2}$ ,$\mathbf{2}$ &  1,1  &  $\mathbf{2}'$,$\mathbf{2}''$ &  1,1  &  0,0  &  1,3  &  0,0  &  1,2  &  1,2  &  0,0\\
$\Sigma_{2''}^A$,$\bar{\Sigma}_{2'}^A$  &  $\mathbf{2}$ ,$\mathbf{2}$ &  1,1  &  $\mathbf{2}''$ ,$\mathbf{2}'$  &  1,1  &  0,0  &  1,3  &  0,0  &  1,2  &  1,2  &  0,0\\
\midrule
$\Delta_{1}^A$,$\bar{\Delta}_{1}^A$ &  $\mathbf{1}$ ,$\mathbf{1}$ &  0,0  &  $\mathbf{1}$ ,$\mathbf{1}$ &  0,2  &  0,0  &  2,2  &  0,0  &  0,0  &  0,0  &  0,0\\
$\Delta_{1}^B$,$\bar{\Delta}_{1}^B$ &  $\mathbf{1}$ ,$\mathbf{1}$ &  0,0  &  $\mathbf{1}$ ,$\mathbf{1}$ &  0,2  &  0,0  &  0,0  &  0,0  &  2,1  &  2,1  &  0,0\\
$\Delta_{1'}^A$ ,$\bar{\Delta}_{1''}^A$ &  $\mathbf{1}$ ,$\mathbf{1}$ &  0,0  &  $\mathbf{1}'$,$\mathbf{1}''$ &  0,2  &  0,0  &  0,0  &  0,0  &  0,0  &  2,1  &  0,0\\
$\Delta_{2'}^A$ ,$\bar{\Delta}_{2''}^A$ &  $\mathbf{1}$ ,$\mathbf{1}$ &  0,0  &  $\mathbf{2}'$,$\mathbf{2}''$ &  0,2  &  0,0  &  0,0  &  3,1  &  1,2  &  1,2  &  1,1\\
$\Delta_{3}^A$,$\bar{\Delta}_{3}^A$ &  $\mathbf{1}$ ,$\mathbf{1}$ &  0,0  &  $\mathbf{3}$ ,$\mathbf{3}$ &  0,2  &  4,4  &  0,0  &  0,0  &  0,0  &  0,0  &  1,1\\
$\Delta_{3}^B$,$\bar{\Delta}_{3}^B$ &  $\mathbf{1}$ ,$\mathbf{1}$ &  0,0  &  $\mathbf{3}$ ,$\mathbf{3}$ &  0,2  &  6,2  &  2,2  &  0,0  &  1,2  &  0,0  &  1,1\\
$\Delta_{3}^C$,$\bar{\Delta}_{3}^C$ &  $\mathbf{1}$ ,$\mathbf{1}$ &  0,0  &  $\mathbf{3}$ ,$\mathbf{3}$ &  0,2  &  6,2  &  1,3  &  2,2  &  0,0  &  1,2  &  0,0\\
$\Delta_{3}^D$,$\bar{\Delta}_{3}^D$ &  $\mathbf{1}$ ,$\mathbf{1}$ &  0,0  &  $\mathbf{3}$ ,$\mathbf{3}$ &  0,2  &  5,3  &  0,0  &  3,1  &  0,0  &  1,2  &  0,0\\
$\Delta_{3}^E$,$\bar{\Delta}_{3}^E$ &  $\mathbf{1}$ ,$\mathbf{1}$ &  0,0  &  $\mathbf{3}$ ,$\mathbf{3}$ &  0,2  &  2,6  &  0,0  &  2,2  &  2,1  &  0,0  &  0,0\\
\bottomrule
\end{tabular}
\caption{
List of the messengers fields and their transformation properties.
\label{tab:MessFields}}
\end{table}

The effective model we have considered so far contains
only non-renormalisable operators allowed by the symmetry
group $G_f \times Z_2 \times Z_3^2
\times Z_4^2 \times Z_8 \times
U(1)_R$. But in fact using only this symmetry there would
be more effective operators allowed which might spoil our model
predictions.

Therefore we discuss in
in this section we a so-called ultraviolet
completion defining a renormalisable theory
which gives the effective model described in the previous
sections after integrating out the heavy messenger superfields.
In this way we can justify why we have chosen only a
certain subset of the effective operators allowed by the symmetries.
The quantum numbers of the messenger fields
are given in Table~\ref{tab:MessFields}.
We label them with $\Sigma$, $\Xi$
and $\Delta$ for the charged
lepton, neutrino and flavon sector
respectively.

\begin{figure}
\begin{center}
\includegraphics[scale=0.7]{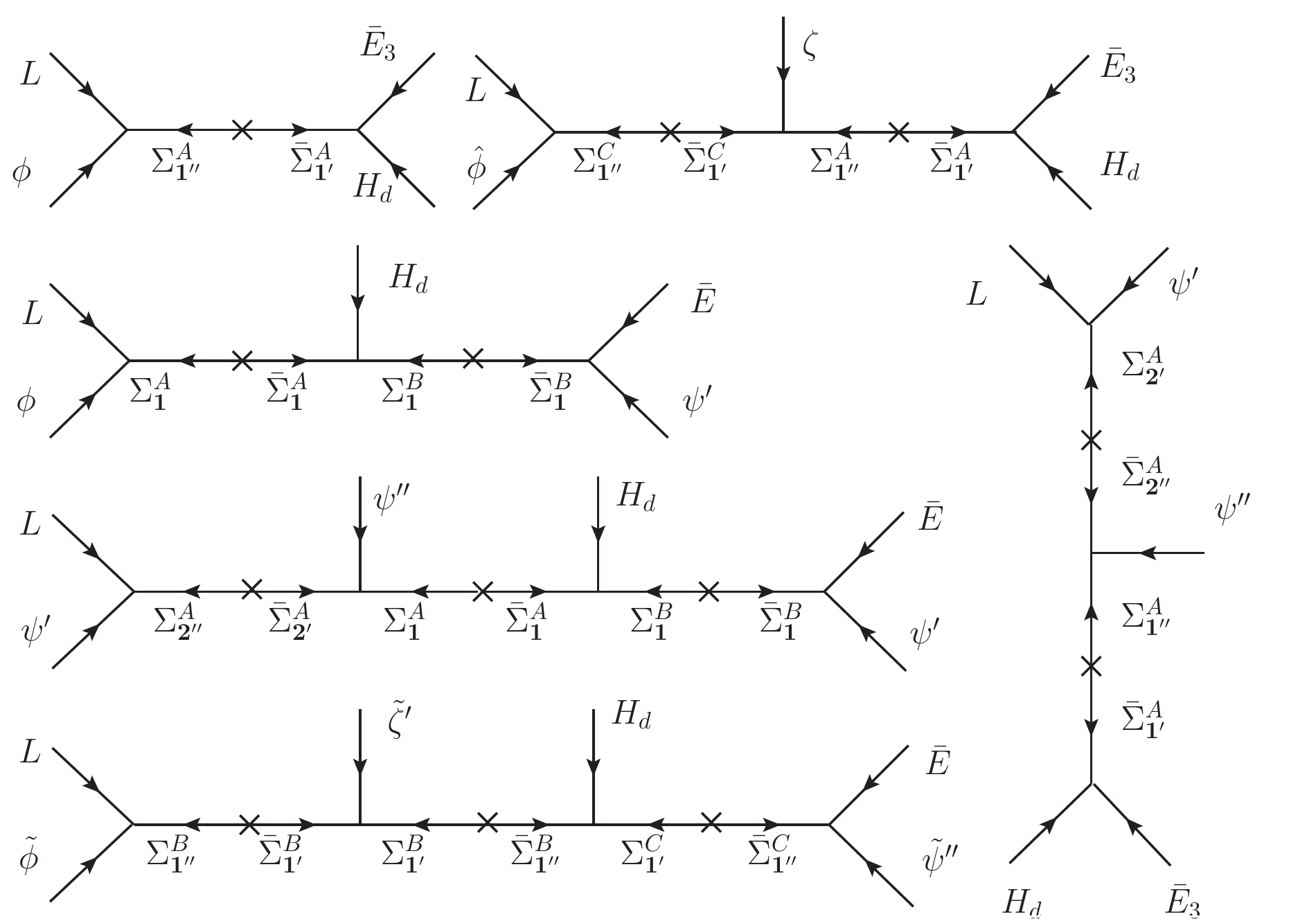}
\caption{The supergraphs before integrating out the messengers for the charged lepton sector.}
\label{Fig:Yesupergraphs}
\end{center}
\end{figure}

For the charged lepton sector we find the renormalisable
superpotential $\mathcal{W}_e^{\rm ren}$
\be
\begin{split}
\mathcal{W}_e^{\text{ren}} & =  L \, \phi \, \Sigma^A_{1^{\sec}} +
 L \, \phi \, \Sigma^A_{1}  + \bar E_3 \, H_d \, \bar \Sigma^A_{1^{\pr}}
 + L \, \hat \phi \, \Sigma^C_{1^{\sec}} + \zeta \,
 \bar \Sigma_{1^{\pr}}^C \, \Sigma_{1^{\sec}}^A  +
 H_d \, \bar \Sigma_{1}^A \, \Sigma_{1}^B\\
& + \bar E \, \psi^{\pr} \, \bar \Sigma_{1}^B +
L \, \psi^{\pr} \, \Sigma_{2^{\sec}}^A +
\psi^{\sec} \, \bar \Sigma_{2^{\pr}}^A \, \Sigma_{1}^A +
L \, \tilde \phi \, \Sigma_{1^{\pr}}^B +
\tilde \zeta^{\pr} \, \bar \Sigma_{1^{\pr}}^B \, \Sigma_{1^{\pr}}^B\\
& + H_d \, \bar \Sigma_{1^{\sec}}^B \, \Sigma_{1^{\pr}}^C +
\bar E \, \tilde \psi^{\sec} \, \bar \Sigma_{1^{\sec}}^C
+ L \, \psi^{\pr} \, \Sigma_{2^{\pr}}^A + \psi^{\sec} \,
\bar \Sigma_{2^{\sec}}^A \, \Sigma_{1^{\sec}}^A \;,
\end{split}
\ee
which through the diagrams of Fig.~\ref{Fig:Yesupergraphs}
generates at low energy the non-renormalisable superpotential
$\mathcal{W}_{{\rm Y}_e}$ of Eq.~\eqref{eq:Yde}.

\begin{figure}
\begin{center}
\includegraphics[scale=0.7]{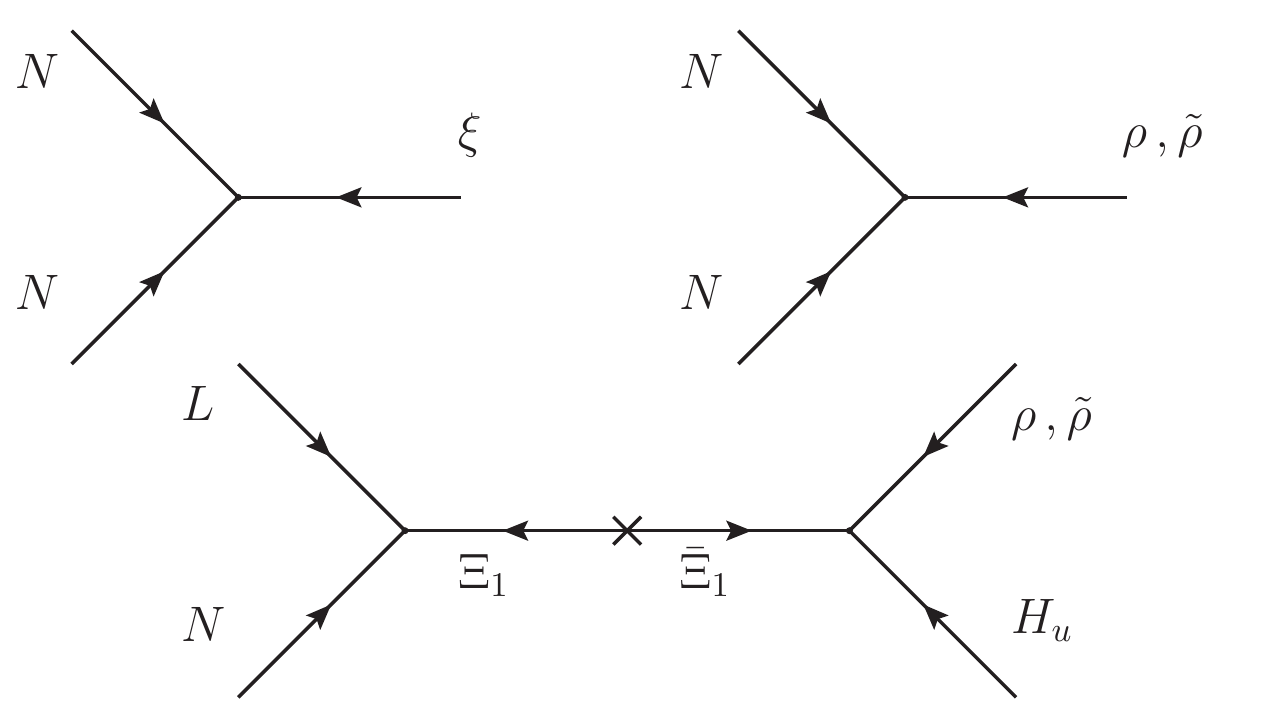}
\caption{The supergraphs before integrating out the messengers for the neutrino sector.}
\label{Fig:nusupergraphs}
\end{center}
\end{figure}

\begin{figure}
\begin{center}
\includegraphics[scale=0.65]{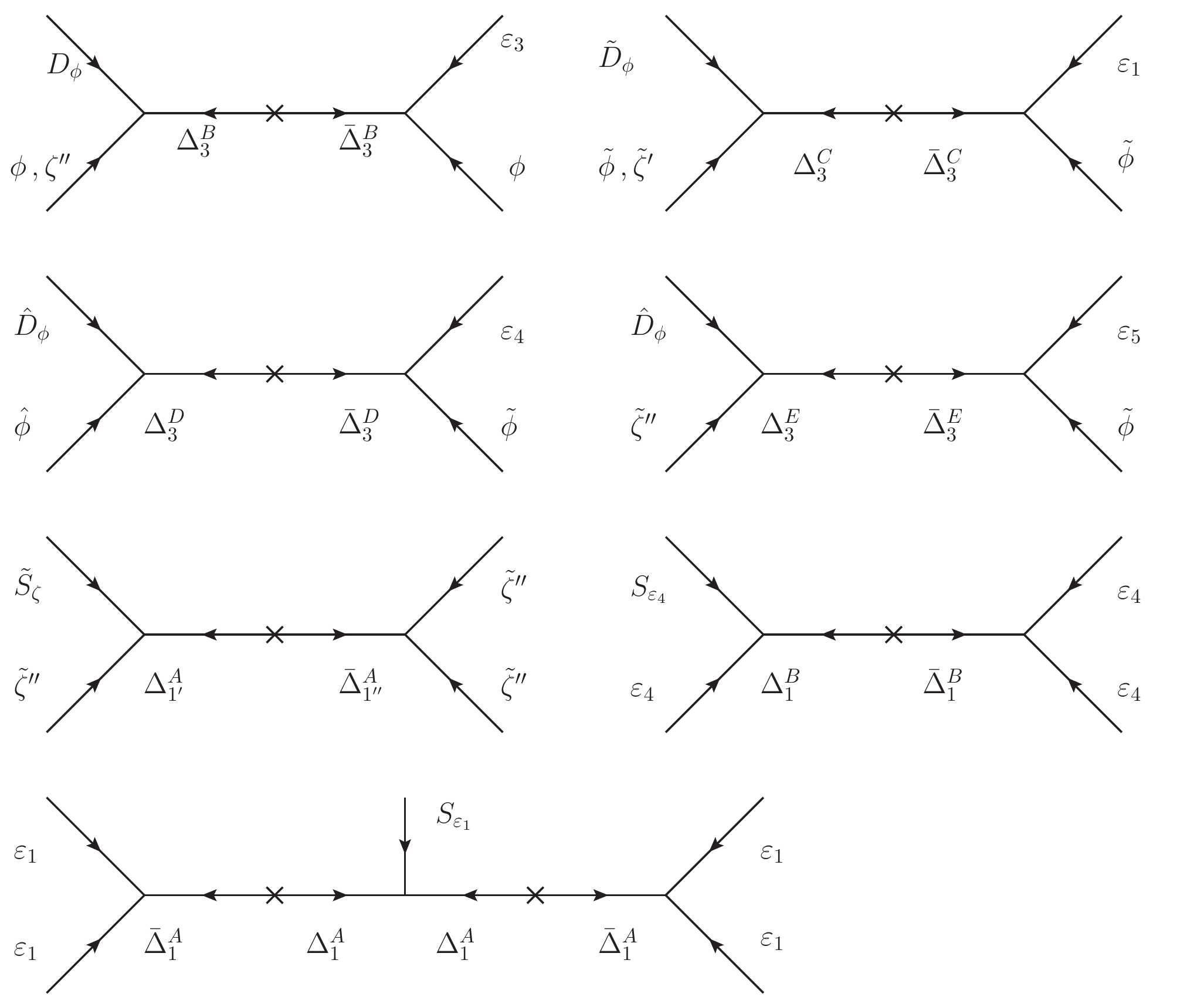}
\caption{The supergraphs before integrating out the messengers for the flavon sector.
We have omitted for simplicity the supergraphs of the higher order
corrections in the flavon superpotential.
}
\label{Fig:flavsupergraphs}
\end{center}
\end{figure}

For the neutrino and the flavon sector we obtained similarly to the previous case
\begin{align}
\mathcal{W}_{\nu}^{\text{ren}} &= N^2 \xi + N^2 \rho + N^2 \tilde \rho + L N \Xi_{1} + H_u \bar \Xi_{1} \rho
+ H_u \bar \Xi_{1} \tilde \rho \;, \\
\begin{split}
\mathcal{W}_{\text{flavon}}^{\text{ren}}  & =  D_{\phi} \, \phi \, \Delta_3^B + \varepsilon_3 \, \phi \, \bar \Delta_3^B + D_{\phi} \, \zeta^{\sec} \, \Delta_3^B
+ \tilde D_{\phi} \, \tilde \phi \, \Delta_3^C + \varepsilon_1 \,\tilde \phi \, \bar \Delta_3^C +  \tilde D_{\phi} \, \tilde \zeta^{\pr} \, \Delta_3^C \\
& + \hat D_{\phi} \, \hat \phi \, \Delta_3^D + \varepsilon_4 \, \hat \phi \, \bar \Delta_3^D + \hat D_{\phi} \, \tilde \zeta^{\sec} \, \Delta_3^E +
\varepsilon_5 \, \tilde \phi \,  \bar \Delta_3^E + \tilde S_{\zeta} \, \tilde \zeta^{\sec} \, \Delta_{1^{\pr}}^A + \tilde \zeta^{\sec} \,  \tilde \zeta^{\sec} \, \bar \Delta_{1^{\sec}}^A \\
& + S_{\varepsilon_4} \, \varepsilon_4 \, \Delta_1^B + \varepsilon_4 \, \varepsilon_4 \, \bar \Delta_1^B + \varepsilon_1 \, \varepsilon_1 \, \bar \Delta_1^A +
S_{\varepsilon_1} \, \Delta_1^A \, \Delta_1^A \;.
\end{split}
\end{align}
The corresponding  diagrams that generate the effective
operators in the neutrino and flavon sector in our model
are given in Figs.~\ref{Fig:nusupergraphs} and
\ref{Fig:flavsupergraphs}.

\begin{thebibliography}{99}

\bibitem{PDG2012}
K. Nakamura and S.~T. Petcov,
in J. Beringer {\it et al.} (Particle Data Group),
Phys. Rev. D {\bf 86} (2012) 010001.


\bibitem{BHP80} S.M.~Bilenky, J.~Hosek and S.T.~Petcov,
              Phys.\ Lett. B {\bf 94} (1980) 495.

\bibitem{EMSPEJP09}
  E.~Molinaro and S.~T.~Petcov,
  Eur.\ Phys.\ J.\  C {\bf 61} (2009) 93.

\bibitem{Ibarra:2011xn}
  A.~Ibarra, E.~Molinaro and S.~T.~Petcov,
Phys. Rev. D {\bf 84} (2011) 013005.   


\bibitem{LW81} L. Wolfenstein,  Phys. Lett. B {\bf 107} (1981) 77;
S.M.\ Bilenky, N.P.\ Nedelcheva and
S.T.\ Petcov,  Nucl. Phys. B\textbf{ 247} (1984) 61;
B. Kayser, Phys. Rev. D {\bf 30} (1984) 1023.


\bibitem{Fogli:2012ua}
  G.~L.~Fogli, E.~Lisi, A.~Marrone, D.~Montanino,
A.~Palazzo and A.~M.~Rotunno,
  Phys.\ Rev.\ D {\bf 86} (2012) 013012.

\bibitem{Gonzales-Garcia:2012}
M. C. Gonzalez-Garcia, M. Maltoni, J. Salvado and T. Schwetz,
JHEP {\bf 1212} (2012) 123.


\bibitem{GTani02}
C.~Giunti and M.~Tanimoto,
Phys.\ Rev.\ D {\bf 66} (2002) 053013,
Phys.\ Rev.\ D {\bf 66} (2002) 113006.

\bibitem{FPR04}
P.H. Frampton, S.T. Petcov and W. Rodejohann,
Nucl. Phys. B {\bf 687} (2004) 31.

\bibitem{SPWR04}
S.T. Petcov and W. Rodejohann,
Phys. Rev. {\bf D71} (2005) 073002.

\bibitem{Romanino:2004ww}
  A.~Romanino,
  Phys.\ Rev.\  D {\bf 70} (2004) 013003.

\bibitem{HPR07}   K.A. Hochmuth, S.T. Petcov and W. Rodejohann,
Phys. Lett B {\bf 654} (2007) 177.

\bibitem{Marzocca:2011dh}
 D.~Marzocca, S.~T.~Petcov, A.~Romanino, M.~Spinrath,
JHEP {\bf 11}  (2011) 009.

\bibitem{Marzocca:2013cr}
  D.~Marzocca, S.~T.~Petcov, A.~Romanino and M.~C.~Sevilla,
  JHEP {\bf 1305} (2013) 073.


\bibitem{Alta} G.~Altarelli, F.~Feruglio and I.~Masina,
  Nucl.\ Phys.\ B {\bf 689} (2004) 157;
 S.~F.~King,
  JHEP {\bf 0508} (2005) 105;
I.~Masina,
  Phys.\ Lett.\  B {\bf 633} (2006) 134;
S.~Antusch and S.~F.~King,
  Phys.\ Lett.\ B {\bf 631} (2005) 42;
  S.~Dev, S.~Gupta and R.~R.~Gautam,
  Phys.\ Lett.\ B {\bf 704} (2011) 527;
  S.~Antusch and V.~Maurer,
  Phys.\ Rev.\ D {\bf 84} (2011) 117301;
 A. Meroni, S.T. Petcov and M. Spinrath,
 Phys. Rev. D {\bf 86} (2012) 113003;
  C.~Duarah, A.~Das and N.~N.~Singh,
 [arXiv:{1210.8265}]. 

\bibitem{Chao:2011sp}
  W.~Chao and Y.~-j.~Zheng,
  JHEP {\bf 1302} (2013) 044;
  D.~Meloni,
  JHEP {\bf 1202} (2012) 090;
  S.~Antusch, C.~Gross, V.~Maurer and C.~Sluka,
  Nucl.\ Phys.\ B {\bf 866} (2013) 255;
  G.~Altarelli, F.~Feruglio, L.~Merlo and E.~Stamou,
  JHEP {\bf 1208} (2012) 021;
  G.~Altarelli, F.~Feruglio and L.~Merlo,
  [arXiv:{1205.5133}]; 
  F.~Bazzocchi and L.~Merlo,
  [arXiv:{1205.5135}]; 
  S.~Gollu, K.~N.~Deepthi and R.~Mohanta,
  [arXiv:{1303.3393}]. 

\bibitem{AlbRode2010}
C. H. Albright, A. Dueck and W. Rodejohann,
Eur.\ Phys.\ J. \ C {\bf 70} (2010) 1099.

\bibitem{TBM}
  P.~F.~Harrison, D.~H.~Perkins and W.~G.~Scott,
  Phys.\ Lett.\ B {\bf 530} (2002) 167;
  Phys.\ Lett.\ B {\bf 535} (2002) 163;
  Z.~Z.~Xing,
  Phys.\ Lett.\ B {\bf 533} (2002) 85;
  X.~G.~He and A.~Zee,
  Phys.\ Lett.\ B {\bf 560} (2003) 87;
see also
L.~Wolfenstein,
  Phys.\ Rev.\ D {\bf 18} (1978) 958.

\bibitem{SPPD82} S.T. Petcov, Phys.\ Lett.\ B {\bf 110} (1982) 245.

\bibitem{BM}
F.~Vissani, 
[arXiv:{hep-ph/9708483}];
V.~D.~Barger, S.~Pakvasa, T.~J.~Weiler and K.~Whisnant,
Phys.\ Lett.\ B {\bf 437} (1998) 107;
A.~J.~Baltz, A.~S.~Goldhaber and M.~Goldhaber,
Phys.\ Rev.\ Lett.\  {\bf 81} (1998) 5730.


\bibitem{King:2013eh}
  S.~F.~King and C.~Luhn,
  Rept.\ Prog.\ Phys.\  {\bf 76} (2013) 056201.

\bibitem{Alta:2010ab}
G. Altarelli and F. Feruglio, 
Rev.\ Mod.\ Phys. {\bf 82} (2010) 2701.


\bibitem{Tani:2010cd} 
 H. Ishimori {\it et al.},
Prog.\ Theor.\ Phys.\ Suppl. {\bf 183} (2010) 1.


\bibitem{Feruglio:2007uu}
  P.~H.~Frampton and T.~W.~Kephart,
  Int.\ J.\ Mod.\ Phys.\ A {\bf 10} (1995) 4689;
 F.~Feruglio, C.~Hagedorn, Y.~Lin and L.~Merlo,
  Nucl.\ Phys.\ B {\bf 775} (2007) 120
   [Erratum-ibid.\  {\bf 836} (2010) 127];
 G.~J.~Ding,
 Phys.\ Rev.\  D {\bf 78} (2008) 036011;
  P.~H.~Frampton, T.~W.~Kephart and S.~Matsuzaki,
  Phys.\ Rev.\ D {\bf 78} (2008) 073004;
  D.~A.~Eby, P.~H.~Frampton and S.~Matsuzaki,
  Phys.\ Lett.\ B {\bf 671} (2009) 386;
  P.~H.~Frampton and S.~Matsuzaki,
  Phys.\ Lett.\ B {\bf 679} (2009) 347.

\bibitem{Chen:2007afa}
  M.-C.~Chen and K.~T.~Mahanthappa,
  Phys.\ Lett.\ B {\bf 652} (2007) 34.

\bibitem{Chen:2009gf}
  M.-C.~Chen and K.~T.~Mahanthappa,
  Phys. Lett. B {\bf 681} (2009) 444.



\bibitem{DayaBay}
  F.~P.~An {\it et al.}  [DAYA-BAY Collaboration],
  Phys.\ Rev.\ Lett.\  {\bf 108} (2012) 171803.
   
\bibitem{RENO}
  J.~K.~Ahn {\it et al.}  [RENO Collaboration],
  Phys.\ Rev.\ Lett.\  {\bf 108} (2012) 191802.


\bibitem{Abe:2011sj}
  K.~Abe {\it et al.}  [T2K Collaboration],
  Phys.\ Rev.\ Lett.\  {\bf 107} (2011) 041801.


\bibitem{Abe:2011fz}
  Y.~Abe {\it et al.}  [DOUBLE-CHOOZ Collaboration],
  Phys.\ Rev.\ Lett.\  {\bf 108} (2012) 131801;
  Y.~Abe {\it et al.}  [DOUBLE-CHOOZ Collaboration],
  Phys.\ Rev.\ D {\bf 86} (2012) 052008.

\bibitem{Adamson:2011qu}
  P.~Adamson {\it et al.}  [MINOS Collaboration],
  Phys.\ Rev.\ Lett.\  {\bf 107} (2011) 181802.

\bibitem{seesaw}
P.~Minkowski,
  Phys.\ Lett.\ B {\bf 67} (1977) 421;
M. Gell-Mann, P. Ramond and R. Slansky in Sanibel Talk,
CALT-68-709, Feb 1979, and in {\it Supergravity} (North Holland,
Amsterdam 1979);
T. Yanagida in {\it Proc. of the Workshop on Unified Theory and
Baryon Number of the Universe}, KEK, Japan, 1979;
S.L.Glashow, Cargese Lectures (1979);
R.~N.~Mohapatra and G.~Senjanovic,
Phys.\ Rev.\ Lett.\  {\bf 44} (1980) 912.

\bibitem{cg}
J.-Q.~Chen and P.-D.~Fan, J.~Math.~Phys.~{\bf 39} (1998) 5519.

\bibitem{Branco:1983tn}
  G.~C.~Branco, J.~M.~Gerard and W.~Grimus,
  Phys.\ Lett.\ B {\bf 136} (1984) 383.

\bibitem{Meroni:2012ty}
  A.~Meroni, S.~T.~Petcov and M.~Spinrath,
  Phys.\ Rev.\ D {\bf 86} (2012) 113003.

\bibitem{Chen:2013wba}
  M.~-C.~Chen, J.~Huang, K.~T.~Mahanthappa and A.~M.~Wijangco,
  arXiv:1307.7711. 
 
\bibitem{Antusch:2011sx}
  S.~Antusch, S.~F.~King, C.~Luhn and M.~Spinrath,
  Nucl.\ Phys.\ B {\bf 850} (2011) 477.

\bibitem{Antusch:2013wn}
  S.~Antusch, S.~F.~King and M.~Spinrath,
  Phys.\  Rev.\  D {\bf 87} (2013) 096018;
  S.~F.~King,
  JHEP {\bf 1307} (2013) 137 and
  arXiv:1305.4846; 
  S.~Antusch, C.~Gross, V.~Maurer and C.~Sluka,
  arXiv:1305.6612 and
  arXiv:1306.3984. 

\bibitem{Antusch:2011qg}
  S.~Antusch and V.~Maurer,
  Phys.\ Rev.\ D {\bf 84} (2011) 117301.


\bibitem{Antusch:2009gu}
  S.~Antusch and M.~Spinrath,
  Phys.\ Rev.\ D {\bf 79} (2009) 095004;
  M.~Spinrath,
  arXiv:1009.2511; 
S.~Antusch, S.~F.~King and M.~Spinrath,
  arXiv:1311.0877. 
 

\bibitem{Holthausen:2012dk}
  M.~Holthausen, M.~Lindner and M.~A.~Schmidt,
  JHEP {\bf 1304} (2013) 122.

\bibitem{Feruglio:2012cw}
  F.~Feruglio, C.~Hagedorn and R.~Ziegler,
  JHEP {\bf 1307} (2013) 027.


\bibitem{Aranda:2000tm}
  A.~Aranda, C.~D.~Carone and R.~F.~Lebed,
  Phys.\ Rev.\ D {\bf 62} (2000) 016009.


\bibitem{Ding:2013bpa}
  G.J.~Ding, S.~F.~King and A.~J.~Stuart,
  arXiv:1307.4212.

\bibitem{Grimus:2011fk}
  W.~Grimus and P.~O.~Ludl,
  J.\ Phys.\ A {\bf 45} (2012) 233001.



\bibitem{PKSP3nu88}
P.I.~Krastev and S.~T.~Petcov,
  Phys.\ Lett.\  B {\bf 205} (1988) 84.

\bibitem{AMEMSPSU5T} A. Meroni, E. Molinaro and S.T. Petcov,
Phys.\ Lett.\ B {\bf 710} (2012) 435;


\bibitem{BiPet87}  S.~M.~Bilenky and S.~T.~Petcov,
Rev. Mod. Phys. {\bf 59} (1987) 671;
%
W. Rodejohann, 
Int. J. Mod. Phys. {\bf E20} (2011) 1833.


\bibitem{Agostini:2013mba}
  M.~Agostini {\it et al.},
  arXiv:1307.4720. 


\bibitem{MainzKATRIN} K. Eitel {\it et al.},
Nucl. Phys. Proc. Suppl. {\bf 143} (2005) 197.

\bibitem{Capozzi:2013csa}
  F.~Capozzi, G.~L.~Fogli, E.~Lisi, A.~Marrone, D.~Montanino and A.~Palazzo,
  arXiv:1312.2878 [hep-ph].
  
\bibitem{TAUP}  
  M.~C.~Gonzales-Garcia et al., talk given at the International Workshop
  TAUP 2013, September 9-13, 2013, Asilomar, USA.
  
\end{thebibliography}
\end{document}